\def\isarxiv{1}

\ifdefined\isarxiv
\documentclass[11pt]{article}
\else
\documentclass{article}
\usepackage{hyperref} 
\usepackage[accepted]{icml2021}
\fi

\usepackage[utf8]{inputenc} 
\usepackage[T1]{fontenc}    
\usepackage{url}            
\usepackage{booktabs}       
\usepackage{amsfonts}       

\usepackage{amsmath}
\usepackage{amsthm}
\usepackage{amssymb}
\usepackage{algorithm}
\usepackage{subfig}
\usepackage{color}
\usepackage[english]{babel}
\usepackage{graphicx}
\usepackage{grffile}
\usepackage{wrapfig,epsfig}
\usepackage{epstopdf}
\usepackage{url}
\usepackage{color}
\usepackage{epstopdf}
\usepackage{algpseudocode}
\usepackage[T1]{fontenc}
\usepackage{bbm}
\usepackage{comment}
\usepackage{dsfont}

\usepackage{tikz}

\usetikzlibrary{arrows}
\ifdefined\isarxiv
\usepackage{hyperref}  
\hypersetup{colorlinks=true,citecolor=red,linkcolor=blue}
\usepackage[margin=1in]{geometry}
\else 

\fi
\graphicspath{{./figs/}}
\usepackage{mathtools}

\newtheorem{theorem}{Theorem}[section]
\newtheorem{lemma}[theorem]{Lemma}
\newtheorem{definition}[theorem]{Definition}

\newtheorem{corollary}[theorem]{Corollary}

\newtheorem{observation}[theorem]{Observation}
\newtheorem{fact}[theorem]{Fact}
\newtheorem{remark}[theorem]{Remark}
\newtheorem{claim}[theorem]{Claim}

\newcommand{\wh}{\widehat}
\newcommand{\wt}{\widetilde}
\newcommand{\ov}{\overline}

\newcommand{\N}{\mathbb{N}}
\newcommand{\R}{\mathbb{R}}

\newcommand{\floor}[1]{\lfloor #1 \rfloor}
\newcommand{\ceil}[1]{\lceil #1 \rceil}
\newcommand{\set}[1]{\left\lbrace #1\right\rbrace}

\renewcommand{\varepsilon}{\epsilon}
\renewcommand{\tilde}{\wt}
\renewcommand{\hat}{\wh}
\renewcommand{\bar}{\overline}

\renewcommand{\d}{\mathrm{d}}
\newcommand{\op}{\mathrm{op}}
\newcommand{\inner}[2]{\left\langle #1, #2\right\rangle}
\newcommand{\ose}{$\mathsf{OSE}$}
\newcommand{\srht}{$\mathsf{SRHT}$}
\newcommand{\tensorsrht}{$\mathsf{TensorSRHT}$}
\newcommand{\ntk}{$\mathsf{NTK}$}
\newcommand{\krr}{$\mathsf{KRR}$}

\DeclareMathOperator*{\E}{{\mathbb{E}}}
\DeclareMathOperator*{\Var}{{\bf {Var}}}

\DeclareMathOperator{\poly}{poly}
\DeclareMathOperator{\Tr}{tr}
\DeclareMathOperator{\nnz}{nnz}

\DeclareMathOperator{\rank}{rank}

\DeclareMathOperator{\tr}{tr}

\DeclareMathOperator{\diag}{diag}

\DeclareMathOperator{\pr}{Pr}

\renewcommand{\k}{\mathsf{K}}

\definecolor{b2}{RGB}{51,153,255}
\definecolor{mygreen}{RGB}{80,180,0}

\makeatletter
\newcommand*{\RN}[1]{\expandafter\@slowromancap\romannumeral #1@}
\makeatother

\ifdefined\isarxiv
\title{Fast Sketching of Polynomial Kernels of Polynomial Degree\thanks{A preliminary version of this paper appeared in the Proceedings of the 38th International Conference on Machine Learning (ICML 2021).}}

\author{
Zhao Song\thanks{\texttt{magic.linuxkde@gmail.com}. Princeton University and Institute for Advanced Study}
\and David P. Woodruff\thanks{\texttt{dwoodruf@andrew.cmu.edu}. Carnegie Mellon University.} 
\and Zheng Yu\thanks{\texttt{zhengy@princeton.edu}. Princeton University.}
\and Lichen Zhang\thanks{\texttt{lichenz@andrew.cmu.edu}. Carnegie Mellon University.}
}
\date{}
\else
\date{
}

\fi

\begin{document}

\ifdefined\isarxiv
\else
\twocolumn[
\icmltitle{Fast Sketching of Polynomial Kernels of Polynomial Degree}




\begin{icmlauthorlist}
\icmlauthor{Zhao Song}{a}
\icmlauthor{David P. Woodruff}{b}
\icmlauthor{Zheng Yu}{c}
\icmlauthor{Lichen Zhang}{d}
\end{icmlauthorlist}

\icmlaffiliation{a}{Princeton University and Institute for Advanced Study }
\icmlaffiliation{b}{Carnegie Mellon University}
\icmlaffiliation{c}{Princeton University}
\icmlaffiliation{d}{Carnegie Mellon University}

\icmlcorrespondingauthor{Zhao Song}{magic.linuxkde@gmail.com}
\icmlcorrespondingauthor{David P. Woodruff}{dwoodruf@andrew.cmu.edu}
\icmlcorrespondingauthor{Zheng Yu}{zhengyu@princeton.edu}
\icmlcorrespondingauthor{Lichen Zhang}{lichenz@andrew.cmu.edu}

\icmlkeywords{Machine Learning, ICML}

\vskip 0.3in
]



\printAffiliationsAndNotice{}  
\fi

\ifdefined\isarxiv
\begin{titlepage}
  \maketitle
  \begin{abstract}
  Kernel methods are fundamental in machine learning, and faster algorithms for kernel approximation provide direct speedups for many core tasks in machine learning. The polynomial kernel is especially important as other kernels can often be approximated by the polynomial kernel via a Taylor series expansion. Recent techniques in oblivious sketching reduce the dependence in the running time on the degree $q$ of the polynomial kernel from exponential to polynomial, which is useful for the Gaussian kernel, for which $q$ can be chosen to be polylogarithmic (for data of bounded radius and ridge parameter $\lambda\geq 1/\poly(n)$). However, for more slowly growing kernels, such as the neural tangent and arc-cosine kernels, $q$ needs to be polynomial, and previous work incurs a polynomial factor slowdown in the running time. We give a new oblivious sketch which greatly improves upon this running time, by removing the dependence on $q$ from the $nd$ term, which is the leading order term when the dimension $d$ is large. For kernels approximated through a truncated Taylor series, our guarantees hold up to a ridge term $\lambda I$. Combined with a sampling scheme over the degrees of the Taylor series, we further reduce the running time for slowly growing kernels such as the neural tangent kernel, for which we obtain time $\tilde O(nd+\epsilon^{-4}\lambda^{-4}n^6)$ on $n$ unit-norm points with accuracy $\epsilon$. Since the neural tangent kernel has a closed form that can be evaluated exactly in $O(n^2d)$ time, this bound is faster only when $d$ is at least about $(n/(\epsilon\lambda))^4$.

  \end{abstract}
  \thispagestyle{empty}
\end{titlepage}

\pagenumbering{roman}
{
\tableofcontents
}

\newpage
\else
\begin{abstract}
Kernel methods are fundamental in machine learning, and faster algorithms for kernel approximation provide direct speedups for many core tasks in machine learning. The polynomial kernel is especially important as other kernels can often be approximated by the polynomial kernel via a Taylor series expansion. Recent techniques in oblivious sketching reduce the dependence in the running time on the degree $q$ of the polynomial kernel from exponential to polynomial, which is useful for the Gaussian kernel, for which $q$ can be chosen to be polylogarithmic (for data of bounded radius and ridge parameter $\lambda\geq 1/\poly(n)$). However, for more slowly growing kernels, such as the neural tangent and arc-cosine kernels, $q$ needs to be polynomial, and previous work incurs a polynomial factor slowdown in the running time. We give a new oblivious sketch which greatly improves upon this running time, by removing the dependence on $q$ from the $nd$ term, which is the leading order term when the dimension $d$ is large. For kernels approximated through a truncated Taylor series, our guarantees hold up to a ridge term $\lambda I$. Combined with a sampling scheme over the degrees of the Taylor series, we further reduce the running time for slowly growing kernels such as the neural tangent kernel, for which we obtain time $\tilde O(nd+\epsilon^{-4}\lambda^{-4}n^6)$ on $n$ unit-norm points with accuracy $\epsilon$. Since the neural tangent kernel has a closed form that can be evaluated exactly in $O(n^2d)$ time, this bound is faster only when $d$ is at least about $(n/(\epsilon\lambda))^4$.

\end{abstract}
\fi

\ifdefined\isarxiv
\pagenumbering{arabic}
\setcounter{page}{1}
\else

\fi

\section{Introduction}

Kernel methods are a powerful tool for solving non-parametric learning problems, such as kernel regression, support vector machines (SVM), principal component analysis (PCA), and many others. A typical burden for kernel methods is that they suffer from scalability, since computing a kernel matrix requires computing a quadratic (in the number of input points) number of entries in the matrix. A direction that has received less attention but still of particular interest is the regime where the dimension $d$ of the data points is large. Typically, applying the kernel function to each pair of data points takes $O(d)$ time. This is especially undesirable in applications for natural language processing \cite{dl20} and computational biology \cite{tpk02}, where $d$ can be as large as $\poly(n)$, with $n$ being the number of data points. To compute the kernel matrix, the algorithm does have to read the $d\times n$ input matrix. Therefore, algorithms that have a nearly linear dependence on $nd$ are of particular interest.

To accelerate the computation of kernel matrices from the na\"ive $O(n^2d)$ time algorithm, a lot of work has focused on finding a good approximation to a kernel matrix efficiently \cite{rr07,am15,mm17,akk+20,wz20}. All of these methods make use of randomized algorithmic primitives such as sampling or sketching. Roughly speaking, the idea is to randomly generate a ``sketching matrix" with a small number of rows, multiply the sketching matrix with the input matrix, and show that the resulting matrix approximately preserves the length of vectors in the row or column space of the original matrix. 

The polynomial kernel is of interest, since any kernel can be written as a sum of polynomial kernels through a Taylor expansion. If we can efficiently approximate the polynomial kernel, then we will be able to efficiently approximate many types of kernels. In \cite{anw14}, Avron, Nguyen and Woodruff approximate the polynomial kernel with a sketching matrix in time that depends exponentially on the degree $p$ of the polynomial kernel. Recent works of \cite{akk+20, wz20} have improved the dependence on $p$ to polynomial. However, their algorithms mainly focus on optimizing the dependence on $n$ and reducing the exponential dependence on $p$ to polynomial. If $X\in \R^{d\times n}$ is the input matrix and is dense, then the algorithm of \cite{akk+20} has runtime $\tilde O(pnd+\epsilon^{-2}n^3p^2)$, or $\tilde O(\epsilon^{-2}p^5(nd+n^2))$ with their Theorem 3, and the algorithm of \cite{wz20} has runtime $\tilde O(p^{2.5}nd+\poly(\epsilon^{-1},p)n^3)$.\footnote{We use $\tilde O(\cdot ), \tilde \Omega(\cdot ), \tilde \Theta(\cdot)$ to suppress $\poly(\log(ndp/(\epsilon\delta)))$ factors. In the Taylor-series results of Section~\ref{sec:application}, where the largest degree is $q$, they suppress $\poly(\log(ndq/(\epsilon\delta)))$ factors, which include $\poly(\log(1/\min(\lambda,1)))$ factors.} Notice these bounds are unsatisfactory when both $d$ and $p$ are large.

Thus, a natural question to ask is:
\begin{center}
{\it Does there exist a sketch for polynomial kernels of degree $p$, such that the runtime is nearly linear in $nd$, and with an improved dependence on $p$?}
\end{center}

Notice this is especially desirable for kernels such as the neural tangent kernel ({\ntk}) \cite{jgh18} and the arc-cosine kernel \cite{cs09}, whose Taylor series have a much slower decay rate ($1/n^c$ for some $c$) compared to the Gaussian kernel (which is $1/n!$).  

We list our contributions as follows:
\begin{itemize}
\item We develop an efficient algorithm that computes a sketch of the polynomial kernel of degree $p$ in time linear in $p^2$ and nearly linear in $nd$. 
\item Our algorithm uses one sketch at the leaves and one independent sketch per level of the tree, plus one for each step that merges two partial results when $p$ is not a power of $2$, so $O(\log p)$ independent sketches in total, compared to the $O(p)$ independent sketches of \cite{akk+20}. Nodes within a level share a sketch, which lets us use repeated powering to compute our sketch very efficiently. 
\item We characterize kernel matrices by considering their Taylor series, and provide two algorithmic schemes for them: one sketches every term of a truncation of the series, and one, for kernels whose Taylor coefficients decay slowly, sketches the low-degree terms and replaces the remaining terms of the truncation by a few randomly sampled degrees. These guarantees are ridge-regularized: for a ridge parameter $\lambda>0$, the sketch $W$ satisfies $(1-\epsilon)(K+\lambda I_n)\preceq W^\top W+\lambda I_n\preceq (1+\epsilon)(K+\lambda I_n)$. Our characterization includes a family of interesting and popular kernels. We also use this sketch as a preconditioner for solving linear systems involving a kernel matrix, and we extend our sketch to solve kernel ridge regression, by composing it with another sketch that depends on the statistical dimension. 
\item For the {\ntk} on unit-norm data, a version of the sampling scheme that groups the degrees into bands of doubling length runs in time $\tilde O(nd+\epsilon^{-4}\lambda^{-4}n^6)$, compared to $\tilde O(nd+\epsilon^{-8}\lambda^{-6}n^8)$ for sketching every term. We also show a matching lower bound, up to polylogarithmic factors, for schemes that choose the degrees and their weights independently of the data, fail with probability at most $1/8$, and except with probability at most $1/8$ sketch every chosen degree to relative accuracy $1/2$, in the cost model of our sketch, when $\epsilon\leq 1/4$ and $\epsilon\lambda$ is at most a small constant times $n$. Since the closed form of the {\ntk} gives the kernel matrix exactly in $O(n^2d)$ time, the sampling scheme is faster only when $d$ is at least about $(n/(\epsilon\lambda))^4$, and when the kernel matrix itself is the output, a random projection of the data followed by the closed form is cheaper still (Section~\ref{sec:app_pconv}).
\end{itemize}

\subsection{Related Work}

\paragraph{Kernel regression}
Classical regression has the form $\min_{w} \| Y - X w \|_2^2$, where $X$ and $Y$ are a given dataset and corresponding labels, respectively. Kernel regression \cite{b13,zdw15,am15,acw17,akmmvz17,znvkr20,lsswy20,acss20} allows for $X$ to be a kernel matrix $K\in \R^{n\times n}$, where each entry is the application of a kernel function to a pair of data points in $X$. Kernel regression $\min_{w} \| Y - K w \|_2^2$ enables fitting non-linear data into a hyperplane by transforming the data into a high-dimensional space.

\paragraph{Sketching techniques for tensor-related problems}
Sketching techniques have been used extensively in tensor-related problems, e.g., for linear-algebraic problems involving polynomial kernels \cite{anw14,akk+20,wz20}, for tensor low-rank approximation \cite{swz19_soda}, and for tensor regression \cite{hlw17,dssw18,djssw19}.

\paragraph{Subspace embeddings}
An (oblivious) subspace embedding is a useful concept in randomized numerical linear algebra introduced by S\'arlos \cite{s06}. Many applications rely on subspace embeddings or their variants, such as linear regression, low-rank approximation \cite{cw13,nn13,mm13,bw14,bwz16,swz17,alszz18}, tensor decomposition \cite{swz19_soda}, cutting plane methods \cite{jlsw20}, and linear programming \cite{lsz19,jswz20,sy21}

\vspace{-0.1in}
\paragraph{Roadmap}
In Section~\ref{sec:preli}, we introduce definitions, notations and some basic facts. In Section~\ref{sec:tech}, we present a technical overview of our results. In Section~\ref{sec:algorithm}, we propose an efficient algorithm to generate a sketch and apply it to a polynomial kernel of arbitrary positive integer degree $p$. In Section~\ref{sec:analysis}, we analyze our algorithm with a specific sketching matrix. In Section~\ref{sec:application}, we present applications to the Gaussian kernel and a more general class of kernels, which can be characterized through the coefficients of their Taylor expansion. We also discuss how to use our sketch as a preconditioner for solving kernel linear systems, and solve sketched kernel ridge regression.
\vspace{-0.1in}
\section{Preliminaries}\label{sec:preli}
For an integer $n$, let $[n]$ denote the set $\{1,2,\cdots, n\}$. For two scalars $a$ and $b$, we say $a \approx_{\epsilon} b$ if $(1-\epsilon) b \leq a \leq (1+\epsilon) b$. We say a square symmetric matrix $A$ is positive semi-definite (PSD) if $\forall x$, $x^\top A x \geq 0$. For two PSD matrices $A$ and $B$, we define $A \approx_{\epsilon} B$ if  $(1-\epsilon)B \preceq A \preceq  (1+\epsilon) B$, where $A\preceq B$ means $B-A$ is PSD. For a matrix $A$, we use $\| A \|_F = (\sum_{i,j} A_{i,j}^2)^{1/2}$ to denote its Frobenius norm and use $\| A \|_{\op}$ to denote its operator (spectral) norm. For a square symmetric matrix $A$, we use $\tr[A]$ to denote the trace of $A$. For a square symmetric matrix $A$, we use $\lambda_{\min}(A)$, $\lambda_{\max}(A)$ to denote its smallest and largest eigenvalues, respectively. For a rectangular matrix $A$, we use $\sigma_{\min}(A),\sigma_{\max}(A)$ to denote its smallest and largest singular values, respectively, and we use $\kappa=\frac{\sigma_{\max}(A)}{\sigma_{\min}(A)}$ to denote its condition number.
\vspace{-3mm}
\subsection{Definitions}
\vspace{-2mm}
We define an oblivious subspace embedding (\cite{s06}) as follows:
\begin{definition}[Oblivious Subspace Embedding(\ose)]\label{def:OSE}
Let $\epsilon,\delta\in (0,1)$ and $d,n\geq 1$ be integers. An $(\epsilon,\delta,d,n)$-Oblivious Subspace Embedding (\ose) is a distribution over $m\times d$ matrices with the guarantee that for any fixed matrix $A\in \R^{d\times n}$, we have
\begin{align*}
\underset{\Pi\sim D}{\pr}\left[  ( (\Pi A)^\top \Pi A ) \approx_{\epsilon} (A^\top A) \right] \geq 1-\delta .
\end{align*}
\end{definition}
We also introduce tensor products of vectors and Kronecker products of matrices.
\begin{definition}[Vector tensor product]
Given two vectors $x\in \R^n$ and $y\in \R^m$, we define the \textit{tensor product between $x$ and $y$}, denoted by $x\times y$, to be $\mathrm{vec}(xy^\top)$. We will use $x^{\otimes p}$ to denote the self-tensoring of the vector $x$ a total of $p$ times.
\end{definition}
The Kronecker product of matrices is a natural extension of the tensor product of vectors:
\begin{definition}
Given $A_1\in \R^{m_1\times n_1},A_2\in\R^{m_2\times n_2},\ldots,A_k\in\R^{m_k\times n_k}$, we define $A_1\times A_2\times \ldots \times A_k$ to be the matrix in $\R^{m_1m_2\ldots m_k\times n_1n_2\ldots n_k}$ whose element at row $(i_1,\ldots,i_k)$ and column $(j_1,\ldots,j_k)$ is $A_1(i_1,j_1)\ldots A_k(i_k,j_k)$.
\end{definition}
An important property of the Kronecker product is the so-called \textit{mixed product} property:
\begin{claim}[Mixed product]
For matrices $A,B,C,D$ with appropriate sizes, the following holds:
\begin{align*}
(A\cdot B)\times (C\cdot D)= (A\times C)\cdot (B\times D).
\end{align*}
\end{claim}
One consequence is the following claim:
\begin{claim}
\label{clm:mat_vec_prod}
Let $A_1\in \R^{m_1\times n_1},A_2\in\R^{m_2\times n_2},\ldots,A_k\in\R^{m_k\times n_k}$ and $v_1\in \R^{n_1},v_2\in \R^{n_2},\ldots,v_k\in\R^{n_k}$. Then, 
\begin{align*}
& ~ \left( A_1 \times A_2 \times \ldots \times A_k \right) \left( v_1 \times v_2 \times \ldots \times v_k \right) \\
= & ~ (A_1 v_1) \times (A_2 v_2)\times \ldots \times (A_k v_k).
\end{align*}
\end{claim}

We will extensively use the following notation: 
\begin{definition}
Given $A_1\in \R^{m_1\times n_1},A_2\in\R^{m_2\times n_2},\ldots,A_k\in\R^{m_k\times n_k}$, we define $A_1\otimes A_2\otimes \ldots \otimes A_k$ to be the matrix in $\R^{m_1m_2\ldots m_k\times n}$ whose $j^{th}$ column is $A^j_1 \times A^j_2 \times \ldots \times A^j_k$ for every $j\in [n]$, where $A^j_l$ is the $j^{th}$ column of $A_l$ for every $l\in [k]$.
\end{definition}
\vspace{-3mm}
\subsection{Sketching Matrices}
\vspace{-2mm}
We recall the Subsampled Randomized Hadamard Transform (\srht), which is a Fast Johnson-Lindenstrauss transform~\cite{ac06}.
\begin{definition}[Subsampled Randomized Hadamard Transform (\srht), see \cite{ldfu13,w14}]
\label{def:SRHT}
The {\srht} matrix $S\in \R^{m\times d}$ is defined as $S =\frac{1}{\sqrt m}PHD$, where each row of matrix $P\in \{0,1\}^{m\times d}$  contains exactly one $1$ at a random position, $H$ is the $d\times d$ Hadamard matrix, and $D$ is a $d\times d$ diagonal matrix with each diagonal entry being a value in $\{-1,+1\}$ with equal probability.
\begin{remark}
Using the Fast Fourier Transform (FFT) \cite{ct65}, $S$ can be applied to a vector in time $O(d\log d)$.
\end{remark}
\end{definition}
We also introduce a sketching matrix for degree-$2$ tensors, which is a generalization of the {\srht}. 
\begin{definition}[Tensor Subsampled Randomized Hadamard Transform (\tensorsrht) \cite{akk+20}]
\label{def:TensorSRHT}
We define the {\tensorsrht} $S:\R^d\times \R^d\rightarrow \R^m$ as $S=\frac{1}{\sqrt m} P \cdot (H D_1\times H D_2)$, where each row of $P\in \{0,1\}^{m\times d^2}$ contains only one $1$ at a random coordinate, one can view $P$ as a sampling matrix. $H$ is a $d\times d$ Hadamard matrix, and $D_1,D_2$ are two $d\times d$ independent diagonal matrices with diagonals that are each independently set to be a Rademacher random variable (uniform in $\{-1,1\}$). 
\end{definition}
\begin{remark}
By leveraging the FFT algorithm in the sketch space, $S(x\times y)$ for any $x,y\in\R^d$, and in particular $S(x^{\otimes 2})$, can be computed in time $O(d\log d+m)$.
\end{remark}

We will use the following properties of the {\srht} and {\tensorsrht}.

\begin{lemma}[{\srht} is an {\ose}, see Lemma 4.5 in~\cite{bg13}]
\label{thm:SRHT}
Let $T$ be an {\srht} matrix defined in Definition \ref{def:SRHT}. If $m=\Omega(\epsilon^{-2}(\sqrt n+\sqrt{\log(\max(d,m)/\delta)})^2\log(n/\delta))$, which holds whenever $m=\Omega(\epsilon^{-2}n\log(n\max(d,m)/\delta)\log(n/\delta))$, then $T$ is an $(\epsilon,\delta,d,n)$-{\ose}.
\end{lemma}
The lemma follows from Lemma 4.5 in~\cite{bg13} (numbered as in arXiv:1204.0062v4), which samples the rows with replacement, as Definition~\ref{def:SRHT} does. We apply the cited lemma, with $\epsilon^2$ in place of $\epsilon$ and $\delta/3$ in place of $\delta$, to an orthonormal basis $V$ of the column span of $A$, which has $k\leq n$ columns. With probability at least $1-\delta$, it gives $(1-\epsilon)I_k\preceq(TV)^\top TV\preceq(1+\epsilon)I_k$. Every $Ax$ equals $Vu$ for some $u\in\R^k$ with $\|u\|_2=\|Ax\|_2$, so this gives $(TA)^\top TA\approx_\epsilon A^\top A$. The cited lemma also assumes that $d$ is a power of $2$ and that $m\leq d$. Otherwise, we pad the input vectors with zeros to length $d'$, the smallest power of $2$ with $d'\geq\max(d,m)$, and use the {\srht} on $\R^{d'}$. Zero padding does not change $A^\top A$, and since $d'<2\max(d,m)$, the condition on $m$ in the lemma implies the condition of the cited lemma on $\R^{d'}$. Applying $T$ to a vector takes $O(d'\log d')=O(d\log d+m\log m)$ time.

\begin{lemma}[Lemma 33 in~\cite{akk+20}]
\label{thm:TensorSRHT}
Let $S$ be a {\tensorsrht} matrix defined in Definition \ref{def:TensorSRHT}. If $m=\Omega(n\log^3(nd/\epsilon\delta)\epsilon^{-2})$, then $S$ is an $(\epsilon,\delta,d^2,n)$-{\ose}, that is, for every fixed matrix $A\in\R^{d^2\times n}$, whose columns lie in the input space $\R^d\times\R^d$, with probability at least $1-\delta$, $(SA)^\top SA\approx_\epsilon A^\top A$.
\end{lemma}
The lemma follows from Lemma 33 in the full version of~\cite{akk+20} (arXiv:1909.01410v5) with $k=1$, $\mu_F=n$ and $\mu_2=1$, applied to $U=A(A^\top A)^{+1/2}$, where $(A^\top A)^{+1/2}$ is the square root of the pseudoinverse of $A^\top A$. This $U$ satisfies $\|U\|_F^2=\rank(A)\leq n$ and $\|U\|_{\op}\leq 1$. The resulting bound $\|U^\top S^\top SU-U^\top U\|_{\op}\leq\epsilon$ gives $(SA)^\top SA\approx_\epsilon A^\top A$. In Algorithm~\ref{alg1}, the input space of each $S_l$ and $S'_k$ is $\R^m\times\R^m$, so $d$ becomes $m$ inside the logarithm, and $\log(nm/(\epsilon\delta))=O(\log(nd/(\epsilon\delta)))$, because $m$, which is also the number of rows of $T$, is at most $\poly(nd/(\epsilon\delta))$ for the $\epsilon$ and $\delta$ of the lemma. We need the lemma for every fixed matrix, and not only for matrices whose columns are self-tensors $x^{\otimes 2}$, because the combination step of Algorithm~\ref{alg1} applies $S'_k$ to $z\times w_{i_{k+1}}$, which is not a self-tensor.
\vspace{-3mm}
\subsection{Kernels}
\vspace{-2mm}
We introduce several kernels that are widely-used in practice, e.g., see \cite{ge08,chc+10} for the polynomial kernel, and see \cite{njw01} for the Gaussian kernel.
\begin{definition}[Polynomial Kernel]
\label{def:poly_kernel}
Given two data points $x,y\in \R^d$, the \textit{degree-$p$ polynomial kernel, $P$,} between $x$ and $y$ is defined as\footnote{A more standard definition is $P(x,y)=\inner{x}{y}^p+c$; we can simulate this by creating an extra dimension on all data points.}:
$
P(x,y)=\inner{x}{y}^p
$. 
Let $X\in \R^{d\times n}$. The \textit{degree-$p$ polynomial kernel $P$,} defined on matrix $X$, is the matrix
$
P_{i,j} = \inner{x_i}{x_j}^p,
$
where $x_i,x_j$ are the $i^{th}$ and $j^{th}$ column of $X$, respectively. 
\end{definition}

\begin{definition}[Gaussian Kernel]
\label{def:gaussian_kernel}
Given two data points $x,y\in \R^d$, the \textit{Gaussian kernel, $G$}, between $x$ and $y$ is defined as
$
G(x,y)  = \exp(- {\|x-y\|_2^2} / {2})
$. 
Let $X\in \R^{d\times n}$. The \textit{Gaussian kernel $G$,} defined on matrix $X$, is the matrix
$
G_{i,j} = \exp(- {\|x_i-x_j\|_2^2} / {2}),
$
where $x_i,x_j$ are the $i^{th}$ and $j^{th}$ column of $X$, respectively. 
\end{definition}

\section{Technical Overview}\label{sec:tech}
We first consider one way to compute the polynomial kernel $P$ via the identity $P=(X^{\otimes p})^\top X^{\otimes p}$. Our algorithm will try to compute $X^{\otimes p}$ quickly.

Suppose we want to compute the $p$-fold tensoring of a vector $x$, and assume for simplicity that $p$ is a power of $2$. Our algorithm is inspired by that of \cite{akk+20}, which generates a complete binary tree with $2p-1$ nodes, and thus $p$ leaves. For the $i$-th leaf node, it picks a sketch $T^i$ and applies it to $x$, obtaining $T^ix$. Each internal node $j$ then does the following: it picks a sketch $S^j$, and applies $S^j$ to the tensor product of its two children. $S^j$ is picked as a map from $\R^{m^2}\rightarrow \R^m$ for each $j$, so at each level of the binary tree, we reduce the number of vectors by half, while remaining in the low-dimensional space $\R^m$. One drawback of this algorithm is the usage of an independent sketch for each node of the tree, thus incurring a linear dependence on $p$ in the runtime. 

Our algorithm instead uses a much smaller amount of randomness. We pick a single $T$ for all leaf nodes, i.e., $T^1 = T^2 = \cdots = T^p = T$, and a single sketch $S_l$ for each level $l$ of the tree, shared by all internal nodes of that level, i.e., $S^j = S_l$ for every internal node $j$ at level $l$. When $p$ is not a power of $2$, we also use a fresh sketch for each combination step that merges two partial results. This gives $O(\log p)$ independent sketches in total. Each level needs fresh randomness because a sketch is only guaranteed to work on inputs chosen independently of it, and the inputs at level $l$ are outputs of the sketches at the lower levels. Sharing $T$ is safe, because its guarantee covers the fixed column span of $X$. The challenge with this approach is of course that we have much less independence in our analysis, and consequently do not obtain the same guarantees for preserving the length of the tensor product of an arbitrary set of vectors as in previous work. We stress that we compute a sketch that preserves the column span of $X^{\otimes p}$, and this is weaker than the guarantee we would get had we used full independence, which gives a sketch that is an oblivious subspace embedding, meaning that it can preserve the column span of \textit{any} matrix in $\R^{d^p\times n}$. However, the key point is that we can preserve the tensor product of a vector with itself $p$ times, and this will suffice for our applications. 

This allows for a much faster way to compute $x^{\otimes p}$: ``square'' a vector by computing the tensor product with itself, then apply the next level's sketch, and repeat this process. By doing so, the leaves apply $T$ only once instead of $p$ times, and the number of sketch applications per input vector drops from $2p-1$ to $O(\log p)$. However, this will incur a $p^2$ factor in the dimension of the sketch, and so we will pay more for $p$ in levels other than the first level. Fortunately, levels other than the first level apply sketches to lower dimensional vectors. By carefully balancing the complexity of applying $T$ and the level sketches $S_l$, we achieve an improved running time, which is useful when the degree $p$ is large. 

Finally, we sketch kernels through their Taylor series $K=\sum_{l\geq 0}C_l(X^{\otimes l})^\top X^{\otimes l}$, truncated at a degree $q$ that is large enough for the dropped terms to be absorbed by a ridge term $\lambda I_n$. Sketching degree $l$ costs about $\epsilon^{-2}n^2l^2$, so sketching every term costs about $\epsilon^{-2}n^2q^3$, and for kernels whose coefficients decay polynomially, such as the {\ntk}, $q$ is polynomial in $n/(\epsilon\lambda)$. The expensive degrees between $q/2$ and $q$ carry little weight, since their coefficients add up to at most about $\epsilon\lambda/n$. So we sketch the low degrees as before, and replace the high degrees by a few degrees sampled with probability proportional to $C_l$ and reweighted so that the estimate stays unbiased. A matrix Bernstein bound, measured relative to $K+\lambda I_n$ rather than to $\|K\|_{\op}$, controls the sampling error, and the expected cost of a sample is governed by $\E[l^2]$ rather than by $(\E[l])^2$. For the {\ntk} we group the degrees into bands $[2^j,2^{j+1})$, sample within each band, and sketch each sampled degree only as accurately as the weight of its band requires. The top bands then need only polylogarithmically many samples, each sketched only to accuracy $1/O(\log q)$, independent of $\epsilon$, and the running time becomes about that of sketching the single degree $q$ once.
\section{Fast Sketching Algorithm for the Polynomial Kernel}\label{sec:algorithm}
We introduce our algorithm that sketches a single vector $x^{\otimes p}$ and extend it to each column of a matrix. In Section~\ref{sec:cs_def} we give some definitions. In Section~\ref{sec:cs_tensor} we prove several technical tools related to tensors. In Section~\ref{sec:cs_poly} we show our sketch preserves the column space of the polynomial kernel. In Section~\ref{sec:cs_main} we prove our main result for this section.
\subsection{Definitions}
\label{sec:cs_def}
We define the sketching matrix formed by Algorithm~\ref{alg1} as follows:
\begin{definition}\label{def:Pi_q}
Let $q$ be a power of $2$ and $\Pi^q:\R^{d^q}\rightarrow \R^m$ be defined as the following matrix:
\begin{align*}
\Pi^q=Q^q\cdot T^q,
\end{align*}
where $T^q=\underbrace{T\times T\times \ldots \times T}_{\text{$q$ times}}$ and $Q^q=S_{\log_2 q}^{(1)}\cdot S_{\log_2 q-1}^{(2)}\cdot\ldots \cdot S_1^{(q/2)}$, and for a matrix $A$ and a positive integer $k$, $A^{(k)}=\underbrace{A\times A\times \ldots \times A}_{\text{$k$ times}}$ denotes its $k$-fold Kronecker power. Here $T$ and $S_1,\ldots,S_{\log_2 q}$ are the base sketches of Algorithm~\ref{alg1}. For $q=1$, $Q^1$ is the identity and $\Pi^1=T$.
\end{definition}

We will design an algorithm that achieves the following goal:

{\bf Case 1} If $p$ is a power of $2$, then it computes $\Pi^pX^{\otimes p}$ efficiently.

{\bf Case 2} If $p$ is not a power of $2$, then let $b$ be its binary representation and let 
\begin{align*}
E= & ~ \set{i:b_i=1,i\in \{0,\ldots,\log_2 p\}}.
\end{align*}
 We will iterate through all indices in $E$ and continue tensoring two vectors where $b_i=1$, and apply a fresh sketch $S'_k$ to them at the $k$-th such step, for $k=1,\ldots,|E|-1$.

\begin{algorithm*}
\caption{Our algorithm for sketching the vector $x^{\otimes p}$ with limited randomness. It uses $O(\log p)$ independent base sketches: $T$, one sketch $S_l\in \R^{m\times m^{2}}$ per squaring level and one sketch $S'_k\in \R^{m\times m^{2}}$ per combination step, collected in ${\cal S}=(S_1,\ldots,S_{\floor{\log_2 p}},S'_1,\ldots,S'_{|E|-1})$.}
\label{alg1}
\begin{algorithmic}[1]
\Procedure{TensorSketchViaLimRand}{$x\in \R^d,p\in \N_+,T\in \R^{m\times d},{\cal S}$} \Comment{Theorem~\ref{thm:main_result} and~\ref{thm:SRHT+TensorSRHT}}
\State Let $q= 2^{\floor{\log_2 p}}$
\State Let $w_0= Tx$ \Comment{$T$ can be {\srht} (Definition \ref{def:SRHT})}
\For{$l=1$ to $\log_2 q$}
\State Compute $w_l= S_l(w_{l-1}^{\otimes 2})$ \Comment{Each $S_l$ can be {\tensorsrht} (Definition \ref{def:TensorSRHT})}
\EndFor
\State Let $b$ be the binary representation of $p$, and let $E=\set{i: b_i=1,i\in \{0,\ldots,\log_2 p\}}$
\State Let $i_1<i_2<\cdots<i_{|E|}$ be the elements of $E$, and let $z=w_{i_1}$
\For{$k=1$ to $|E|-1$}
\State $z=  S'_k(z\times w_{i_{k+1}})$ \Comment{Each $S'_k$ can be {\tensorsrht}}
\EndFor
\State return $z$ \Comment{$z\in \R^m$}
\EndProcedure
\end{algorithmic}
\end{algorithm*}
\begin{definition}
\label{def:alg1_mat}
Let $p$ be a positive integer, and let $E$ be the set of positions of the $1$ bits in the binary representation of $p$, as in Algorithm~\ref{alg1}. Let $T:\R^d\rightarrow \R^m$ and $S_1,\ldots,S_{\floor{\log_2 p}},S'_1,\ldots,S'_{|E|-1}:\R^{m^2}\rightarrow \R^m$ be base sketches, and write ${\cal S}=(S_1,\ldots,S_{\floor{\log_2 p}},S'_1,\ldots,S'_{|E|-1})$. Let $X\in \R^{d\times n}$ be an input matrix. We define ${\cal Z}({\cal S},T,X)$ to be the matrix for which we apply Algorithm \ref{alg1} on each column of $X$, with base sketches ${\cal S}$ and $T$.
\end{definition}
\subsection{Equivalence Results for Tensors}
\label{sec:cs_tensor}
We provide two technical tools for handling tensors.
\begin{lemma}[The output guarantee of Algorithm~\ref{alg1}]
\label{lem:alg_to_sketch}
Let $p$ be a power of $2$ and $\Pi^p$ be defined as in Definition~\ref{def:Pi_q}. Let $x\in \R^d$. Then the output vector $z$ generated by Algorithm~\ref{alg1} satisfies $z=\Pi^p(x^{\otimes p})$. 
\end{lemma}
\begin{proof}
If $p$ is a power of $2$, then $E=\{\log_2 p\}$, so Algorithm \ref{alg1} sets $z=w_{\log_2 p}$ on line 8, skips the loop on line 9, and returns $z$ on line 12. We will exploit the fact that although Algorithm \ref{alg1} only computes one vector at a time, we can view it as computing $p/2^i$ identical vectors in the $i^{\text{th}}$ iteration. On line 3, we can treat it as computing $p$ copies of $w_0$, and therefore, by Claim \ref{clm:mat_vec_prod}, we have
\begin{align*}
w_0^{\otimes p} & = (Tx)^{\otimes p} \\
& = (T\times T\times \ldots \times T)(x\times x\times \ldots \times x) \\
& = T^p x^{\otimes p}.
\end{align*}
We can apply the same line of reasoning to line 5 of the algorithm. In the $i^{\text{th}}$ iteration, we can treat it as
\begin{align*}
S_i(w_{i-1}^{\otimes 2})\times S_i(w_{i-1}^{\otimes 2})\times \ldots \times S_i(w_{i-1}^{\otimes 2})
\end{align*}
a total of $p/2^i$ times. Again, using Claim \ref{clm:mat_vec_prod}, we have
\begin{align*}
w_i^{\otimes p/2^i} & = S_i(w_{i-1}^{\otimes 2})\times S_i(w_{i-1}^{\otimes 2})\times \ldots \times S_i(w_{i-1}^{\otimes 2}) \\
& = \left(S_i(w_{i-1}^{\otimes 2}) \right)^{\otimes p/2^i} \\
& = \underbrace{(S_i\times S_i\times \ldots \times S_i)}_{\text{$p/2^i$ times}} (w_{i-1}^{\otimes 2})^{\otimes p/2^i} \\
& = S_i^{(p/2^i)} w_{i-1}^{\otimes p/2^{i-1}}.
\end{align*}
Recursively applying this identity for $i=k,k-1,\ldots,1$, where $k=\log_2 p$, we will end up with
\begin{align*}
z  = S_k^{(1)} \cdot S_{k-1}^{(2)}\cdot \ldots \cdot S_1^{(p/2)}\cdot T^p x^{\otimes p} = \Pi^p(x^{\otimes p}). ~~~~ \qedhere
\end{align*}
\end{proof}
Next, we wish to show that if $p$ is a power of $2$, then $\Pi^p$ preserves the squared length of every vector in the subspace spanned by the columns of $X^{\otimes p}$ within a factor of $(1\pm \epsilon)^{2p}$, where $\epsilon$ is the accuracy of the base sketches. With base sketches of accuracy $\frac{\epsilon}{4p}$ instead, this factor lies in $[1-\epsilon,1+\epsilon]$, since $(1-\frac{\epsilon}{4p})^{2p}\geq 1-\frac{\epsilon}{2}$ and $(1+\frac{\epsilon}{4p})^{2p}\leq e^{\epsilon/2}\leq 1+\epsilon$ for $\epsilon\in(0,1)$. Notice this is weaker than $\Pi^p$ being an {\ose}, but sufficient for our application to polynomial kernels. We will show that 
\begin{align*}
(\Pi^p X^{\otimes p})^\top \Pi^p X^{\otimes p}\approx_\epsilon (X^{\otimes p})^\top X^{\otimes p}.
\end{align*}

The following lemma outlines the main technique in our proof of this property. It establishes a one-to-one mapping between a vector in the column span of $X^{\otimes p}$ and a matrix. We then use this equivalence to inductively prove that $\Pi^p$ preserves the target subspace.
\begin{lemma}[From Vector Tensor to Matrix Tensor]
\label{lem:vec to mat}
Let $X\in \R^{d\times n}$. Consider $u=X^{\otimes p}y$ for some $y\in \R^n$ and positive integer $p$. Then, 
\begin{align*}
\|u\|_2& =\|X^{\otimes (p-1)} YX^\top \|_F, 
\end{align*}
where $Y=\diag(y)\in \R^{n\times n}$ is a diagonal matrix where the $i$-th entry on the diagonal is $y_i$, $\forall i \in [n]$.
\end{lemma}
\begin{proof}
We will use $X_{i,j}$ to denote the $i^{\mathrm{th}}$ row and $j^{\mathrm{th}}$ column of $X$. We observe that a fixed column $X^{\otimes p}_{*,j}$ is equivalent to taking the outer product $X_{*,j}^{\otimes p-1}X_{*,j}^\top$, then flattening the matrix into a long vector. If we use a double index to indicate an entry of $X^{\otimes p}_{*,j}(a,b)$, then we have $X^{\otimes p}_{*,j}(a,b)=X^{\otimes (p-1)}(a)X(b)$, where $a$ ranges from $1$ to $d^{p-1}$ and $b$ ranges from $1$ to $d$. Consider the $\ell_2$ norm $u$:
\begin{align*}
\|u\|_2^2 & = \|X^{\otimes p}y\|_2^2 \\
& = \sum_{a=1}^{d^{p-1}}\sum_{b=1}^d \left(\sum_{i=1}^n y_i(X^{\otimes p}_{*,i})(a,b)\right)^2.
\end{align*}
On the other hand, we can write $X^{\otimes (p-1)}$ in its column form:
\begin{align*}
\begin{bmatrix}
\vert & \vert & \ldots & \vert \\
X^{\otimes (p-1)}_{*,1} & X^{\otimes (p-1)}_{*,2} & \ldots & X^{\otimes (p-1)}_{*,n} \\
\vert & \vert & \ldots & \vert
\end{bmatrix}.
\end{align*}
Recall that $Y$ is a diagonal matrix and therefore, the product $X^{\otimes (p-1)}Y$ can be expressed as
\begin{align*}
\begin{bmatrix}
\vert & \vert & \ldots & \vert \\
y_1X^{\otimes (p-1)}_{*,1} & y_2X^{\otimes (p-1)}_{*,2} & \ldots & y_nX^{\otimes (p-1)}_{*,n} \\
\vert & \vert & \ldots & \vert
\end{bmatrix}.
\end{align*}
Using the outer product definition of matrix product, we have
\begin{align*}
X^{\otimes (p-1)}YX^\top = \sum_{i=1}^n y_iX_{*,i}^{\otimes (p-1)}X_{*,i}^\top .
\end{align*}
This is a matrix of size $d^{p-1}\times d$. Therefore, we can write its Frobenius norm as
\begin{align*}
\|X^{\otimes (p-1)}YX^\top\|_F^2 & = \|\sum_{i=1}^n y_iX_{*,i}^{\otimes (p-1)}X_{*,i}^\top\|_F^2 \\
& = \sum_{a=1}^{d^{p-1}}\sum_{b=1}^d \left(\sum_{i=1}^n y_iX_{*,i}^{\otimes (p-1)}(a)X_{*,i}(b)\right)^2 \\
& = \sum_{a=1}^{d^{p-1}}\sum_{b=1}^d \left(\sum_{i=1}^n y_i(X^{\otimes p}_{*,i})(a,b)\right)^2 \\
& = \|u\|_2^2.
\end{align*}
This completes the proof.
\end{proof}
\vspace{-6mm}
\subsection{Preserving the Polynomial Kernel Subspace}
\label{sec:cs_poly}
We prove the sketch generated by Algorithm~\ref{alg1} preserves the column space of the polynomial kernel.
\begin{lemma}[$T^p$ Preserves Polynomial Kernel Subspace]
\label{lem: power q sub emb}
Let $X\in \R^{d\times n}$ and let $p$ be a positive integer. If $T$ is an $(\epsilon,\delta,d,n)$ {\ose}, then with probability at least $1-\delta$, simultaneously for all $y\in \R^n$,
\begin{align*}
\|(TX)^{\otimes p}y\|_2^2=(1 \pm \epsilon)^p\|X^{\otimes p}y\|_2^2.
\end{align*}
\end{lemma}
\begin{proof}
Let ${\cal E}_T$ be the event $(TX)^\top TX\approx_\epsilon X^\top X$. By Definition \ref{def:OSE}, ${\cal E}_T$ holds with probability at least $1-\delta$. This is a spectral bound, so on ${\cal E}_T$ we have $\|TXv\|_2^2=(1\pm\epsilon)\|Xv\|_2^2$ for all $v\in\R^n$ at once, including vectors $v$ that depend on $T$. We show that on ${\cal E}_T$ the claim holds for all $y\in\R^n$. The argument uses nothing else about $T$.

We proceed by induction on $p$. For $p=1$, on ${\cal E}_T$, we have for all $y\in\R^n$
\begin{align*}
\|TXy\|_2^2=(1 \pm \epsilon)\|Xy\|_2^2.
\end{align*}
For the inductive step, we will prove this for a general positive integer $p>1$. We break $p$ into $p-1$ and $1$. By Lemma \ref{lem:vec to mat}, we have 
\begin{align*}
\|(TX)^{\otimes p}y\|_2^2& =\|(TX)^{\otimes (p-1)}Y(TX)^\top\|_F^2 \\
& = \|(TX)^{\otimes (p-1)}YX^\top T^\top\|_F^2.
\end{align*}
Recall that on ${\cal E}_T$, $T$ preserves the length of every vector in the column span of $X$. Every row of $(TX)^{\otimes (p-1)}YX^\top$ has the form $(Xv)^\top$ for a vector $v\in\R^n$, which may depend on $T$. This means right multiplying by $T^\top$ preserves the length of all rows of $(TX)^{\otimes (p-1)}YX^\top$, and therefore we have
\begin{align*}
\|(TX)^{\otimes (p-1)}YX^\top T^\top\|_F^2 = (1 \pm \epsilon)\|(TX)^{\otimes (p-1)}YX^\top\|_F^2.
\end{align*}
Using the inductive hypothesis, on ${\cal E}_T$, for any vector $z\in \R^n$, we have
\begin{align*}
\|(TX)^{\otimes (p-1)}z\|_2^2=(1 \pm \epsilon)^{p-1}\|X^{\otimes (p-1)}z\|_2^2.
\end{align*}
Applying this to each column of $YX^\top$, we have 
\begin{align*}
\|(TX)^{\otimes (p-1)}YX^\top\|_F^2&=(1 \pm \epsilon)^{p-1}\|X^{\otimes (p-1)}YX^\top\|_F^2 \\
& = (1 \pm \epsilon)^{p-1}\|X^{\otimes p}y\|_2^2.
\end{align*}
This enables us to conclude that
\begin{align*}
\|(TX)^{\otimes p}y\|_2^2=(1 \pm \epsilon)^p\|X^{\otimes p}y\|_2^2 
\end{align*}
\end{proof}
\begin{remark}
As we motivated in Section \ref{sec:tech}, if we view Algorithm~\ref{alg1} as a binary tree, Lemma~\ref{lem: power q sub emb} effectively proves that the bottom layer of the tree preserves the column space of $X^{\otimes p}$. We then pick each level sketch $S_l$ to be an {\ose} on $\R^{m^2}$, independent of the sketches at the lower levels, and inductively establish our embedding.
\end{remark}

\begin{lemma}[$\Pi^p$ Preserves the Polynomial Kernel Subspace]\label{lem:power_2_thm}
Let $X\in \R^{d\times n}$, let $p$ be a power of $2$ and let $k=\log_2 p$. Let $T:\R^d\rightarrow \R^m$ be an $(\epsilon,\delta',d,n)$-{\ose}, and let $S_1,\ldots,S_k:\R^{m^2}\rightarrow \R^m$ each be an $(\epsilon,\delta',m^2,n)$-{\ose} (Definition~\ref{def:OSE}), where $T,S_1,\ldots,S_k$ are mutually independent. Let $\Pi^p$ be the sketching matrix defined as in Definition~\ref{def:Pi_q}. Then with probability at least $1-(k+1)\delta'$, simultaneously for all $y\in \R^n$,
\begin{align*}
(1-\epsilon)^{2p} \|X^{\otimes p}y\|_2^2\leq \|\Pi^pX^{\otimes p}y\|_2^2\leq (1+\epsilon)^{2p} \|X^{\otimes p}y\|_2^2.
\end{align*}
\end{lemma}

\begin{proof}
For $i\in[n]$ and $l\in\{0,\ldots,k\}$, let $w^i_l\in\R^m$ denote the vector $w_l$ that Algorithm~\ref{alg1} computes from the $i^{\text{th}}$ column of $X$, that is, the $i^{\text{th}}$ column after $l$ iterations. For $l\in[k]$, let $Z_l\in\R^{m^2\times n}$ be the matrix
\begin{align*}
Z_l=\begin{bmatrix}
\vert & \vert & \ldots & \vert \\
(w^1_{l-1})^{\otimes 2} &(w^2_{l-1})^{\otimes 2} & \ldots & (w^n_{l-1})^{\otimes 2} \\
\vert & \vert & \ldots & \vert
\end{bmatrix},
\end{align*}
so that the $i^{\text{th}}$ column of $S_lZ_l$ is $w^i_l$. Let ${\cal E}_T$ be the event $(TX)^\top TX\approx_\epsilon X^\top X$, and for $l\in[k]$ let ${\cal E}_l$ be the event $(S_lZ_l)^\top S_lZ_l\approx_\epsilon Z_l^\top Z_l$. By Definition~\ref{def:OSE}, $\Pr[{\cal E}_T]\geq 1-\delta'$. The matrix $Z_l$ is a function of $X,T,S_1,\ldots,S_{l-1}$ only. So conditioned on $T,S_1,\ldots,S_{l-1}$, the matrix $Z_l$ is fixed and $S_l$ is independent of it, and Definition~\ref{def:OSE} gives $\Pr[{\cal E}_l\mid T,S_1,\ldots,S_{l-1}]\geq 1-\delta'$. Hence $\Pr[{\cal E}_l]\geq 1-\delta'$, and by a union bound the events ${\cal E}_T,{\cal E}_1,\ldots,{\cal E}_k$ all hold with probability at least $1-(k+1)\delta'$. Each ${\cal E}_l$ is a spectral bound, so on ${\cal E}_l$ we have $\|S_lZ_lv\|_2^2=(1\pm\epsilon)\|Z_lv\|_2^2$ for all $v\in\R^n$ at once, including vectors $v$ that depend on $S_l$. We show that the conclusion holds on ${\cal E}_T\cap{\cal E}_1\cap\cdots\cap{\cal E}_k$.

This is where the independence across levels is used. If a single sketch $S$ were used at every level, then from level $2$ on the columns of $Z_l$ would be outputs of $S$ itself. The matrix $Z_l$ would then depend on $S$, and Definition~\ref{def:OSE}, which only covers fixed matrices, would give no guarantee that $S$ preserves the column span of $Z_l$.

We will prove the claim by induction on the number of iterations of Algorithm~\ref{alg1}, that is, on the level $l$ from $0$ to $k$. Let $\Upsilon^{2^l}$ denote the sketching matrix at level $l$: $\Upsilon^{2^l}=S_l^{(p/2^l)}\cdot S_{l-1}^{(p/2^{l-1})}\cdot \ldots \cdot S_1^{(p/2)}\cdot T^p$ for $l\in [k]$, and $\Upsilon^{1}=T^p$. We will prove the following statement: for every $l\in\{0,\ldots,k\}$, on ${\cal E}_T\cap{\cal E}_1\cap\cdots\cap{\cal E}_l$, simultaneously for all $y\in\R^n$,
\begin{align*}
\|\Upsilon^{2^l} X^{\otimes p}y\|_2^2 = & ~ (1\pm \epsilon)^{\sum_{i=0}^l\frac{p}{2^{i}}} \|X^{\otimes p}y\|_2^2.
\end{align*}
Note that when $l=k$, we have $\Upsilon^{2^k}=\Pi^p$, and the exponent is
\begin{align*}
\sum_{i=0}^k\frac{p}{2^i} = p+\frac{p}{2}+\cdots+1 = 2p-1\leq 2p.
\end{align*}
Since $(1-\epsilon)^{2p}\leq(1-\epsilon)^{2p-1}$ and $(1+\epsilon)^{2p-1}\leq(1+\epsilon)^{2p}$, this gives us the desired result.

For the base case $l=0$, Claim~\ref{clm:mat_vec_prod} applied to each column gives $\Upsilon^1X^{\otimes p}=T^pX^{\otimes p}=(TX)^{\otimes p}$, and the proof of Lemma~\ref{lem: power q sub emb} shows that on ${\cal E}_T$, simultaneously for all $y\in\R^n$, $\|(TX)^{\otimes p}y\|_2^2=(1\pm\epsilon)^p\|X^{\otimes p}y\|_2^2$. For the inductive step, let $l\in[k]$ and assume the statement holds for $l-1$, so on ${\cal E}_T\cap{\cal E}_1\cap\cdots\cap{\cal E}_{l-1}$, for all $y\in\R^n$,
\begin{align*}
\|\Upsilon^{2^{l-1}}X^{\otimes p}y\|_2^2=(1 \pm \epsilon)^{\sum_{i=0}^{l-1}p/2^i} \|X^{\otimes p}y\|_2^2.
\end{align*}
Let $r=p/2^l$. By the proof of Lemma~\ref{lem:alg_to_sketch}, the $i^{\text{th}}$ column of $\Upsilon^{2^{l-1}}X^{\otimes p}$ is $(w^i_{l-1})^{\otimes 2r}=((w^i_{l-1})^{\otimes 2})^{\otimes r}$, so $\Upsilon^{2^{l-1}}X^{\otimes p}=Z_l^{\otimes r}$. Since $\Upsilon^{2^l}=S_l^{(r)}\cdot \Upsilon^{2^{l-1}}$, Claim~\ref{clm:mat_vec_prod} applied to each column gives
\begin{align*}
\Upsilon^{2^l}X^{\otimes p}=S_l^{(r)}\Upsilon^{2^{l-1}}X^{\otimes p}=S_l^{(r)}Z_l^{\otimes r}=(S_lZ_l)^{\otimes r}.
\end{align*}

If $r>1$, then similar to Lemma~\ref{lem: power q sub emb}, on ${\cal E}_T\cap{\cal E}_1\cap\cdots\cap{\cal E}_l$, for all $y\in\R^n$,
\begin{align}
& ~ \|(S_lZ_l)^{\otimes r}y\|_2^2 \notag \\
= & ~ \|(S_lZ_l)^{\otimes (r-1)}\diag(y)Z_l^\top S_l^\top \|_F^2 \notag  \\
= & ~ (1 \pm \epsilon)\|(S_lZ_l)^{\otimes (r-1)}\diag(y)Z_l^\top\|_F^2 \notag \\
= & ~ (1 \pm \epsilon)^{r}\|Z_l^{\otimes (r-1)}\diag(y)Z_l^\top\|_F^2 \notag \\
= & ~ (1 \pm \epsilon)^{r}\|Z_l^{\otimes r}y\|_2^2 \notag \\
= & ~ (1 \pm \epsilon)^{r}\|\Upsilon^{2^{l-1}}X^{\otimes p}y\|_2^2 \notag \\
= & ~ (1 \pm \epsilon)^{\sum_{i=0}^l p/2^i}\|X^{\otimes p}y\|_2^2. \notag
\end{align}
The first and fourth steps use Lemma~\ref{lem:vec to mat}. The second step holds on ${\cal E}_l$. Every row of the matrix that multiplies $S_l^\top$ in the second line has the form $(Z_lv)^\top$ for a vector $v\in\R^n$, and right multiplying by $S_l^\top$ preserves its length, even though $v$ depends on $S_l$. The third step is the proof of Lemma~\ref{lem: power q sub emb} with $S_l$, $Z_l$, ${\cal E}_l$ and $r-1$ in place of $T$, $X$, ${\cal E}_T$ and $p$, applied to each column of $\diag(y)Z_l^\top$. It pulls out the remaining $r-1$ copies of $S_l$ at a cost of $(1\pm\epsilon)^{r-1}$. The fifth step uses $Z_l^{\otimes r}=\Upsilon^{2^{l-1}}X^{\otimes p}$. The last step uses the inductive hypothesis and $r+\sum_{i=0}^{l-1}p/2^i=\sum_{i=0}^{l}p/2^i$.

If $r=1$, that is, $l=k$, then $\Upsilon^{2^k}X^{\otimes p}y=S_kZ_ky$ with $Z_k=\Upsilon^{2^{k-1}}X^{\otimes p}$. On ${\cal E}_k$ we have $\|S_kZ_ky\|_2^2=(1\pm\epsilon)\|Z_ky\|_2^2$, and the inductive hypothesis on $Z_ky$ gives the exponent $1+\sum_{i=0}^{k-1}p/2^i=\sum_{i=0}^{k}p/2^i$.
\end{proof}

Below, we state and prove a theorem that establishes the correctness of Algorithm~\ref{alg1} without instantiating the sketching matrices $T$, $S_l$ and $S'_k$. This enables us to use different sketches with various trade-offs.
\subsection{Main Result}
\label{sec:cs_main}
We prove the main result of this section, which establishes the correctness of Algorithm~\ref{alg1}.
\begin{theorem}[Main Result, Correctness Part]
\label{thm:main_result}
Let $X\in \R^{d\times n}$ and let $p$ be a positive integer. Let $E$ be the set of positions of the $1$ bits in the binary representation of $p$. Let $T:\R^d\rightarrow \R^m$ be an $(\epsilon,\delta',d,n)$-{\ose}, and let $S_1,\ldots,S_{\floor{\log_2 p}}$ and $S'_1,\ldots,S'_{|E|-1}$ be maps from $\R^{m^2}$ to $\R^m$, each an $(\epsilon,\delta',m^2,n)$-{\ose}, where all of these base sketches are mutually independent. Let ${\cal S}=(S_1,\ldots,S_{\floor{\log_2 p}},S'_1,\ldots,S'_{|E|-1})$ and let $Z = {\cal Z}({\cal S},T,X)$ be the matrix as defined in Def. \ref{def:alg1_mat}. Then with probability at least $1-(1+\floor{\log_2 p}+|E|-1)\delta'$, simultaneously for all $y\in \R^n$,
\begin{align*}
(1-\epsilon)^{3p}\|X^{\otimes p}y\|_2^2\leq \|Zy\|_2^2\leq (1+\epsilon)^{3p} \|X^{\otimes p}y\|_2^2.
\end{align*}
\end{theorem}
\begin{proof}
Let $b$ be the binary representation of $p$, and let $E=\set{i:b_i=1,i\in \{0,1,\ldots,\log_2 p\}}$, with elements $i_1<i_2<\cdots<i_{|E|}$. For $j\in[|E|]$, let $q_j=2^{i_j}$ and $p_j=\sum_{i=1}^j q_i$, so that $p_{|E|}=p$. Let $K=\floor{\log_2 p}=i_{|E|}$. Since $E\subseteq\{0,\ldots,K\}$, we have $|E|-1\leq K$. For $l\in\{0,\ldots,K\}$, let $W_l\in\R^{m\times n}$ be the matrix whose $i^{\text{th}}$ column is the vector $w_l$ that Algorithm~\ref{alg1} computes from the $i^{\text{th}}$ column of $X$. By Lemma~\ref{lem:alg_to_sketch} with $q_j$ in place of $p$, applied to each column, $W_{i_j}=\Pi^{q_j}X^{\otimes q_j}$ for every $j\in[|E|]$.

We first handle the powers of $2$. Let ${\cal E}_T,{\cal E}_1,\ldots,{\cal E}_K$ be the events defined in the proof of Lemma~\ref{lem:power_2_thm} with $k=K$. The matrix $Z_l$ and the event ${\cal E}_l$ depend only on $X,T,S_1,\ldots,S_l$ and not on the target degree, because Algorithm~\ref{alg1} computes $w_0,\ldots,w_K$ once. So for every $j\in[|E|]$, the proof of Lemma~\ref{lem:power_2_thm} with $q_j$ in place of $p$ shows that on ${\cal E}_T\cap{\cal E}_1\cap\cdots\cap{\cal E}_K$, simultaneously for all $v\in\R^n$,
\begin{align}\label{eq:power_2_all}
(1-\epsilon)^{2q_j}\|X^{\otimes q_j}v\|_2^2\leq \|\Pi^{q_j}X^{\otimes q_j}v\|_2^2\leq (1+\epsilon)^{2q_j}\|X^{\otimes q_j}v\|_2^2.
\end{align}

Next we handle the combination steps. Let $Z^{(1)}=W_{i_1}$ and $Z^{(j)}=S'_{j-1}(Z^{(j-1)}\otimes W_{i_j})$ for $j=2,\ldots,|E|$. These are the values of $z$ in Algorithm~\ref{alg1}, column by column, after $j-1$ combination steps, so $Z=Z^{(|E|)}$. For $k\in[|E|-1]$, let $A_k=Z^{(k)}\otimes W_{i_{k+1}}\in\R^{m^2\times n}$, and let ${\cal E}'_k$ be the event $(S'_kA_k)^\top S'_kA_k\approx_\epsilon A_k^\top A_k$. This is where a fresh sketch is needed at each combination step. The matrix $A_k$ is a function of $X,T,S_1,\ldots,S_K$ and $S'_1,\ldots,S'_{k-1}$ only, so conditioned on these sketches it is fixed and $S'_k$ is independent of it. Definition~\ref{def:OSE} then gives $\Pr[{\cal E}'_k]\geq 1-\delta'$. Let ${\cal E}={\cal E}_T\cap{\cal E}_1\cap\cdots\cap{\cal E}_K\cap{\cal E}'_1\cap\cdots\cap{\cal E}'_{|E|-1}$. By a union bound over these $1+\floor{\log_2 p}+|E|-1$ events, ${\cal E}$ holds with probability at least $1-(1+\floor{\log_2 p}+|E|-1)\delta'$.

We will prove by induction on $j$ that on ${\cal E}$, simultaneously for all $y\in\R^n$,
\begin{align}\label{eq:induct_step}
 \|Z^{(j)}y\|_2^2 
 =  (1 \pm \epsilon)^{j-1+\sum_{i=1}^j 2q_i}\|X^{\otimes p_j}y\|_2^2 .
\end{align}
For the base case $j=1$, we have $Z^{(1)}=\Pi^{q_1}X^{\otimes q_1}$, and Eq.~\eqref{eq:power_2_all} gives Eq.~\eqref{eq:induct_step} with exponent $2q_1$. For the inductive step, let $j\geq 2$ and suppose Eq.~\eqref{eq:induct_step} holds for $j-1$. Since $Z^{(j)}y=S'_{j-1}A_{j-1}y$ and $W_{i_j}=\Pi^{q_j}X^{\otimes q_j}$, on ${\cal E}'_{j-1}$ we have
\begin{align}\label{eq:one_sketch_induct}
 \|Z^{(j)}y\|_2^2 
 =  (1 \pm \epsilon)\|\left(Z^{(j-1)}\otimes(\Pi^{q_j}X^{\otimes q_j})\right)y\|_2^2 .
\end{align}
By Lemma \ref{lem:vec to mat}, we have
\begin{align}\label{eq:two_sketches_induct}
& ~ \|\left(Z^{(j-1)}\otimes(\Pi^{q_j}X^{\otimes q_j})\right)y\|_2^2 \notag \\
= & ~ \|(\Pi^{q_j}X^{\otimes q_j})\diag(y)(Z^{(j-1)})^\top\|_F^2 \notag \\
= & ~ (1 \pm \epsilon)^{2q_j}\|X^{\otimes q_j}\diag(y)(Z^{(j-1)})^\top\|_F^2 \notag \\
= & ~ (1 \pm \epsilon)^{2q_j+j-2+\sum_{i=1}^{j-1}2q_i} 
 \cdot \| X^{\otimes q_j}\diag(y)(X^{\otimes p_{j-1}})^\top \|_F^2 \notag \\
= & ~ (1 \pm \epsilon)^{j-2+\sum_{i=1}^j2q_i}\|X^{\otimes p_j}y\|_2^2.
\end{align}
The first step is the computation in Lemma~\ref{lem:vec to mat}, with $Z^{(j-1)}$ and $\Pi^{q_j}X^{\otimes q_j}$ in place of $X^{\otimes (p-1)}$ and $X$, together with $\|M\|_F=\|M^\top\|_F$. The second step applies Eq.~\eqref{eq:power_2_all} to each column of $\diag(y)(Z^{(j-1)})^\top$. These columns depend on the sketches, which is allowed because Eq.~\eqref{eq:power_2_all} holds for all $v\in\R^n$ at once on ${\cal E}$. The third step writes $\|X^{\otimes q_j}\diag(y)(Z^{(j-1)})^\top\|_F=\|Z^{(j-1)}\diag(y)(X^{\otimes q_j})^\top\|_F$ and applies the inductive hypothesis, which also holds for all vectors at once, to each column of $\diag(y)(X^{\otimes q_j})^\top$. The last step is the computation in Lemma~\ref{lem:vec to mat}, with $X^{\otimes q_j}$ and $X^{\otimes p_{j-1}}$ in place of $X^{\otimes (p-1)}$ and $X$, and $q_j+p_{j-1}=p_j$. Combining Eq.~\eqref{eq:one_sketch_induct} and Eq.~\eqref{eq:two_sketches_induct}, the exponent is $1+(j-2)+\sum_{i=1}^j 2q_i=j-1+\sum_{i=1}^j 2q_i$, which is Eq.~\eqref{eq:induct_step}.

Finally, take $j=|E|$. The exponent in Eq.~\eqref{eq:induct_step} is
\begin{align*}
|E|-1+\sum_{i=1}^{|E|}2q_i = |E|-1+2p\leq 2p+\log_2 p\leq 3p,
\end{align*}
using $|E|-1\leq\floor{\log_2 p}$ and $\log_2 p\leq p$. Since $(1-\epsilon)^{3p}\leq (1-\epsilon)^{a}$ and $(1+\epsilon)^{a}\leq(1+\epsilon)^{3p}$ for every $a\leq 3p$, on ${\cal E}$ we get the claimed bound for all $y\in\R^n$. If $p$ is a power of $2$, then $|E|=1$, there are no combination steps, and this is Lemma~\ref{lem:power_2_thm}.
\end{proof}
\section{Analysis of Sketching Matrices: {\srht} and {\tensorsrht}}\label{sec:analysis}

In this section, we analyze the runtime of Algorithm \ref{alg1} with $T$ being an {\srht} sketch (Definition \ref{def:SRHT}) and each $S_l$ and $S'_k$ being a {\tensorsrht} sketch (Definition \ref{def:TensorSRHT}).
\vspace{-3mm}
\subsection{Main Result}
\vspace{-2mm}
The goal of this section is to give a runtime analysis of Algorithm~\ref{alg1} using {\srht} as $T$ and independent {\tensorsrht}s as the $S_l$ and the $S'_k$.
\begin{theorem}[Main Result, Running Time]
\label{thm:SRHT+TensorSRHT}
Let $p\in \N_+$ and $\epsilon,\delta\in (0,1)$. There exists a distribution over oblivious linear sketches $\Pi:\R^{d^p}\rightarrow \R^m$ with $m=\tilde \Theta(\epsilon^{-2}np^2)$, independent of $X$, such that for every $X\in \R^{d\times n}$, with probability at least $1-\delta$,
\begin{align*}
(\Pi X^{\otimes p})^\top \Pi X^{\otimes p}\approx_\epsilon (X^{\otimes p})^\top X^{\otimes p}.
\end{align*}
Moreover, using Algorithm \ref{alg1}, $\Pi X^{\otimes p}={\cal Z}({\cal S},T,X)$ can be computed in time $\tilde O(nd+\epsilon^{-2}n^2p^2)$.
\end{theorem}
\begin{proof}
Let $E$ be the set of positions of the $1$ bits in the binary representation of $p$. We will use an {\srht} for $T$ and a {\tensorsrht} for each of $S_1,\ldots,S_{\floor{\log_2 p}}$ and $S'_1,\ldots,S'_{|E|-1}$, all mutually independent, and write ${\cal S}=(S_1,\ldots,S_{\floor{\log_2 p}},S'_1,\ldots,S'_{|E|-1})$. Let $\wh \epsilon=\frac{\epsilon}{6p}$ and $\delta'=\frac{\delta}{2+2\log_2 p}$. We pick $T$ to be an $(\wh\epsilon,\delta',d,n)$-{\ose} and each $S_l$ and $S'_k$ to be an $(\wh\epsilon,\delta',m^2,n)$-{\ose}. The sketch $\Pi$ is the linear map that Algorithm \ref{alg1} applies to $x^{\otimes p}$ with these base sketches, and its distribution does not depend on $X$. Let $Z={\cal Z}({\cal S},T,X)=\Pi X^{\otimes p}$ be the matrix generated by Algorithm \ref{alg1} with these parameters. Since $E\subseteq\{0,\ldots,\floor{\log_2 p}\}$, we have $|E|-1\leq \floor{\log_2 p}$, so the number of base sketches is
\begin{align*}
1+\floor{\log_2 p}+(|E|-1)\leq 1+2\log_2 p< 2+2\log_2 p.
\end{align*}
By Theorem \ref{thm:main_result}, with probability at least $1-(1+\floor{\log_2 p}+|E|-1)\delta'\geq 1-\delta$, simultaneously for all $y\in\R^n$ we have $(1-\wh\epsilon)^{3p}\|X^{\otimes p}y\|_2^2\leq\|Zy\|_2^2\leq(1+\wh\epsilon)^{3p}\|X^{\otimes p}y\|_2^2$, that is,
\begin{align*}
(1-\wh\epsilon)^{3p}(X^{\otimes p})^\top X^{\otimes p}\preceq Z^\top Z\preceq (1+\wh\epsilon)^{3p}(X^{\otimes p})^\top X^{\otimes p}.
\end{align*}
Using $1+x\leq e^x$, then $e^x\leq 1+x+x^2$ for $x\in[0,1]$, and finally $\epsilon\leq 1$, we get
\begin{align*}
(1+\wh\epsilon)^{3p}\leq e^{3p\wh\epsilon}=e^{\epsilon/2}\leq 1+\frac{\epsilon}{2}+\frac{\epsilon^2}{4}\leq 1+\epsilon.
\end{align*}
By Bernoulli's inequality, $(1-\wh\epsilon)^{3p}\geq 1-3p\wh\epsilon=1-\frac{\epsilon}{2}\geq 1-\epsilon$. Thus, with probability at least $1-\delta$,
\begin{align*}
Z^\top Z\approx_\epsilon (X^{\otimes p})^\top X^{\otimes p}.
\end{align*}
By Lemma \ref{thm:SRHT} and Lemma \ref{thm:TensorSRHT}, for {\srht} and {\tensorsrht} to be $(\epsilon/(6p),\delta')$-{\ose}s on their input spaces, we need $m=\tilde \Theta\left(n/(\epsilon/(6p))^2\right)=\tilde \Theta(p^2n/\epsilon^2)$. The logarithmic factors depend on $p$ through $\log(1/\wh\epsilon)=\log(1/\epsilon)+\log(6p)$ and $\log(1/\delta')=\log(1/\delta)+O(\log\log p)$, so $\tilde\Theta$ here and $\tilde O$ below also absorb $\poly(\log p)$ factors.

We now analyze the runtime of Algorithm \ref{alg1} under {\srht} and {\tensorsrht}. On line 3, we compute $TX$ in time $\tilde O(n(d+m))=\tilde O(nd+p^2n^2/\epsilon^2)$ since $T$ is an {\srht} (see the discussion after Lemma~\ref{thm:SRHT}). We then enter a loop with $O(\log p)$ iterations, where in iteration $l$ we apply the level sketch $S_l$ to the tensor product of a column with itself resulting from the previous iteration. Each level applies its own $S_l$ at the same cost. Since $S_l$ is a {\tensorsrht}, this takes $O(m)=\tilde O(p^2n/\epsilon^2)$ time per column, and there are $n$ columns, so $\tilde O(p^2n^2/\epsilon^2)$ time in total for this step. We also compute each bit in the binary representation, and each combination step applies its own $S'_k$ at the same cost, which incurs an $O(\log p)$ factor in the final runtime.
%
So it takes Algorithm \ref{alg1} $\tilde O(nd+p^2n^2/\epsilon^2)$ time to compute ${\cal Z}({\cal S},T,X)$. This completes the proof.
\end{proof}
\subsection{Discussion}
We compare our result with the results obtained in~\cite{akk+20,wz20}. The setting we are considering is 1) matrix $X$ is dense, i.e., $\nnz(X)\approx nd$, and 2) $d\gg n$. In such a scenario, Theorem 1 of~\cite{akk+20}, which succeeds with probability $9/10$, obtains a sketching dimension $m=\Omega(\epsilon^{-2}n^2p)$ and the runtime of applying the sketch to $X$ is $\tilde O(pnd+\epsilon^{-2}n^3p^2)$. Theorem 3 of~\cite{akk+20} fails with probability only $1/\poly(n)$, and obtains $m=\tilde\Theta(\epsilon^{-2}np^4)$ with runtime $\tilde O(\epsilon^{-2}p^5(nd+n^2))$. Another result from~\cite{wz20} has $m=\tilde \Theta(\epsilon^{-2}n)$ but the time to apply sketching is $\tilde O(p^{2.5}nd+\poly(\epsilon^{-1},p)n^3)$, which is much worse in the leading $nd$ term, compared to our result. Compared with Theorems 1 and 2 of~\cite{akk+20} and with~\cite{wz20}, our result improves the dependence on the $nd$ term and pays only $n^2$ instead of $n^3$ on the second term. Theorem 3 of~\cite{akk+20} already has an $n^2$ second term. Compared with it, our result improves the first term from $\epsilon^{-2}p^5nd$ to $nd$, and the dependence on $p$ in the $n^2$ term from $p^5$ to $p^2$. So, up to logarithmic factors, our running time is at most that of each of Theorems 1--3 of~\cite{akk+20}, and it improves the leading $nd$ term of~\cite{wz20}. However, we also point out the results obtained in these two works are more general than ours in the sense that their sketches have the {\ose} property, while our sketch only preserves the column space of $X^{\otimes p}$. Nevertheless, the latter suffices for our applications. We use Table~\ref{tab:comp_main} to summarize and compare the different results.

\begin{table*}[ht]
\begin{center}
\begin{tabular}{|l|l|l|}
\hline
{\bf Reference} & {\bf Sketch Dimension} & {\bf Running Time} \\
\hline
\begin{tabular}[c]{@{}l@{}}\cite[Theorem 1]{akk+20}\\ success probability $9/10$\end{tabular} & $\Theta(\epsilon^{-2}n^2p)$ & $\tilde O(pnd+\epsilon^{-2}n^3p^2)$ \\
\hline
\begin{tabular}[c]{@{}l@{}}\cite[Theorem 3]{akk+20}\\ failure probability $1/\poly(n)$\end{tabular} & $\tilde \Theta(\epsilon^{-2}np^4)$ & $\tilde O(\epsilon^{-2}p^5(nd+n^2))$ \\
\hline
\cite{wz20}  & $\tilde \Theta(\epsilon^{-2}n)$ & $\tilde O(p^{2.5}nd+\poly(\epsilon^{-1},p)n^3)$ \\
\hline
Theorem~\ref{thm:SRHT+TensorSRHT} & $\tilde \Theta(\epsilon^{-2}np^2)$ & $\tilde O(nd+\epsilon^{-2}n^2p^2)$ \\
\hline
\end{tabular}
\end{center}
\caption{Comparison of different algorithms. We assume $\nnz(X)\approx nd$ and $d\gg n$. We also assume there is no regularization, i.e., $\lambda=0$. The bound of~\cite{wz20} is stated for $\lambda>0$.}
\label{tab:comp_main}
\end{table*}
Also, the prior results mentioned are stated in terms of the statistical dimension, while we do not directly obtain bounds in terms of the statistical dimension, though our sketches can be composed with sketches that do. Therefore, we consider the case when there is no regularization $(\lambda = 0)$ and $X^{\otimes p}$ is of full rank. In this case, the statistical dimension reduces to $n$. 

\section{Applications}\label{sec:application}
In this section, we introduce various applications using our sketch.
In Section~\ref{sec:app_gaussian}, we study approximating the Gaussian kernel using our algorithm. In Section~\ref{sec:app_pconv} we extend the analysis to a class of slow-decaying kernels. In Section~\ref{sec:app_precondition} we illustrate an efficient algorithm to solve kernel linear systems. In Section~\ref{sec:app_krr} we show how to solve kernel ridge regression using our sketch.

\subsection{Gaussian Kernels}\label{sec:app_gaussian}
We provide a fast algorithm to approximate a Gaussian kernel matrix $G$ up to a ridge term when $d$ is large, that is, to compute $W_g(X)$ with $W_g(X)^\top W_g(X)+\lambda I_n\approx_\epsilon G+\lambda I_n$ for a given $\lambda>0$.
\begin{theorem}[Gaussian Kernel, informal version of Theorem~\ref{thm:gaussian_kernel:app}]\label{thm:gaussian_kernel}
Let $X\in \R^{d\times n}$ be the data matrix and $r\in \R_+$ be the radius of $X$, i.e., $\|x_i\|_2\leq r$ for all $i\in [n]$, where $x_i$ is the $i^{th}$ column of $X$. Suppose $G\in \R^{n\times n}$ is the Gaussian kernel matrix given in Definition \ref{def:gaussian_kernel}. For any $\epsilon\in(0,1)$ and any ridge parameter $\lambda>0$, there exists an algorithm that computes a matrix $W_g(X)\in \R^{m\times n}$ in time $\tilde O(\epsilon^{-2}n^2q^3+nd)$, such that with probability at least $1-1/\poly(n)$,
\begin{align*}
(1-\epsilon)(G+\lambda I_n)\preceq W_g(X)^\top W_g(X)+\lambda I_n\preceq (1+\epsilon)(G+\lambda I_n),
\end{align*}
where $m=\tilde \Theta(q^3n/\epsilon^2)$ and $q=\Theta(r^2+\log(n/(\epsilon\min(\lambda,1))))$.
\end{theorem}
We provide a sketch of the proof here, and further details can be found in the appendix. The Taylor expansion of the Gaussian kernel can be written as
\begin{align*}
G & = \sum_{l=0}^\infty \frac{(X^{\otimes l}D)^\top X^{\otimes l}D}{l!}
\end{align*}
where $D$ is a diagonal matrix with $D_{i,i}=\exp(-\|x_i\|_2^2/2)$. Let
\begin{align*}
K & = \sum_{l=0}^\infty \frac{(X^{\otimes l})^\top X^{\otimes l}}{l!}.
\end{align*}
If we set $q=\Omega(r^2+\log(n/(\epsilon\min(\lambda,1))))$ and just use the first $q$ terms of $K$:
\begin{align*}
Q & = \sum_{l=0}^q \frac{(X^{\otimes l})^\top X^{\otimes l}}{l!},
\end{align*}
then we have that $\|K-Q\|_{\mathrm{op}}\leq \frac{\epsilon\lambda}{2}$. This additive error is absorbed by the ridge term $\lambda I_n$, which gives the guarantee above. Our algorithm applies Algorithm \ref{alg1} on each term of $Q$. This gives us the desired runtime and dimension. The complete proof is in Appendix \ref{sec:gaussian}.

\subsection{General $p$-convergent Kernels}\label{sec:app_pconv}
A key advantage of Algorithm \ref{alg1} is its moderate dependence on the degree of the polynomial kernel, which gives it more leverage when that degree is large. We introduce a characterization of kernels, based on the series of the coefficients in the Taylor expansion of the kernel. As we will later see in the proof of Theorem \ref{thm:general_sequence:app}, the decay exponent $p$ of the coefficients determines the truncation degree $q$ we need for approximating a kernel.
\begin{definition}[$p$-convergent kernel, informal version of Definition~\ref{def:p_convergent_restate}]\label{def:p_convergent}
We say the kernel matrix $K$ for data matrix $X$ is $p$-convergent if its corresponding Taylor expansion series can be written as follows:
$K = \sum_{l=0}^{\infty} C_l \cdot ( X^{\otimes l} )^\top X^{\otimes l}$, 
where $p>1$ is a real number and the coefficients satisfy $0\leq C_l\leq c(l+1)^{-p}$ for all $l\geq 0$, for an absolute constant $c>0$.
\end{definition}

\begin{theorem}[Sketch $p$-convergent Kernels, informal version of Theorem~\ref{thm:general_sequence:app}]\label{thm:general_sequence}
Let $X\in \R^{d\times n}$ be the data matrix with radius $r\leq 1$. Let $p> 1$ be a real number, suppose $K$ is a $p$-convergent matrix with constant $c$, let $\epsilon\in(0,1)$, and let $\lambda>0$ be a ridge parameter. Let $m =  \tilde \Theta(\epsilon^{-2}nq^3)$ and $q = \ceil{(2cn/((p-1)\epsilon\lambda))^{1/(p-1)}}$. There exists an algorithm that computes a matrix $W_k(X)\in \R^{m\times n}$ in time $\tilde O(\epsilon^{-2}n^2q^3+nd)$ such that with probability at least $1-1/\poly(n)$,
	\begin{align*}
		(1-\epsilon)(K+\lambda I_n)\preceq W_k(X)^\top W_k(X)+\lambda I_n\preceq (1+\epsilon)(K+\lambda I_n).
	\end{align*}
\end{theorem}
For the sake of illustration, when $n/(\epsilon\lambda)$ is large, the first term in the running time becomes $\epsilon^{-2}n^2(n/(\epsilon\lambda))^{3/(p-1)}$, up to the factor $(2c/(p-1))^{3/(p-1)}$. When $p$ is large, Theorem~\ref{thm:general_sequence} gives a fast algorithm for approximating the kernel, but the runtime becomes much slower when $p\in(1,4)$. Therefore, for $1<p<3$ we also give a sampling scheme. Roughly speaking, we sketch the first $s$ terms of the Taylor expansion, and replace the remaining terms up to degree $q$ by a few degrees sampled with probability proportional to their coefficients $C_l$, reweighted so that the estimate stays unbiased. Using a matrix Bernstein bound (Theorem~\ref{thm:bernstein}) relative to the ridge term, we obtain a faster algorithm. With $N=n/(\epsilon\lambda)$, Corollary~\ref{cor:sampling_p12} gives the guarantee of Theorem~\ref{thm:general_sequence} in expected time
\begin{align*}
\tilde O\big(nd+\epsilon^{-2}n^2+\epsilon^{-3}n^2N^{2/(p-1)}\min\{1,(\epsilon^{p+2}N)^{1/p}\}\big),
\end{align*}
where $\tilde O$ also hides factors that depend only on $c$ and $p$. Up to such factors, this expected-time bound is never larger than the time bound of Theorem~\ref{thm:general_sequence}, and apart from the common $nd$ term it is smaller by a factor polynomial in $N$ when $N$ is large. Evaluating the truncated series directly, by computing $X^\top X$ in $O(n^2d)$ time, the entries $\sum_{l=0}^{q}C_l(x_i^\top x_j)^l$ in $O(n^2q)$ time and a factorization in $O(n^\omega)$ time, where $\omega$ is the exponent of matrix multiplication, gives the same guarantee, so this bound is smaller only when $n^2d+n^\omega$ is larger than about $\epsilon^{-3}n^2N^{2/(p-1)}\min\{1,(\epsilon^{p+2}N)^{1/p}\}$, up to factors depending on $c$ and $p$ and polylogarithmic factors. We apply our results to the neural tangent kernel ({\ntk}), which is a $1.5$-convergent kernel on unit-norm data.
\begin{corollary}[Approximate {\ntk}, informal version of Corollary~\ref{cor:ntk}]
Let $X\in \R^{d\times n}$ be a data matrix with unit-norm columns, that is, $\|x_i\|_2=1$ for all $i\in[n]$. Suppose $\k\in \R^{n\times n}$ is the {\ntk} matrix, and let $\epsilon\in(0,1)$ and $\lambda>0$. There exist two algorithms that compute a matrix $W_k(X)$ such that with probability at least $1-1/\poly(n)$,
\begin{align*}
(1-\epsilon)(\k+\lambda I_n)\preceq W_k(X)^\top W_k(X)+\lambda I_n\preceq (1+\epsilon)(\k+\lambda I_n).
\end{align*}
The first sketches every term of the truncated series and runs in time
\begin{align*}
\tilde O(\epsilon^{-2}n^2q^3+nd), \quad\text{where } q=\ceil{16n^2/(\epsilon^2\lambda^2)},
\end{align*}
which is $\tilde O(\epsilon^{-8}\lambda^{-6}n^8+nd)$ when $\lambda\leq 4n/\epsilon$. The second samples the degrees of the series in bands of doubling length (Theorem~\ref{thm:ntk_banded}) and runs in time
\begin{align*}
\tilde O(\epsilon^{-4}\lambda^{-4}n^6+nd).
\end{align*}
\end{corollary}
The {\ntk} has a closed form that gives $\k$ exactly in $O(n^2d)$ time, plus $O(n^\omega)$ time to factor it (with fast rectangular matrix multiplication, $X^\top X$ takes $O(dn^{\omega-1})$ time when $d\geq n$, which raises the threshold for the second bound to about $n^{7-\omega}/(\epsilon\lambda)^4$). So, up to polylogarithmic factors, the first bound is faster only when $d$ is larger than about $\epsilon^{-8}\lambda^{-6}n^6$ or $\epsilon^{-8}\lambda^{-6}n^8$ is below $n^\omega$, and the second only when $d$ is larger than about $(n/(\epsilon\lambda))^4$ or $\epsilon\lambda$ is larger than about $n^{(6-\omega)/4}$. The second alternative in each case accounts for the fact that both algorithms already output a factor. When the kernel matrix itself is the output, projecting the data to $\tilde O(n^2/(\epsilon\lambda)^2)$ dimensions and evaluating the closed form exactly on the normalized projected points gives the same guarantee and is cheaper still (Remark~\ref{rem:ntk_jl}). In the cost model of our sketch, when $\epsilon\leq 1/4$ and $\epsilon\lambda$ is at most a small constant times $n$, a scheme whose output is a nonnegatively weighted sum of per-degree sketches, whose degrees and weights are drawn from a distribution that does not depend on the data, which fails with probability at most $1/8$, and which, except with probability at most $1/8$, sketches every used degree to within a factor in $[\frac12,\frac32]$, cannot improve on the second bound by more than polylogarithmic factors (Observation~\ref{obs:ntk_floor}).
\begin{remark}
We remark that our definition of $p$-convergent kernels captures a wide range of kernels that have slow decay rate in their coefficients in their Taylor expansion, such as {\ntk} and arc-cosine kernels. Typically, the coefficients are of the form $1/n^c$ for some $c>1$. In contrast, Gaussian kernels enjoy a much faster decay rate, and therefore, designing algorithm for the Gaussian kernel is considerably simpler, since the number of terms we need to approximate it with in its Taylor expansion is small, and no sampling is necessary.
\end{remark}

\subsection{Kernel Linear Systems}\label{sec:app_precondition}

Another interesting application of our sketching scheme is to constructing a preconditioner for solving PSD systems involving a kernel matrix \cite{cocf16}. In order to apply algorithms such as Conjugate Gradient \cite{s94}, one has to obtain a good preconditioner for a potentially ill-conditioned kernel system.
\begin{theorem}[Sketching as a Preconditioner, informal version of Theorem \ref{thm:precondition_main}]
\label{thm:precondition}
Let $G\in \R^{n\times n}$ be the Gaussian kernel matrix for $X\in \R^{d\times n}$ and suppose $\|x_i\|_2\leq 1,\forall i\in [n]$, where $x_i$ is the $i^{\text{th}}$ column of $X$. Let $G=Z^\top Z$ and $\kappa$ denote the condition number of $Z$, and write $\|v\|_{G^{-1}}=\sqrt{v^\top G^{-1}v}$ for $v\in\R^n$. 
There exists an algorithm that, given an upper bound $\ov\kappa\geq\kappa$, with probability at least $1-\delta$, computes an $\epsilon$-approximate solution $\wh x$ satisfying
\begin{align*}
\|G\wh x-y\|_{G^{-1}}\leq & ~ \epsilon\|y\|_{G^{-1}},
\end{align*}
and hence $\|G\wh x-y\|_2\leq\epsilon\kappa\|y\|_2$, in $\tilde O(\epsilon^{-2}n^2+n^\omega+nd)$ time, where $\tilde O$ also hides $\poly(\log\ov\kappa)$ factors and $\omega$ is the exponent of matrix multiplication (currently $\omega \approx 2.373$~\cite{w12,l14}).
\end{theorem}
\begin{remark}
In certain NLP~\cite{dl20} and biological tasks~\cite{tpk02} where $d=n^c$ for a positive integer $c$, Theorem \ref{thm:precondition} provides a fast algorithm for which the running time depends nearly linearly on $nd$. We also remark that the algorithm we use for Theorem \ref{thm:precondition} is inspired by the idea of \cite{bpsw21} (their situation involves $c=4$). It is also interesting that in their applications, regularization is not needed since solving a kernel system is equivalent to training an over-parametrized ReLU network without regularization.
\end{remark}

\subsection{Kernel Ridge Regression}\label{sec:app_krr}
Kernel ridge regression ({\krr}) is a popular method to model the relationship between data points and labels. Let $K=A^\top A$ denote the kernel matrix. Instead of solving the ordinary regression $\min_{x\in \R^n} \|Kx-y\|_2^2$, which is equivalent to solving a linear system, we focus on solving the following ridge regression problem:
\begin{align*}
\min_{x\in \R^n}~\|Kx-y\|_2^2+\lambda \|Ax\|_2^2,
\end{align*}
for $\lambda>0$. A relevant notion is the \emph{statistical dimension}:
\begin{definition}[Statistical Dimension]
Let $\lambda>0$, and $K\in \R^{n\times n}$ be a positive semi-definite matrix. We define the $\lambda$-statistical dimension of $K$ to be
\begin{align*}
s_\lambda(K) := & ~ \Tr[K(K+\lambda I_n)^{-1}].
\end{align*}
\end{definition}
One drawback of our sketch is that we cannot obtain a dimension depending on $s_\lambda(K)$ instead of $n$, since it does not have the approximate matrix product property. To mitigate this effect, we propose the following \emph{composition of sketches}. 
\begin{theorem}[Kernel Ridge Regression, informal version of Theorem~\ref{thm:KRR:app}]
\label{thm:KRR}
Let $\epsilon\in (0,1/2]$, $\delta\in(0,1)$, $p>1$ be an integer and $X\in \R^{d\times n}$. Let $K$ be its degree-$p$ polynomial kernel, assumed to have full rank, with statistical dimension $s_{\lambda}(K)$, where $0<\lambda<\epsilon^{-1}(1-\epsilon)\lambda_{\max}(K)$, and let $t=\tilde O(\max(\epsilon^{-2},\epsilon^{-1}\lambda_{\max}(K)/\lambda)p^2n)$, where $\tilde O$ also hides $\poly(\log(\lambda_{\max}(K)/\lambda))$ factors. Then we can compute $Z$ such that, with probability at least $1-\delta/2$, $Z$ has at most $t$ rows, is computed in $\tilde O(tn+nd+n^\omega)$ time, and $Z^\top Z$ is a $(1\pm\epsilon)$-spectral approximation to $K$. Moreover, let $m=\tilde O(\epsilon^{-1}(s_\lambda(K)+1))$, and let $\wh x$ be the best, in the value of $\|Z^\top Zx-y\|_2^2+\lambda \|Zx\|_2^2$, of the optimal solutions to $\|S(Z^\top Zx-y)\|_2^2+\lambda \|Zx\|_2^2$ for $O(\log(1/\delta))$ independent {\srht} matrices $S$ with $m$ rows. Then with probability at least $1-\delta$,
\begin{align*}
\|K\wh x-y\|_2^2+\lambda \|X^{\otimes p}\wh x\|_2^2 
\leq & ~ (1+\epsilon)\cdot \mathrm{OPT},
\end{align*} 
where $\mathrm{OPT}$ is $\min_{x\in \R^n}\|Kx-y\|_2^2+\lambda \|X^{\otimes p}x\|_2^2$. The time to compute $\wh x$ from $Z$ is 
\begin{align*}
\tilde O(t(n+m^2)+n^\omega).
\end{align*}
\end{theorem}

{\bf Acknowledgments:} D. Woodruff would like to thank partial support from NSF grant No. CCF-1815840, Office of Naval Research grant N00014-18-1-2562, and a Simons Investigator Award. The corrections in this version were found and written with the help of Claude Opus 5.5. The authors have verified the corrections and take full responsibility for the content of this paper.

\newpage
\addcontentsline{toc}{section}{References}
\ifdefined\isarxiv
\bibliographystyle{alpha}
\else
\bibliographystyle{icml2021}
\fi
\bibliography{ref}
\newpage
\appendix
\onecolumn

\section*{Appendix}
\paragraph{Roadmap}

In Section~\ref{sec:gaussian}, we give an algorithm to compute a subspace embedding for the Gaussian kernel up to a ridge term using Theorem~\ref{thm:main_result}. In Section~\ref{sec:general}, we characterize a large class of kernels based on the coefficients in their Taylor expansion, and develop a fast algorithm for them that sketches a truncation of the Taylor series, together with a sampling scheme over the degrees of the series for $1<p<3$. In Section~\ref{sec:ntk}, we apply our results in Section~\ref{sec:general} to the Neural Tangent kernel, and give a sampling scheme for it that samples the degrees in bands. In Section~\ref{sec:precondition}, we use our sketch in conjunction with another sketch to compute a good preconditioner for the Gaussian kernel. In Section~\ref{sec:KRR}, we compose our sketch with our sketching matrices to solve Kernel Ridge Regression.

\paragraph{Notation}
We use $\tilde{O}(f)$ to denote $f \poly (\log f)$ and use $\tilde{\Omega}(f)$ to denote $f/\poly(\log f)$.

For an integer $n$, let $[n]$ denote the set $\{1,2,\cdots, n\}$. For two scalars $a$ and $b$, we say $a \approx_{\epsilon} b$ if $(1-\epsilon) b \leq a \leq (1+\epsilon) b$. We say a square symmetric matrix $A$ is positive semidefinite (PSD) if $\forall x$, $x^\top A x \geq 0$. For two PSD matrices $A$ and $B$, we define $A \approx_{\epsilon} B$ if  $(1-\epsilon)B \preceq A \preceq  (1+\epsilon) B$. For a matrix $A$, we use $\| A \|_F = (\sum_{i,j} A_{i,j}^2)^{1/2}$ to denote its Frobenius norm and use $\| A \|_{\op}$ to denote its operator (spectral) norm. For a square matrix $A$, we use $\Tr[A]$ to denote the trace of $A$. For a square symmetric matrix $A$, we use $\lambda_{\min}(A)$, $\lambda_{\max}(A)$ to denote its smallest and largest eigenvalues, respectively. For a rectangular matrix $A$, we use $\sigma_{\min}(A),\sigma_{\max}(A)$ to denote its smallest and largest singular values.

\section{Gaussian Kernel}\label{sec:gaussian}
We apply Algorithm~\ref{alg1} to compute a subspace embedding, up to a ridge term, to the Gaussian kernel matrix $G\in\R^{n\times n}$ defined over $n$ data points of dimension $d$, denoted by $X\in\R^{d\times n}$. Our method has the advantage that when $d$ is large and the matrix $X$ is dense, its leading factor depends nearly linearly on $nd$, which makes it useful for certain biological and NLP tasks.

We remark that our construction of a sketch for the Gaussian kernel and its corresponding analysis is inspired by~\cite{akk+20}, and is thus similar to the proof of Theorem 5 in their paper. For completeness, we include a proof here.
\begin{theorem}[Gaussian Kernel, formal version of Theorem~\ref{thm:gaussian_kernel}]\label{thm:gaussian_kernel:app}
Let $X\in \R^{d\times n}$ and $r\in \R_+$ be the radius of $X$. Suppose $G\in \R^{n\times n}$ is the Gaussian kernel matrix given in Definition \ref{def:gaussian_kernel}. For any accuracy parameter $\epsilon \in (0,1)$, for any ridge parameter $\lambda>0$ and for any failure probability $\delta \in (0,1)$, there exists an algorithm running in time:
\begin{align*}
O(\epsilon^{-2} n^2 q^3 \cdot \log^3(ndq/\epsilon\delta)+nd\log(nd/\epsilon\delta) )
\end{align*}
and outputting a matrix $W_g(X)\in \R^{m\times n}$ such that
\begin{align*}
\pr{[(1-\epsilon)(G+\lambda I_n)\preceq W_g(X)^\top W_g(X)+\lambda I_n\preceq (1+\epsilon)(G+\lambda I_n)]}\geq 1-\delta
\end{align*}
where $m=\Omega( \epsilon^{-2} n q^3\log^3(ndq/\epsilon\delta))$ and $q=\Theta(r^2+\log(n/(\epsilon\min(\lambda,1))))$.
\end{theorem}

\begin{proof}
By definition of the Gaussian kernel matrix $G_{i,j} = \exp( -\|x_i - x_j\|_2^2/2 )$, we can rewrite it as $G = DKD$, where $D$ is an $n \times n$ diagonal matrix with $i^{\text{th}}$ diagonal entry equal to $\exp( - \|x_i\|_2^2 / 2 )$ and $K \in \R^{n \times n}$ is a positive definite kernel matrix defined as $K_{i,j} = \exp( x_i^\top x_j )$.
Note the Taylor series expansion for kernel $K$ gives
\begin{align*}
K = \sum_{l=0}^\infty \frac{(X^{\otimes l})^\top X^{\otimes l}}{l!}.
\end{align*}
Let $q=\ceil{2er^2+\log_2(2n/(\epsilon\min(\lambda,1)))}$, so that $q=\Theta(r^2+\log(n/(\epsilon\min(\lambda,1))))$, and let $Q = \sum_{l=0}^q\frac{(X^{\otimes l})^\top X^{\otimes l}}{l!}$ be the first $q$ terms of $K$. Every entry of $(X^{\otimes l})^\top X^{\otimes l}$ is $(x_i^\top x_j)^l$, whose absolute value is at most $r^{2l}$, so $\|(X^{\otimes l})^\top X^{\otimes l}\|_F\leq n r^{2l}$. For $l>q$ we have $l\geq 2er^2$, and $l!\geq (l/e)^l$ gives
\begin{align*}
\frac{r^{2l}}{l!}\leq \Big(\frac{er^2}{l}\Big)^l\leq 2^{-l}.
\end{align*}
By the triangle inequality we have:
\begin{align*}
\|K-Q\|_{\op} &\leq \sum_{l>q} \left\|\frac{(X^{\otimes l})^\top X^{\otimes l}}{l!}\right\|_{\op}\\
&\leq \sum_{l>q} \left\|\frac{(X^{\otimes l})^\top X^{\otimes l}}{l!}\right\|_{F}\\
&\leq \sum_{l>q} \frac{n\cdot r^{2l}}{l!}\\
&\leq n\sum_{l>q} 2^{-l} = n 2^{-q}\\
&\leq \epsilon\min(\lambda,1)/2\leq \epsilon\lambda/2,
\end{align*}
where the last line uses $q\geq \log_2(2n/(\epsilon\min(\lambda,1)))$. The matrix $K-Q=\sum_{l>q}\frac{(X^{\otimes l})^\top X^{\otimes l}}{l!}$ is a sum of PSD matrices, and $D$ is diagonal with entries in $(0,1]$, so $\|D\|_{\op} \leq 1$. Since $G=DKD$, we get
\begin{align}\label{eq:gauss_trunc}
0\preceq G-DQD = D(K-Q)D\preceq \|K-Q\|_{\op}\cdot D^2\preceq \frac{\epsilon\lambda}{2} I_n.
\end{align}
We will show that the following holds with probability at least $1-\delta$:
\begin{align*}
(1-\epsilon/2) \cdot DQD \preceq W_g(X)^\top W_g(X) \preceq (1 + \epsilon/2) \cdot DQD.
\end{align*}
For $l=0$ we have $X^{\otimes 0}=\mathbf{1}_n^\top$, and we set $Z_0=\mathbf{1}_n^\top$, which is exact, has one row and takes $O(n)$ time. For each term $(X^{\otimes l})^\top X^{\otimes l}$ with $1\leq l\leq q$, we run Algorithm~\ref{alg1} to approximate $X^{\otimes l}$. Let $Z_l\in \R^{m_l\times n}$ be the resulting matrix ${\cal Z}({\cal S},T,X)$ for degree $l$, where
\begin{align*}
m_l=\Omega( \epsilon^{-2} n q^2 \cdot \log^2(ndq/\epsilon\delta) \cdot \log(nq/\delta) )
\end{align*}
(the dimension is set by degree $q$ because the degrees share sketches, see below). The factor $q$ inside the logarithms comes from the failure probability $\frac{\delta}{q+1}$ that we use for each degree. Then by Theorem~\ref{thm:SRHT+TensorSRHT}, we get
\begin{align}\label{eq:gauss_sketch_l}
(1-\epsilon/2)(X^{\otimes l} D)^\top X^{\otimes l} D \preceq (\Pi^l X^{\otimes l} D)^\top \Pi^l X^{\otimes l} D \preceq (1 + \epsilon/2) (X^{\otimes l}D)^\top X^{\otimes l} D
\end{align}
with probability at least $1-\frac{\delta}{q+1}$. Moreover, $Z_l$ can be computed in time
\begin{align*}
O( \epsilon^{-2} n^2 q^2 \cdot \log^2(ndq/\epsilon\delta) \cdot \log(nq/\delta) ).
\end{align*}
Our algorithm will simply compute $Z_l$ from $l=0$ to $q$, normalize each $Z_l$ by $\frac{1}{\sqrt{l!}}$, and then multiply by $D$. 
More precisely, the approximation $W_g(X)$ will be
\begin{align*}
W_g(X) & = \Big( \bigoplus_{l=0}^q \frac{Z_l}{\sqrt{l!}} \Big) D
\end{align*}
where we use $A\oplus B$ to denote the matrix 
$
\begin{bmatrix}
A \\
B
\end{bmatrix}
$ 
if $A$ and $B$ have the same number of columns. Notice $W_g(X)\in \R^{m\times n}$. The following holds for $W_g(X)^\top W_g(X)$:
\begin{align*}
W_g(X)^\top W_g(X) & = D \Big( \sum_{l=0}^q \frac{Z_l^\top Z_l}{l!} \Big) D \\
& = \sum_{l=0}^q \frac{(Z_lD)^\top Z_lD}{l!}. 
\end{align*}
By combining terms in \eqref{eq:gauss_sketch_l} and using a union bound over all $0\leq l\leq q$, we obtain that with probability at least $1-\delta$, we have the following:
\begin{align*}
(1-\epsilon/2) \cdot DQD \preceq W_g(X)^\top W_g(X) \preceq (1+\epsilon/2) \cdot DQD.
\end{align*}
The union bound needs no independence between different degrees. So the degrees may share $T$ and the level sketches, as described at the end of this proof, provided all base sketches are taken with the accuracy, failure probability and dimension for degree $q$, which only strengthens the guarantee for each degree $l\leq q$. For each fixed $l$, the base sketches that produce $Z_l$ are still mutually independent, so \eqref{eq:gauss_sketch_l} still holds with probability at least $1-\frac{\delta}{q+1}$.

We now add the ridge term. Since $G\succeq 0$, we have $\frac{\epsilon\lambda}{2}I_n\preceq\frac{\epsilon}{2}(G+\lambda I_n)$. Together with \eqref{eq:gauss_trunc}, this gives
\begin{align*}
W_g(X)^\top W_g(X)+\lambda I_n & \succeq (1-\epsilon/2)(DQD+\lambda I_n)\\
& \succeq (1-\epsilon/2)\Big(G+\lambda I_n-\frac{\epsilon\lambda}{2}I_n\Big)\\
& \succeq (1-\epsilon/2)^2(G+\lambda I_n)\\
& \succeq (1-\epsilon)(G+\lambda I_n),
\end{align*}
where the first step also uses $\lambda\geq(1-\epsilon/2)\lambda$, and the last step uses $(1-\epsilon/2)^2=1-\epsilon+\epsilon^2/4\geq 1-\epsilon$. For the upper bound, $DQD\preceq G$ by \eqref{eq:gauss_trunc}, so
\begin{align*}
W_g(X)^\top W_g(X)+\lambda I_n\preceq (1+\epsilon/2)DQD+\lambda I_n\preceq (1+\epsilon/2)(G+\lambda I_n)\preceq (1+\epsilon)(G+\lambda I_n).
\end{align*}
Thus, we conclude that with probability at least $1-\delta$,
\begin{align*}
(1-\epsilon)(G+\lambda I_n) \preceq W_g(X)^\top W_g(X)+\lambda I_n\preceq (1+\epsilon)(G+\lambda I_n).
\end{align*}
Note the target dimension of $W_g$ is
\begin{align*}
m = & ~ m_0+m_1+\cdots + m_q \\
= & ~ \Omega( \epsilon^{-2} n q^3 \cdot \log^2(ndq/\epsilon\delta) \cdot \log(nq/\delta) ).
\end{align*}
 Also, by Theorem~\ref{thm:SRHT+TensorSRHT}, the time to compute $W_g(X)$ is
\begin{align*}
t = & ~ t_0 + t_1 + \cdots + t_q \\
= & ~ O(\epsilon^{-2} n^2 q^3 \cdot \log^2(ndq/\epsilon\delta) \cdot \log(nq/\delta) ).
\end{align*}
Notice we will have to pay an additive $nd\log(nd/\epsilon\delta)$ due to line 3 of Algorithm~\ref{alg1}, when applying the SRHT to $X$. However, we only need to perform this operation once for the term with the highest degree, or the terms with lower degree that can be formed by combining nodes computed with the highest degree. Thus, the final runtime is
\begin{align*}
O(\epsilon^{-2} n^2 q^3 \cdot \log^2(ndq/\epsilon\delta) \cdot \log(nq/\delta) +nd\log(nd/\epsilon\delta) ).
\end{align*}
\end{proof}
\section{General $p$-Convergent Sequences}\label{sec:general}
We consider general $p$-convergent kernels defined below in Definition~\ref{def:p_convergent_restate}. We apply our proposed Algorithm~\ref{alg1} to compute a subspace embedding of the kernel matrix, up to a ridge term.

\subsection{General Theorem for $p>1$}
In this section, we state a general theorem for $p>1$. The proof is similar to the proof for Theorem~\ref{thm:gaussian_kernel:app}. We start by restating the definition of a $p$-convergent kernel.

\begin{definition}[$p$-convergent kernel matrix, formal version of Definition~\ref{def:p_convergent}]\label{def:p_convergent_restate}
Given an input matrix $X \in \R^{d \times n}$, we say the kernel matrix $K \in \R^{n \times n}$ is $p$-convergent if its corresponding Taylor expansion series can be written as follows:
\begin{align*}
K = \sum_{l=0}^{\infty} C_l \cdot ( X^{\otimes l} )^\top X^{\otimes l},
\end{align*}
where $p>1$ is a real number, and the coefficients $C_l$ are a function of $l$ that satisfy
\begin{align*}
0\leq C_l \leq c\cdot (l+1)^{-p} \quad \text{for all } l\geq 0
\end{align*}
for an absolute constant $c>0$.
\end{definition}

\begin{theorem}[Sketch for $p$-convergent Kernels, formal version of Theorem~\ref{thm:general_sequence}]\label{thm:general_sequence:app}
Let $X\in \R^{d\times n}$ with radius $r\leq 1$. Suppose that $K$ is a $p$-convergent kernel matrix where $p>1$ is a real number, and let $c$ be the constant in Definition~\ref{def:p_convergent_restate}. Let $\epsilon\in(0,1)$ be an accuracy parameter, $\lambda>0$ a ridge parameter and $\delta\in(0,1)$ a failure probability. Further, let
\begin{align*}
q = \ceil{\Big(\frac{2cn}{(p-1)\epsilon\lambda}\Big)^{1/(p-1)}}
\end{align*}
and choose $m =   \Omega(\epsilon^{-2}nq^3\log^3(ndq/\epsilon\delta))$. There exists an algorithm which computes a matrix $W_k(X)\in \R^{m\times n}$ in time
\begin{align*}
O(\epsilon^{-2}n^2q^3 \cdot \log^3(ndq/\epsilon\delta) +nd \cdot \log(nd/\epsilon\delta))
\end{align*}
 such that
	\begin{align*}
		\pr{[(1-\epsilon)(K+\lambda I_n)\preceq W_k(X)^\top W_k(X)+\lambda I_n\preceq (1+\epsilon)(K+\lambda I_n)]}\geq 1-\delta.
	\end{align*}
\end{theorem}

\begin{proof}
Similar to the Gaussian kernel, here we use the first $q$ terms to approximate the kernel matrix $K$.

Let $Q = \sum_{l=0}^q C_l(X^{\otimes l})^\top X^{\otimes l}$ be the first $q$ terms of $K$. Since $r\leq 1$, every entry of $(X^{\otimes l})^\top X^{\otimes l}$ has absolute value at most $1$, so $\|(X^{\otimes l})^\top X^{\otimes l}\|_F\leq n$. By the triangle inequality and $0\leq C_l\leq c(l+1)^{-p}$, we have
\begin{align*}
\|K-Q\|_{\op} &\leq \sum_{l>q} C_l\left\|(X^{\otimes l})^\top X^{\otimes l}\right\|_{\op}\\
&\leq \sum_{l>q} C_l\left\|(X^{\otimes l})^\top X^{\otimes l}\right\|_{F}\\
&\leq cn\sum_{l\geq q+1} (l+1)^{-p}\\
&\leq cn\int_{q+1}^\infty x^{-p}\d x\\
&= \frac{cn(q+1)^{1-p}}{p-1}\\
&\leq \frac{cnq^{1-p}}{p-1}\\
&\leq \epsilon\lambda/2.
\end{align*}
Here the fourth step uses $(l+1)^{-p}\leq\int_{l}^{l+1}x^{-p}\d x$, which holds since $x^{-p}$ is decreasing, the sixth step uses $1-p<0$, and the last step uses $q^{p-1}\geq\frac{2cn}{(p-1)\epsilon\lambda}$. Since $C_l\geq 0$, the matrix $K-Q$ is a sum of PSD matrices, so
\begin{align*}
0\preceq K-Q\preceq\frac{\epsilon\lambda}{2}I_n.
\end{align*}

The rest of the proof is identical to the proof of Theorem~\ref{thm:gaussian_kernel:app}, with $D=I_n$, $G$ replaced by $K$ and $W_g$ replaced by $W_k$. Degrees $l$ with $C_l=0$ do not contribute to $Q$, and we skip them. If $C_0>0$, we use the exact row $\sqrt{C_0}\,\mathbf{1}_n^\top$. For each other $1\leq l\leq q$, we compute $Z_l$ as there and use $\sqrt{C_l}\,Z_l$ in place of $Z_l/\sqrt{l!}$. By the union bound over at most $q+1$ degrees, with probability at least $1-\delta$ we have $(1-\epsilon/2)Q\preceq W_k(X)^\top W_k(X)\preceq(1+\epsilon/2)Q$. Using $\frac{\epsilon\lambda}{2}I_n\preceq\frac{\epsilon}{2}(K+\lambda I_n)$ and $Q\preceq K$, the same chain as in the proof of Theorem~\ref{thm:gaussian_kernel:app} gives
\begin{align*}
W_k(X)^\top W_k(X)+\lambda I_n & \succeq (1-\epsilon/2)(Q+\lambda I_n)\\
& \succeq (1-\epsilon/2)\Big(K+\lambda I_n-\frac{\epsilon\lambda}{2}I_n\Big)\\
& \succeq (1-\epsilon/2)^2(K+\lambda I_n)\\
& \succeq (1-\epsilon)(K+\lambda I_n),\\
W_k(X)^\top W_k(X)+\lambda I_n & \preceq (1+\epsilon/2)Q+\lambda I_n\\
& \preceq (1+\epsilon/2)(K+\lambda I_n)\\
& \preceq (1+\epsilon)(K+\lambda I_n).
\end{align*}
The target dimension of $W_k$ is at most $m_0+m_1+\cdots + m_q = O(\epsilon^{-2}nq^3\log^2(ndq/\delta\epsilon)\log(nq/\delta))$, with $m_l$ as in the proof of Theorem~\ref{thm:gaussian_kernel:app}.

Similar to Theorem~\ref{thm:gaussian_kernel:app}, we have to pay an extra $nd\log(nd/\epsilon\delta)$ term to apply the {\srht} to $X$, so the final running time is
\begin{align*}
t_0 + t_1 + \cdots + t_q+nd\log(nd/\epsilon\delta) & = O(\epsilon^{-2}n^2q^3\log^2(ndq/\delta\epsilon)\log(nq/\delta)+nd\log(nd/\epsilon\delta)).
\end{align*}
\end{proof}
\begin{remark}
Recall our setting is when $d=\poly(n)$. Since $\ceil{y}\leq 2\max(y,1)$ for $y>0$, the degree in Theorem~\ref{thm:general_sequence:app} satisfies
\begin{align*}
q^3 = O\Big(1+\Big(\frac{2c}{p-1}\Big)^{3/(p-1)}\Big(\frac{n}{\epsilon\lambda}\Big)^{3/(p-1)}\Big),
\end{align*}
so its running time is $\tilde O(\epsilon^{-2}n^2+(\frac{2c}{p-1})^{3/(p-1)}\epsilon^{-2-3/(p-1)}\lambda^{-3/(p-1)}n^{2+3/(p-1)}+nd)$. If $p\geq 4$, then $3/(p-1)\leq 1$. So for a constant $\lambda$, the running time is $\tilde O(\epsilon^{-2-3/(p-1)}n^3+nd)$, where the hidden constant depends on $c$, $p$ and $\lambda$. This is better than the classical result of $O(n^2d)$ as long as $d$ is larger than about $\epsilon^{-2-3/(p-1)}n$. If $p\in (1,4)$, the exponent of $n$ is larger than $3$, and as $p$ approaches $1$ the exponent $3/(p-1)$ grows without bound, and so does the factor $(\frac{2c}{p-1})^{3/(p-1)}$. For $1<p<3$, the sampling scheme of Section~\ref{sec:sampling} improves this running time.
\end{remark}

\subsection{Sampling Scheme for $1<p<3$}\label{sec:sampling}
Theorem~\ref{thm:general_sequence:app} sketches every degree up to $q$, and sketching degree $l$ costs about $\epsilon^{-2}n^2l^2$, so its running time grows like $q^3$. When $1<p<3$, the degrees close to $q$ are the most expensive, but they carry little weight. We sketch the first $s$ degrees as before, and replace the remaining degrees up to $q$ by a few degrees sampled with probability proportional to $C_l$, reweighted so that the estimate is unbiased. Three points matter in the analysis. The sampling error is measured relative to the ridge term, through the normalization $(R+\lambda I_n)^{-1/2}$ below, so no lower bound on $\|K\|_{\op}$ is needed. A sampled degree $l$ costs about $l^2$, so the expected cost of a sample involves $\E[l^2]$ and not $(\E[l])^2$. And when the sampled part is light, each sample needs only a coarse accuracy. We use the following concentration bound.

\begin{theorem}[Matrix Bernstein Inequality, Theorem 6.1.1 in~\cite{t15}]\label{thm:bernstein}
Let $S_1,\ldots,S_m$ be independent, zero-mean random matrices with common size $d_1\times d_2$, and assume each one is uniformly bounded:
\begin{align*}
\E[S_k]=0,\quad \|S_k\|_{\op}\leq L,\quad k\in [m].
\end{align*}
Let $Z=\sum_{k=1}^m S_k$, and let $\Var[Z]=\max\{ \|\E[Z^\top Z]\|_{\op}, \|\E[ZZ^\top]\|_{\op} \}$. Then for all $t>0$,
\begin{align*}
\pr{[\|Z\|_{\op}\geq t]}\leq (d_1+d_2)\cdot \exp\left(\frac{-t^2/2}{\Var[Z]+Lt/3}\right).
\end{align*}
\end{theorem}
If the $S_k$ are symmetric $n\times n$ matrices, then $d_1+d_2=2n$, and since they are independent with mean zero, $\Var[Z]=\|\sum_{k=1}^m\E[S_k^2]\|_{\op}$.

\begin{algorithm}[!ht]
\caption{Degree sampling for a $p$-convergent kernel with $1<p<3$. For $j\geq 1$, let $G_j=\{l:2^j\leq l<2^{j+1}\}$. All degrees in $G_j$ share one set of base sketches $T^{(j)},{\cal S}^{(j)}$ for Algorithm~\ref{alg1}, namely a leaf sketch, $j$ level sketches and $j$ combination sketches, each taken with the accuracy, failure probability and dimension that the proof of Theorem~\ref{thm:SRHT+TensorSRHT} uses for degree $2^{j+1}$, accuracy $\epsilon'$ and failure probability $\delta/(3q)$.}
\label{alg:sampling}
\begin{algorithmic}[1]
\Procedure{DegreeSampling}{$X\in \R^{d\times n},(C_l)_{l\geq 0},c,p,s,\epsilon,\lambda,\delta$} \Comment{Theorem~\ref{thm:sampling_p23}}
\State Let $q$, $\beta_s$ and $\rho_s$ be as in Theorem~\ref{thm:sampling_p23}
\State Let $H$ be the vertical stack of $\sqrt{C_0}\,\mathbf{1}_n^\top$ and of $\sqrt{C_l}\,Z_l$ for $1\leq l\leq s$, computed as in the proof of Theorem~\ref{thm:general_sequence:app} with $s$ in place of $q$, accuracy $\epsilon/4$ and failure probability $\delta/3$
\If{$\rho_s<\epsilon/4$}
\State return $W=H$
\EndIf
\State Let $\gamma_s=\min\{1,\rho_s\}$, $m=\ceil{43\epsilon^{-2}\rho_s\gamma_s\ln(6n/\delta)}$ and $\epsilon'=\epsilon/(18\gamma_s)$
\State Draw $l_1,\ldots,l_m$ independently from $\{s+1,\ldots,q\}$ with $\Pr[l_i=l]=C_l/\beta_s$
\For{each distinct value $l$ among $l_1,\ldots,l_m$}
\State Let $w_l=(\beta_s/m)\cdot|\{i:l_i=l\}|$ and $Y_l={\cal Z}({\cal S}^{(j)},T^{(j)},X)$ for degree $l$, where $l\in G_j$
\EndFor
\State return $W$, the vertical stack of $H$ and of the matrices $\sqrt{w_l}\,Y_l$
\EndProcedure
\end{algorithmic}
\end{algorithm}

\begin{theorem}[Sampling scheme for $1<p<3$]\label{thm:sampling_p23}
Let $X\in \R^{d\times n}$ with radius $r\leq 1$, and let $K$ be a $p$-convergent kernel matrix with $1<p<3$ and constant $c$ (Definition~\ref{def:p_convergent_restate}). Let $\epsilon,\delta\in(0,1)$ and $\lambda>0$, and let
\begin{align*}
q=\ceil{\Big(\frac{4cn}{(p-1)\epsilon\lambda}\Big)^{1/(p-1)}}.
\end{align*}
For an integer $1\leq s\leq q$, let $\beta_s=\sum_{l=s+1}^qC_l$, $V_s=\sum_{l=s+1}^qC_l\,l^2$, $\rho_s=n\beta_s/\lambda$ and $\gamma_s=\min\{1,\rho_s\}$. Assume that $c$ and $p$ are known and that each $C_l$ can be evaluated in $O(1)$ time. Then Algorithm~\ref{alg:sampling} computes a matrix $W$ with $n$ columns such that
\begin{align*}
\pr{[(1-\epsilon)(K+\lambda I_n)\preceq W^\top W+\lambda I_n\preceq (1+\epsilon)(K+\lambda I_n)]}\geq 1-\delta,
\end{align*}
in expected time
\begin{align*}
\tilde O(nd+q+\epsilon^{-2}n^2s^3+\epsilon^{-4}\lambda^{-1}n^3\gamma_s^3V_s),
\end{align*}
where $\tilde O$ hides factors polylogarithmic in $ndq/(\epsilon\delta)$. Moreover, $\beta_s\leq c(s+1)^{1-p}/(p-1)$ and $V_s\leq c(q+1)^{3-p}/(3-p)$.
\end{theorem}

\begin{proof}
Write $A_l=(X^{\otimes l})^\top X^{\otimes l}$. Each $A_l$ is PSD, and since $r\leq 1$ its trace is $\sum_{i=1}^n\|x_i\|_2^{2l}\leq n$, so $A_l\preceq nI_n$. Let $P_s=\sum_{l=0}^sC_lA_l$, $R=\sum_{l=s+1}^qC_lA_l$ and $K_q=P_s+R$. These three matrices are PSD and at most $K$ in the Loewner order.

{\bf Tail sums.} Since $x^{-p}$ is decreasing, $(l+1)^{-p}\leq\int_l^{l+1}x^{-p}\d x$, so
\begin{align*}
\beta_s\leq c\sum_{l\geq s+1}(l+1)^{-p}\leq c\int_{s+1}^\infty x^{-p}\d x=\frac{c(s+1)^{1-p}}{p-1}.
\end{align*}
For $V_s$, we use $C_ll^2\leq cl^{2-p}$. If $p\leq 2$, then $x^{2-p}$ is nondecreasing and $\sum_{l=1}^ql^{2-p}\leq\int_1^{q+1}x^{2-p}\d x\leq\frac{(q+1)^{3-p}}{3-p}$. If $2<p<3$, then $x^{2-p}$ is decreasing and $\sum_{l=1}^ql^{2-p}\leq\int_0^qx^{2-p}\d x=\frac{q^{3-p}}{3-p}$. In both cases $V_s\leq c(q+1)^{3-p}/(3-p)$.

{\bf Truncation.} As in the proof of Theorem~\ref{thm:general_sequence:app},
\begin{align*}
\|K-K_q\|_{\op}\leq cn\sum_{l\geq q+1}(l+1)^{-p}\leq\frac{cnq^{1-p}}{p-1}\leq\frac{\epsilon\lambda}{4},
\end{align*}
where the last step uses $q^{p-1}\geq\frac{4cn}{(p-1)\epsilon\lambda}$. So $0\preceq K-K_q\preceq\frac{\epsilon\lambda}{4}I_n\preceq\frac{\epsilon}{4}(K+\lambda I_n)$.

{\bf The first $s$ degrees.} By the proof of Theorem~\ref{thm:general_sequence:app}, with $s$ in place of $q$, accuracy $\epsilon/4$ and failure probability $\delta/3$, we have $(1-\epsilon/4)P_s\preceq H^\top H\preceq(1+\epsilon/4)P_s$ with probability at least $1-\delta/3$, and $H$ takes $O(\epsilon^{-2}n^2s^3\log^3(nds/\epsilon\delta)+nd\log(nd/\epsilon\delta))$ time.

{\bf A light tail.} Suppose $\rho_s<\epsilon/4$, so that $W=H$. Then $\|R\|_{\op}\leq n\beta_s<\epsilon\lambda/4$, so $0\preceq K-P_s\preceq\frac{\epsilon\lambda}{2}I_n\preceq\frac{\epsilon}{2}(K+\lambda I_n)$. On the event above,
\begin{align*}
W^\top W+\lambda I_n\succeq(1-\epsilon/4)(P_s+\lambda I_n)\succeq(1-\epsilon/4)(1-\epsilon/2)(K+\lambda I_n)\succeq(1-\epsilon)(K+\lambda I_n),
\end{align*}
and $W^\top W+\lambda I_n\preceq(1+\epsilon/4)P_s+\lambda I_n\preceq(1+\epsilon)(K+\lambda I_n)$. From now on we assume $\rho_s\geq\epsilon/4$, so that $\beta_s>0$ and $\gamma_s\geq\epsilon/4$.

{\bf Concentration relative to the ridge.} Let $\wh R=\frac{\beta_s}{m}\sum_{i=1}^mA_{l_i}=\sum_lw_lA_l$, so that $\E[\wh R]=R$. Let $M=(R+\lambda I_n)^{-1/2}$, $U_i=\beta_sMA_{l_i}M$ and $\ov P=MRM=\E[U_i]$. Since $A_l\preceq nI_n$ and $M^2\preceq\lambda^{-1}I_n$, we have $0\preceq U_i\preceq\rho_sI_n$. Also $\ov P=R(R+\lambda I_n)^{-1}\preceq\min\{1,\|R\|_{\op}/\lambda\}I_n\preceq\gamma_sI_n$. The matrix $U_i-\ov P$ is a difference of two PSD matrices of norm at most $\rho_s$, so $\|U_i-\ov P\|_{\op}\leq\rho_s$. Using $B^2\preceq\|B\|_{\op}B$ for PSD $B$, we get
\begin{align*}
\E[(U_i-\ov P)^2]=\E[U_i^2]-\ov P^2\preceq\rho_s\E[U_i]=\rho_s\ov P\preceq\rho_s\gamma_sI_n.
\end{align*}
We apply Theorem~\ref{thm:bernstein} to $S_i=U_i-\ov P$, $i\in[m]$, so that $Z=\sum_{i=1}^mS_i$, with $L=\rho_s$, $\Var[Z]\leq m\rho_s\gamma_s$ and $t=m\epsilon/4$. Since $\epsilon/12\leq\gamma_s/3$,
\begin{align*}
\frac{t^2/2}{\Var[Z]+Lt/3}\geq\frac{m\epsilon^2/32}{\rho_s(\gamma_s+\epsilon/12)}\geq\frac{3m\epsilon^2}{128\rho_s\gamma_s}\geq\ln(6n/\delta)
\end{align*}
by the choice of $m$. So with probability at least $1-2n\cdot\frac{\delta}{6n}=1-\frac{\delta}{3}$, we have $\|\sum_{i=1}^m(U_i-\ov P)\|_{\op}<m\epsilon/4$, that is, $\|M(\wh R-R)M\|_{\op}<\epsilon/4$. Multiplying by $M^{-1}=(R+\lambda I_n)^{1/2}$ on both sides gives
\begin{align}\label{eq:sampling_bernstein}
-\frac{\epsilon}{4}(R+\lambda I_n)\preceq\wh R-R\preceq\frac{\epsilon}{4}(R+\lambda I_n).
\end{align}

{\bf Sketching the sampled degrees.} A degree $l\in G_j$ satisfies $l<2^{j+1}$, and the base sketches of $G_j$ are chosen for degree $2^{j+1}$. As in the proof of Theorem~\ref{thm:gaussian_kernel:app}, for each fixed $l$ the base sketches that produce $Y_l$ are mutually independent and at least as accurate as Theorem~\ref{thm:SRHT+TensorSRHT} requires for degree $l$, so $(1-\epsilon')A_l\preceq Y_l^\top Y_l\preceq(1+\epsilon')A_l$ with probability at least $1-\delta/(3q)$. Here $\epsilon'\leq 2/9$, since $\gamma_s\geq\epsilon/4$. By a union bound over all $l\in\{s+1,\ldots,q\}$, which needs no independence between degrees, or between the sketches and the sampled degrees, this holds for all such $l$ with probability at least $1-\delta/3$. Let $\wh S=\sum_lw_lY_l^\top Y_l$. Since $\sum_lw_lA_l=\wh R$ and the weights are nonnegative, $-\epsilon'\wh R\preceq\wh S-\wh R\preceq\epsilon'\wh R$. We claim that $\epsilon'\wh R\preceq\frac{\epsilon}{8}(K+\lambda I_n)$. By \eqref{eq:sampling_bernstein}, $\wh R\preceq\frac54R+\frac{\epsilon}{4}\lambda I_n$.
\begin{itemize}
\item If $\rho_s\geq 1$, then $\epsilon'=\epsilon/18$ and $\epsilon'\wh R\preceq\frac{5\epsilon}{72}(R+\lambda I_n)\preceq\frac{\epsilon}{8}(K+\lambda I_n)$.
\item If $\epsilon/4\leq\rho_s<1$, then $R\preceq\|R\|_{\op}I_n\preceq\rho_s\lambda I_n$ and $\epsilon\leq 4\rho_s$, so $\wh R\preceq\frac94\rho_s\lambda I_n$. Since $\epsilon'=\epsilon/(18\rho_s)$, we get $\epsilon'\wh R\preceq\frac{\epsilon}{8}\lambda I_n\preceq\frac{\epsilon}{8}(K+\lambda I_n)$.
\end{itemize}

{\bf The ridge guarantee.} The three events above hold together with probability at least $1-\delta$, and on them $W^\top W=H^\top H+\wh S$. Using $P_s+R=K_q\preceq K$,
\begin{align*}
H^\top H+\wh S&\preceq(1+\epsilon/4)P_s+\wh R+\frac{\epsilon}{8}(K+\lambda I_n)\\
&\preceq K_q+\frac{\epsilon}{4}(P_s+R+\lambda I_n)+\frac{\epsilon}{8}(K+\lambda I_n)\preceq K+\frac{3\epsilon}{8}(K+\lambda I_n),
\end{align*}
and, using also $K_q\succeq K-\frac{\epsilon}{4}(K+\lambda I_n)$,
\begin{align*}
H^\top H+\wh S&\succeq(1-\epsilon/4)P_s+\wh R-\frac{\epsilon}{8}(K+\lambda I_n)\\
&\succeq K_q-\frac{\epsilon}{4}(P_s+R+\lambda I_n)-\frac{\epsilon}{8}(K+\lambda I_n)\succeq K-\frac{5\epsilon}{8}(K+\lambda I_n).
\end{align*}
Adding $\lambda I_n$ gives the guarantee.

{\bf Running time.} Given the values $C_1,\ldots,C_q$, computing $\beta_s$ and the sampling distribution takes $O(q)$ time, and drawing the degrees takes $O(m\log q)$ time. For a group $G_j$ that contains a sampled degree, the leaf product $T^{(j)}X$ and the level outputs of Algorithm~\ref{alg1} are computed once, in $\tilde O(nd+nm_j)$ time, where $m_j=\tilde O(\epsilon'^{-2}n4^{j+1})$ is the dimension from Theorem~\ref{thm:SRHT+TensorSRHT} for degree $2^{j+1}$. After that, the combination steps for each distinct sampled degree in $G_j$ take $\tilde O(nm_j)$ time. Since $4^{j+1}\leq 4l^2$ for $l\in G_j$, charging each group and each distinct degree to one of its samples shows that the sampled part takes
\begin{align*}
\tilde O(nd)+\sum_{i=1}^m\tilde O(\epsilon'^{-2}n^2l_i^2)
\end{align*}
time, where the first term counts the $O(\log q)$ groups. Now $\E[l_i^2]=V_s/\beta_s$, $m=O(\epsilon^{-2}\rho_s\gamma_s\log(n/\delta))$, $\epsilon'^{-2}=O(\epsilon^{-2}\gamma_s^2)$ and $\rho_s/\beta_s=n/\lambda$. So the expected time of the sampled part is
\begin{align*}
\tilde O(nd)+m\cdot\tilde O(\epsilon'^{-2}n^2)\cdot\frac{V_s}{\beta_s}=\tilde O(nd+\epsilon^{-4}\lambda^{-1}n^3\gamma_s^3V_s).
\end{align*}
Adding the time for $H$ gives the bound. Writing a block with $k$ rows and $n$ columns takes $kn$ time, so $W$ has at most $1/n$ times as many rows as the running time.
\end{proof}

The best choice of $s$ depends on how fast the $C_l$ decay. The next corollary uses only the upper bounds of Theorem~\ref{thm:sampling_p23}.

\begin{corollary}[Sampling scheme for $1<p<3$ with a chosen $s$]\label{cor:sampling_p12}
Under the assumptions of Theorem~\ref{thm:sampling_p23}, let $N=n/(\epsilon\lambda)$. For a choice of $s$ that depends only on $n$, $\epsilon$, $\lambda$, $c$ and $p$, Algorithm~\ref{alg:sampling} computes a matrix $W$ with
\begin{align*}
\pr{[(1-\epsilon)(K+\lambda I_n)\preceq W^\top W+\lambda I_n\preceq (1+\epsilon)(K+\lambda I_n)]}\geq 1-\delta
\end{align*}
in expected time
\begin{align*}
\tilde O\Big(nd+\epsilon^{-2}n^2+\epsilon^{-3}n^2N^{2/(p-1)}\cdot\min\big\{1,(\epsilon^{p+2}N)^{1/p}\big\}\Big),
\end{align*}
where $\tilde O$ hides factors polylogarithmic in $ndq/(\epsilon\delta)$ and factors that depend only on $c$ and $p$. Up to such factors, this expected-time bound is never larger than the time bound of Theorem~\ref{thm:general_sequence:app}.
\end{corollary}
\begin{proof}
Let $a=c/(p-1)$ and $Q=(4aN)^{1/(p-1)}$, so that $q=\ceil{Q}$. If $Q\leq 1$, then $q=1$, and with $s=1$ the tail is empty, so Theorem~\ref{thm:sampling_p23} gives time $\tilde O(nd+\epsilon^{-2}n^2)$. Suppose now that $Q>1$, so that $q\leq 2Q$ and $q+1\leq 3Q$. In the rest of this proof, $\lesssim$ hides factors that depend only on $c$ and $p$. Theorem~\ref{thm:sampling_p23} gives $V_s\lesssim Q^{3-p}$ and $\rho_s\leq a(n/\lambda)s^{1-p}=a\epsilon Ns^{1-p}$. Since $\epsilon^{-4}\lambda^{-1}n^3=\epsilon^{-3}n^2N$, $N\lesssim Q^{p-1}$ and $\gamma_s\leq\min\{1,\rho_s\}$, the last term in the running time of Theorem~\ref{thm:sampling_p23} is
\begin{align*}
\epsilon^{-3}n^2N\gamma_s^3V_s\lesssim\epsilon^{-3}n^2Q^2\cdot\min\{1,\epsilon^3Q^{3p-3}s^{3-3p}\}.
\end{align*}
We consider two choices of $s$.
\begin{itemize}
\item With $s=1$, the running time is $\tilde O(nd+q+\epsilon^{-2}n^2+\epsilon^{-3}n^2Q^2)$, up to factors depending on $c$ and $p$, and $q\leq 2Q^2$.
\item Let $\sigma=(\epsilon^2Q^{3p-1})^{1/(3p)}$. If $\ceil{\sigma}\leq q$, take $s=\ceil{\sigma}$. Then the last term is $\lesssim n^2Q^{3p-1}\sigma^{3-3p}=\epsilon^{-2}n^2\sigma^3$, and $\epsilon^{-2}n^2s^3\leq 8\epsilon^{-2}n^2(1+\sigma^3)$. If $\ceil{\sigma}>q$, take $s=q$. Then the tail is empty, and $\epsilon^{-2}n^2s^3<\epsilon^{-2}n^2\sigma^3$. In both cases the running time is $\tilde O(nd+q+\epsilon^{-2}n^2(1+\sigma^3))$, up to factors depending on $c$ and $p$, and $q\leq 2Q\leq 2\epsilon^{-2}\sigma^3$, since $\epsilon^{-2}\sigma^3=\epsilon^{-2+2/p}Q^{3-1/p}\geq Q$.
\end{itemize}
Finally, $\epsilon^{-2}\sigma^3=\epsilon^{-3}Q^2\cdot(\epsilon^{p+2}Q^{p-1})^{1/p}$, and $Q^{p-1}=4aN$, so $Q^2\lesssim N^{2/(p-1)}$ and $(\epsilon^{p+2}Q^{p-1})^{1/p}\lesssim(\epsilon^{p+2}N)^{1/p}$. Taking the better of the two choices gives the bound. For the comparison, $\sigma^3=\epsilon^{2/p}Q^{3-1/p}\leq Q^3$, and $Q$ is at most $2^{1/(p-1)}$ times the degree used in Theorem~\ref{thm:general_sequence:app}, so the second choice is never slower than Theorem~\ref{thm:general_sequence:app}, up to factors depending on $c$ and $p$, and its main term $\epsilon^{-2}n^2\sigma^3$ is smaller than $\epsilon^{-2}n^2Q^3$ by the factor $(Q/\epsilon^2)^{1/p}$.
\end{proof}
Evaluating the truncated series directly gives the same guarantee. This computes $X^\top X$ in $O(n^2d)$ time, the entries $\sum_{l=0}^{q}C_l(x_i^\top x_j)^l$ in $O(n^2q)$ time and a factorization in $O(n^\omega)$ time, where $\omega$ is the exponent of matrix multiplication. So the bound of Corollary~\ref{cor:sampling_p12} is smaller only when $n^2d+n^\omega$ is larger than about $\epsilon^{-3}n^2N^{2/(p-1)}\min\{1,(\epsilon^{p+2}N)^{1/p}\}$, up to factors depending on $c$ and $p$ and polylogarithmic factors.
\section{Properties of the Neural Tangent Kernel}
\label{sec:ntk}
We discuss an application of Theorem~\ref{thm:general_sequence:app} to the Neural Tangent Kernel ({\ntk}). We will first formally define the {\ntk}, 
then consider its Taylor expansion, and then show that it is a $3/2$-convergent kernel on unit-norm data. Finally, we give a sampling scheme for it that samples the degrees in bands of doubling length.

\subsection{Taylor Expansion of {\ntk}}
In this section, we give the Taylor expansion of the {\ntk}, by first examining its corresponding function in a single variable, 
and then extend it to the matrix case.

Consider a simple two-layer (an alternative name is one-hidden-layer) ReLU network with input layer initialized to standard Gaussians, activation function ReLU, and output layer initialized to uniform and independent Rademacher ($\{-1,1\}$) random variables. Suppose we fix the output layer. Then the  neural network can be characterized by a function
\begin{align*}
f ( W , x )=\frac{1}{\sqrt m}\sum_{r=1}^m a_r\sigma ( w_r^\top x )
\end{align*} 
where $W \in \R^{d \times m}$, and $w_r \in \R^d$ denotes the $r$-th column of $W$, for each $r\in [m]$.

The above formulation is standard for convergence analysis of neural networks \cite{ll18,dzps19,als19a,als19b,sy19,bpsw21,hlsy21,mosw21}.

The {\ntk} (\cite{jgh18}) is
\begin{align*}
\k (x,z )=\E \Big[ \big\langle \frac{\partial f (W,x )}{\partial W},\frac{\partial f (W,z )}{\partial W} \big\rangle \Big]. 
\end{align*}
For the sake of simplicity, assume all $|a_r|=1$, $\forall r \in [m]$, and consider an individual summand, which gives rise to
\begin{align*}
\k (x,z )=\int_{w \sim N (0,I )} \sigma' (w^\top x )\sigma' (w^\top z )x^\top z~\d w. 
\end{align*}
If $w \in \R^d$ is chosen uniformly on a sphere, then for unit vectors $x,z\in\R^d$ we will get the following closed-form for this kernel \cite{cs09,xls17}:
\begin{align*}
\k (x,z )= \Big( \frac{1}{2}-\frac{\arccos x^\top z}{2\pi} \Big) \cdot x^\top z. 
\end{align*}

\begin{fact}
Let function $f : \R \rightarrow \R$ be defined as $f(x) := (\frac{1}{2}-\frac{\arccos x}{2\pi} )\cdot x$.
Then the Taylor expansion of $f$ is
\begin{align*}
f(x)& =\frac{x}{4}+\left(\sum_{n=0}^\infty \frac{(2n)!}{2^{2n}(n!)^2}\frac{x^{2n+2}}{(2n+1)(2\pi)}\right) \\
& =\frac{x}{4}+\sum_{n=0}^\infty \binom{2n}{n}\frac{1}{2^{2n}}\frac{x^{2n+2}}{(2n+1)(2\pi)}.
\end{align*}
\end{fact}

\begin{fact}
If the columns of $X$ are unit vectors, the Taylor expansion of the {\ntk} is
\begin{align*}
\k=\frac{X^\top X}{4}+\sum_{l=0}^\infty \binom{2l}{l}\frac{1}{2^{2l}}\frac{(X^{\otimes 2l+2})^\top X^{\otimes 2l+2}}{(2l+1)2\pi}.
\end{align*}
\end{fact}

\subsection{Approximating the {\ntk}}
In this section, we show that the {\ntk} on unit-norm data is $3/2$-convergent, then apply Theorem~\ref{thm:general_sequence:app} and Theorem~\ref{thm:ntk_banded} to approximate it.

\begin{corollary}[Subspace Embedding for the {\ntk} up to a Ridge Term]
\label{cor:ntk}
Let $X\in \R^{d\times n}$ have unit-norm columns, that is, $\|x_i\|_2=1$ for all $i\in[n]$. Suppose $\k\in \R^{n\times n}$ is the {\ntk} matrix. Let $\epsilon\in(0,1)$ be an accuracy parameter, $\lambda>0$ a ridge parameter and $\delta\in(0,1)$ a failure probability, and let $q=\ceil{16n^2/(\epsilon^2\lambda^2)}$. Then there exists an algorithm which computes a matrix $W_k(X)$ in time
\begin{align*}
O(\epsilon^{-2}n^2q^3\cdot\log^3(ndq/\epsilon\delta)+nd\log(nd/\epsilon\delta)),
\end{align*}
which is $\tilde O(\epsilon^{-8}\lambda^{-6}n^8+nd)$ when $\lambda\leq 4n/\epsilon$, such that
\begin{align*}
\pr{\left[(1-\epsilon)(\k+\lambda I_n)\preceq W_k(X)^\top W_k(X)+\lambda I_n\preceq (1+\epsilon)(\k+\lambda I_n)\right]}\geq 1-\delta.
\end{align*}
There also exists an algorithm, Algorithm~\ref{alg:ntk_banded}, which computes a matrix $W_k(X)$ with the same guarantee in time
\begin{align*}
\tilde O(\epsilon^{-4}\lambda^{-4}n^6+nd).
\end{align*}
\end{corollary}
\begin{proof}
Since the columns of $X$ are unit vectors, the {\ntk} matrix is given by the closed form above, and its Taylor expansion is $\k=\sum_{k=0}^\infty C_k (X^{\otimes k})^\top X^{\otimes k}$, where $C_k$ denotes the coefficient of degree $k$:
\begin{align*}
C_0=0,\quad C_1=\frac{1}{4},\quad C_k=0 \text{ for odd } k\geq 3,\quad C_{2l+2}=\binom{2l}{l}\frac{1}{2^{2l}}\frac{1}{(2l+1)2\pi} \text{ for } l\geq 0.
\end{align*}
We show that $C_k\leq (k+1)^{-3/2}$ for all $k\geq 0$.
\begin{itemize}
\item If $k=0$, or if $k\geq 3$ is odd, then $C_k=0$.
\item If $k=1$, then $C_1=\frac{1}{4}\leq 2^{-3/2}$, since $2^{3/2}=2\sqrt 2<4$.
\item If $k=2$, then $C_2=\frac{1}{2\pi}\leq 3^{-3/2}$, since $3^{3/2}=3\sqrt 3<6<2\pi$.
\item If $k=2l+2$ with $l\geq 1$, we use the following bound on the central binomial coefficient $\binom{2l}{l}$, which holds for $l\geq 1$:
\begin{align*}
\binom{2l}{l}\leq \frac{4^l}{\sqrt{3l+1}}.
\end{align*}
This gives
\begin{align*}
C_k \leq \frac{4^l}{\sqrt{3l+1}}\frac{1}{4^l}\frac{1}{(2l+1)2\pi} = \frac{1}{\sqrt{3l+1}(2l+1)2\pi}.
\end{align*}
For $l\geq 1$ we have $3(2l+3)\leq 5(2l+1)$ and $4(2l+3)\leq 5(3l+1)$, so
\begin{align*}
(k+1)^3=(2l+3)^3\leq \Big(\frac{5}{3}\Big)^2(2l+1)^2\cdot\frac{5}{4}(3l+1)=\frac{125}{36}(2l+1)^2(3l+1).
\end{align*}
Hence
\begin{align*}
C_k(k+1)^{3/2}\leq \frac{\sqrt{125/36}\,(2l+1)\sqrt{3l+1}}{\sqrt{3l+1}\,(2l+1)2\pi}=\frac{\sqrt{125/36}}{2\pi}<1,
\end{align*}
since $\sqrt{125/36}<2<2\pi$.
\end{itemize}
Thus the {\ntk} matrix is $p$-convergent (Definition~\ref{def:p_convergent_restate}) with $p=3/2$ and $c=1$. The columns of $X$ have norm $1$, so $X$ has radius $r=1$. We apply Theorem~\ref{thm:general_sequence:app} with $p=3/2$, $c=1$ and $r=1$. Since $p-1=1/2$, its degree is
\begin{align*}
q=\ceil{\Big(\frac{2n}{(1/2)\epsilon\lambda}\Big)^{2}}=\ceil{\frac{16n^2}{\epsilon^2\lambda^2}},
\end{align*}
and the theorem gives the stated guarantee in the stated time. If $\lambda\leq 4n/\epsilon$, then $\frac{16n^2}{\epsilon^2\lambda^2}\geq 1$, so $q\leq\frac{32n^2}{\epsilon^2\lambda^2}$ and $\epsilon^{-2}n^2q^3\leq 32^3\epsilon^{-8}\lambda^{-6}n^8$. The second algorithm and its bound are given by Theorem~\ref{thm:ntk_banded} below.
\end{proof}
The closed form gives $\k$ exactly in $O(n^2d)$ time, plus $O(n^\omega)$ time to factor it, where $\omega$ is the exponent of matrix multiplication (with fast rectangular matrix multiplication, $X^\top X$ takes $O(dn^{\omega-1})$ time when $d\geq n$, which raises the threshold for the second bound to about $n^{7-\omega}/(\epsilon\lambda)^4$). So, up to polylogarithmic factors, the first bound of Corollary~\ref{cor:ntk} is faster only when $d$ is larger than about $\epsilon^{-8}\lambda^{-6}n^6$ or $\epsilon^{-8}\lambda^{-6}n^8$ is below $n^\omega$, and the second only when $d$ is larger than about $\epsilon^{-4}\lambda^{-4}n^4$ or $\epsilon\lambda$ is larger than about $n^{(6-\omega)/4}$. The second alternative in each case accounts for the fact that both algorithms already output a factor. Remark~\ref{rem:ntk_jl} discusses this comparison.

\subsection{Banded Degree Sampling for the {\ntk}}\label{sec:ntk_banded}
Corollary~\ref{cor:sampling_p12} applies to the {\ntk} with $p=3/2$ and $c=1$, and gives expected time $\tilde O(nd+\epsilon^{-2}n^2+\epsilon^{-7}\lambda^{-4}n^6\min\{1,(\epsilon^{5/2}n/\lambda)^{2/3}\})$. When $\lambda\leq\epsilon^{5/2}n$, this is $\tilde O(nd+\epsilon^{-7}\lambda^{-4}n^6)$. We now improve the dependence on $\epsilon$. The degrees in $[2^j,2^{j+1})$ have coefficients that add up to at most $2\cdot 2^{-j/2}$, so the expensive high degrees carry little weight. We sample a few degrees from each such band, and sketch each sampled degree only as accurately as the weight of its band requires.

\begin{algorithm}[!ht]
\caption{Banded degree sampling for the {\ntk}, with the coefficients $C_l$ of Corollary~\ref{cor:ntk}. All degrees of band $j$ share one set of base sketches $T^{(j)},{\cal S}^{(j)}$ for Algorithm~\ref{alg1}, as in Algorithm~\ref{alg:sampling}, with the parameters that the proof of Theorem~\ref{thm:SRHT+TensorSRHT} uses for degree $2^{j+1}$, accuracy $\epsilon_j$ and failure probability $\delta/(2q)$.}
\label{alg:ntk_banded}
\begin{algorithmic}[1]
\Procedure{BandedDegreeSampling}{$X\in \R^{d\times n},\epsilon,\lambda,\delta$} \Comment{Theorem~\ref{thm:ntk_banded}}
\State Let $q$ be the least power of $2$ with $q\geq 256n^2/(\epsilon^2\lambda^2)$
\State If $q=1$, return the empty matrix
\State Let $J=\log_2 q$ and $L=\ln(4nJ/\delta)$
\For{$j=0$ to $J-1$} \Comment{Band $B_j=\{l:2^j\leq l<2^{j+1}\}$}
\State Let $\mu_j=\sum_{l\in B_j}C_l$, $\tau_j=n\mu_j/\lambda$, $\rho_j=J\tau_j/\epsilon$ and $\epsilon_j=\min\{\frac12,\frac{\epsilon}{32J\tau_j}\}$
\State Let $N_j=\ceil{2L(64\rho_j^2+8\rho_j/3)}$
\State Draw $l_1,\ldots,l_{N_j}$ independently from $B_j$ with $\Pr[l_i=l]=C_l/\mu_j$
\For{each distinct value $l$ among $l_1,\ldots,l_{N_j}$}
\State Let $w_l=(\mu_j/N_j)\cdot|\{i:l_i=l\}|$ and $Z_l={\cal Z}({\cal S}^{(j)},T^{(j)},X)$ for degree $l$ \Comment{Algorithm~\ref{alg1}}
\EndFor
\EndFor
\State return $W_k(X)$, the vertical stack of the matrices $\sqrt{w_l}\,Z_l$
\EndProcedure
\end{algorithmic}
\end{algorithm}

\begin{theorem}[Banded degree sampling for the {\ntk}]\label{thm:ntk_banded}
Let $X\in \R^{d\times n}$ have unit-norm columns, let $\k\in\R^{n\times n}$ be the {\ntk} matrix, and let $\epsilon,\delta\in(0,1)$ and $\lambda>0$. Algorithm~\ref{alg:ntk_banded} computes a matrix $W_k(X)\in\R^{m\times n}$ with $m=\tilde O(\epsilon^{-4}\lambda^{-4}n^5)$ in time
\begin{align*}
\tilde O(\epsilon^{-4}\lambda^{-4}n^6+nd)
\end{align*}
such that
\begin{align*}
\pr{\left[(1-\epsilon)(\k+\lambda I_n)\preceq W_k(X)^\top W_k(X)+\lambda I_n\preceq (1+\epsilon)(\k+\lambda I_n)\right]}\geq 1-\delta.
\end{align*}
The time bound holds for every run, and $\tilde O$ hides factors polylogarithmic in $ndq/(\epsilon\delta)$, where $q$ is the truncation degree of Algorithm~\ref{alg:ntk_banded}.
\end{theorem}

\begin{proof}
Write $G_l=(X^{\otimes l})^\top X^{\otimes l}$. Since the columns of $X$ are unit vectors, $G_l$ is PSD with unit diagonal, so $\|G_l\|_{\op}\leq\Tr[G_l]=n$. By the proof of Corollary~\ref{cor:ntk}, $\k=\sum_{l\geq 1}C_lG_l$ with $0\leq C_l\leq(l+1)^{-3/2}$. We first note a fact that we use twice. If a symmetric matrix $A$ satisfies $\|A-\k\|_{\op}\leq\theta\lambda$ with $\theta\leq\epsilon$, then $-\theta\lambda I_n\preceq A-\k\preceq\theta\lambda I_n\preceq\theta(\k+\lambda I_n)$, and so
\begin{align}\label{eq:ridge_from_additive}
(1-\epsilon)(\k+\lambda I_n)\preceq A+\lambda I_n\preceq(1+\epsilon)(\k+\lambda I_n).
\end{align}
So it suffices to show that $\|W_k(X)^\top W_k(X)-\k\|_{\op}<\epsilon\lambda/2$ with probability at least $1-\delta$.

{\bf Truncation.} Since $(l+1)^{-3/2}\leq\int_l^{l+1}x^{-3/2}\d x$, we have $\sum_{l\geq q}C_l\leq\int_q^\infty x^{-3/2}\d x=2q^{-1/2}\leq\frac{\epsilon\lambda}{8n}$ by the choice of $q$. So $\|\k-Q\|_{\op}\leq\epsilon\lambda/8$, where $Q=\sum_{l<q}C_lG_l=\sum_{j<J}R_j$ and $R_j=\sum_{l\in B_j}C_lG_l$. If $q=1$, then $Q=0=W_k(X)^\top W_k(X)$, and we are done. Every $\mu_j$ is positive, since $C_1=1/4$ and every band $B_j$ with $j\geq 1$ contains the even degree $2^j$.

{\bf Sampling within a band.} Fix $j$, and let $\wh R_j=\frac{\mu_j}{N_j}\sum_{i=1}^{N_j}G_{l_i}=\sum_{l\in B_j}w_lG_l$, so that $\E[\wh R_j]=R_j$. Let $S_i=\frac{\mu_j}{N_j}G_{l_i}-\frac{1}{N_j}R_j$. Then $\E[S_i]=0$, and $S_i$ is a difference of two PSD matrices of norm at most $n\mu_j/N_j$, so $\|S_i\|_{\op}\leq n\mu_j/N_j$. Since $G_l^2\preceq nG_l$ and $\E[G_{l_i}]=R_j/\mu_j$,
\begin{align*}
\sum_{i=1}^{N_j}\E[S_i^2]\preceq N_j\Big(\frac{\mu_j}{N_j}\Big)^2n\,\E[G_{l_i}]=\frac{n\mu_j}{N_j}R_j\preceq\frac{(n\mu_j)^2}{N_j}I_n.
\end{align*}
We apply Theorem~\ref{thm:bernstein} to $S_1,\ldots,S_{N_j}$, with uniform bound $n\mu_j/N_j$, variance at most $(n\mu_j)^2/N_j$ and $t=\epsilon\lambda/(8J)$. Then $n\mu_j/t=8\rho_j$, and
\begin{align*}
\frac{t^2/2}{(n\mu_j)^2/N_j+(n\mu_j/N_j)t/3}=\frac{N_j}{2(64\rho_j^2+8\rho_j/3)}\geq L.
\end{align*}
So $\|\wh R_j-R_j\|_{\op}<\epsilon\lambda/(8J)$, except with probability at most $2ne^{-L}=\delta/(2J)$. By a union bound, this holds for all $J$ bands with probability at least $1-\delta/2$.

{\bf Sketching.} As in the proof of Theorem~\ref{thm:sampling_p23}, for each degree $l<q$, with $l$ in band $j$, we have $(1-\epsilon_j)G_l\preceq Z_l^\top Z_l\preceq(1+\epsilon_j)G_l$ with probability at least $1-\delta/(2q)$. By a union bound, this holds for all $l<q$ with probability at least $1-\delta/2$. On this event, since the weights are nonnegative, $-\epsilon_j\wh R_j\preceq\sum_{l\in B_j}w_lZ_l^\top Z_l-\wh R_j\preceq\epsilon_j\wh R_j$. On the event of the previous step, $\|\wh R_j\|_{\op}\leq n\mu_j+\epsilon\lambda/(8J)$. Since $\epsilon_jn\mu_j\leq\epsilon\lambda/(32J)$ and $\epsilon_j\leq 1/2$, the sketching error of band $j$ is at most $\frac{\epsilon\lambda}{32J}+\frac{\epsilon\lambda}{16J}=\frac{3\epsilon\lambda}{32J}$ in operator norm.

{\bf Total error.} On both events, which hold together with probability at least $1-\delta$,
\begin{align*}
\|W_k(X)^\top W_k(X)-\k\|_{\op}\leq\frac{\epsilon\lambda}{8}+J\cdot\frac{\epsilon\lambda}{8J}+J\cdot\frac{3\epsilon\lambda}{32J}=\frac{11\epsilon\lambda}{32}<\frac{\epsilon\lambda}{2}.
\end{align*}

{\bf Running time.} Computing the $\mu_j$ takes $O(q)$ time, and each draw takes $O(\log q)$ time. Both costs are dominated by the bounds below. In band $j$, the leaf product $T^{(j)}X$ and the level outputs of Algorithm~\ref{alg1} are shared by all degrees of the band, and take $\tilde O(nd+nm_j)$ time, where $m_j=\tilde O(\epsilon_j^{-2}n4^{j+1})$ is the dimension from Theorem~\ref{thm:SRHT+TensorSRHT}. Each of the at most $N_j$ distinct degrees then takes $\tilde O(nm_j)$ time. So band $j$ takes $\tilde O(nd+N_j\epsilon_j^{-2}n^24^j)$ time, and contributes at most $N_jm_j$ rows. We have $N_j\leq 1+2L(64\rho_j^2+3\rho_j)$ and $\epsilon_j^{-2}\leq 4+1024\rho_j^2$, so $N_j\epsilon_j^{-2}=O(L(1+\rho_j^4))$. Also $\mu_j\leq\int_{2^j}^\infty x^{-3/2}\d x=2\cdot 2^{-j/2}$, so $\rho_j\leq 2Jn2^{-j/2}/(\epsilon\lambda)$ and $4^j\rho_j^4\leq 16J^4(n/(\epsilon\lambda))^4$. Hence band $j$ takes $\tilde O(nd+n^2(4^j+(n/(\epsilon\lambda))^4))$ time. Since $q\geq 2$ is the least power of $2$ that is at least $256n^2/(\epsilon\lambda)^2$, we have $q<512n^2/(\epsilon\lambda)^2$, so $\sum_{j<J}4^j<q^2=O((n/(\epsilon\lambda))^4)$. Summing over the $J$ bands, the total time is $\tilde O(nd+\epsilon^{-4}\lambda^{-4}n^6)$, and in the same way the number of rows is $\tilde O(\epsilon^{-4}\lambda^{-4}n^5)$. These bounds use only the $N_j$, so they hold for every run.
\end{proof}

The truncation degree $q=\Theta(n^2/(\epsilon\lambda)^2)$ is needed by every scheme of this kind. Sketching a degree $l$ with Algorithm~\ref{alg1} costs about $n^2l^2$ even at constant accuracy, so the bound of Theorem~\ref{thm:ntk_banded} is about the cost of sketching the single degree $q$ once. The following observation makes this precise for schemes that choose the degrees and their weights without looking at the data, in the cost model of our sketch.

\begin{observation}[Degree-based schemes for the {\ntk}]\label{obs:ntk_floor}
Consider a randomized algorithm whose output satisfies $W^\top W=\sum_{l\geq 0}w_lZ_l^\top Z_l$, where the weights $w_l\geq 0$ are random with a joint law that does not depend on $X$, and where, on an event of probability at least $1-\delta_s$ for each input, every $Z_l$ with $w_l>0$ satisfies $\frac12(X^{\otimes l})^\top X^{\otimes l}\preceq Z_l^\top Z_l\preceq\frac32(X^{\otimes l})^\top X^{\otimes l}$. Both algorithms of Corollary~\ref{cor:ntk} have this form. Let $n\geq 2$ and $n'=2\floor{n/2}$, and let $d\geq 2$, $\epsilon\leq 1/4$ and $\epsilon\lambda\leq n'/(680\pi)$, and suppose that on every input with unit-norm columns, the algorithm achieves $(1-\epsilon)(\k+\lambda I_n)\preceq W^\top W+\lambda I_n\preceq(1+\epsilon)(\k+\lambda I_n)$ with probability at least $1-\delta$. Let $q_1=\floor{\frac25(\frac{n'}{340\pi\epsilon\lambda})^2}$, so that $q_1=\Theta(n^2/(\epsilon\lambda)^2)$. Then with probability at least $1-2\delta-2\delta_s$, some $l>q_1$ has $w_l>0$. If each $Z_l$ is computed by Algorithm~\ref{alg1} with the base sketches of Theorem~\ref{thm:SRHT+TensorSRHT}, then $Z_l$ has $\Omega(nl^2)$ rows, and so the expected running time is $\Omega((1-2\delta-2\delta_s)\epsilon^{-4}\lambda^{-4}n^6)$.
\end{observation}
\begin{proof}
Let $\theta_1=17\pi\epsilon\lambda/n'$ and $\theta_2=20\theta_1$, so that $\theta_2\leq 1/2$ and $\theta_1\leq 1/40$. For $i\in\{1,2\}$, let $X_i$ have $n'/2$ columns equal to $e_1$ and $n'/2$ columns equal to $\cos\theta_i\,e_1+\sin\theta_i\,e_2$, plus one more column equal to $e_1$ when $n$ is odd, let $\k^{(i)}$ be its {\ntk} matrix and $G_l^{(i)}=(X_i^{\otimes l})^\top X_i^{\otimes l}$, and let $v\in\R^n$ have entries $1/\sqrt{n'}$ on the first group of columns, $-1/\sqrt{n'}$ on the second, and $0$ on the extra column. The entries of $G_l^{(i)}$ are $1$ within a group and $\cos^l\theta_i$ across groups, so $v^\top G_l^{(i)}v=\frac{n'}2(1-(1-u_i)^l)$ with $u_i=1-\cos\theta_i$. By the closed form, $a_i:=v^\top\k^{(i)}v=\frac{n'}2g(\theta_i)$, where $g(\theta)=\frac{1-\cos\theta}{2}+\frac{\theta\cos\theta}{2\pi}$ satisfies $\frac{\theta\cos\theta}{2\pi}\leq g(\theta)\leq\frac{\theta}{2\pi}+\frac{\theta^2}{4}$.

First, $a_1\geq\frac{n'\theta_1\cos\theta_1}{4\pi}=\frac{17}{4}\epsilon\lambda\cos\theta_1\geq 4\epsilon\lambda$, and similarly $a_2\geq\frac{n'\theta_2\cos\theta_2}{4\pi}=85\epsilon\lambda\cos\theta_2\geq 4\epsilon\lambda$. If the guarantee holds on $X_i$, then $s_i:=v^\top W^\top Wv$ satisfies $(1-\epsilon)(a_i+\lambda)-\lambda\leq s_i\leq(1+\epsilon)(a_i+\lambda)-\lambda$, and with $\epsilon\leq 1/4$ and $\epsilon\lambda\leq a_i/4$ this gives $a_i/2\leq s_i\leq 3a_i/2$.

Next, suppose that every $l$ with $w_l>0$ satisfies $l\leq q_1$, and that the sketch bounds hold on $X_i$. For $l\leq q_1$ we have $lu_i\leq q_1\theta_2^2/2\leq 1/5$. Bernoulli's inequality and $(1-u)^l\leq e^{-lu}\leq 1-lu+(lu)^2/2$ give $lu(1-lu/2)\leq 1-(1-u)^l\leq lu$, so $0.9\,lu_i\leq 1-(1-u_i)^l\leq lu_i$. With $\bar l=\sum_llw_l$, the sketch bounds then give $0.225\,n'\bar lu_i\leq s_i\leq 0.75\,n'\bar lu_i$.

If both steps apply to $X_i$, then $\bar l\in I_i:=[\frac{a_i}{1.5\,n'u_i},\frac{a_i}{0.15\,n'u_i}]$. Using $\frac{\theta^2}{2}-\frac{\theta^4}{24}\leq 1-\cos\theta\leq\frac{\theta^2}{2}$, we get $u_2/u_1\geq 400(1-\theta_2^2/12)>391$, while $a_2/a_1\leq 20(1+\pi\theta_2/2)/\cos\theta_1<36$. So $\frac{a_1/u_1}{a_2/u_2}>10$, and $I_1$ and $I_2$ are disjoint. Let $F_i$ be the event that every $l$ with $w_l>0$ satisfies $l\leq q_1$ and $\bar l\notin I_i$. It depends only on the weights, whose law does not depend on the input, and on input $X_i$ it forces a failure of the guarantee or of a sketch bound, so $\Pr[F_i]\leq\delta+\delta_s$. Since $I_1\cap I_2=\emptyset$, the event that every used degree is at most $q_1$ is contained in $F_1\cup F_2$, so its probability is at most $2\delta+2\delta_s$.

Finally, the proof of Theorem~\ref{thm:SRHT+TensorSRHT} uses base sketches with accuracy at most $1/(6l)$, so by the condition on $m$ in Lemma~\ref{thm:SRHT}, $Z_l$ has $\Omega(nl^2)$ rows and writing it down takes $\Omega(n^2l^2)$ time. Since some degree above $q_1$ is used with probability at least $1-2\delta-2\delta_s$, the expected running time is at least $(1-2\delta-2\delta_s)\cdot\Omega(n^2q_1^2)$. Since $n'\geq 2n/3$, this is $\Omega((1-2\delta-2\delta_s)\epsilon^{-4}\lambda^{-4}n^6)$.
\end{proof}

\begin{remark}[Comparison with exact evaluation]\label{rem:ntk_jl}
As noted after Corollary~\ref{cor:ntk}, Theorem~\ref{thm:ntk_banded} beats exact evaluation of the closed form only when $d$ is at least about $(n/(\epsilon\lambda))^4$, or a factor is needed and $\epsilon\lambda$ is at least about $n^{(6-\omega)/4}$, up to polylogarithmic factors. When the kernel matrix itself is the output, projecting the data to fewer dimensions and then evaluating the closed form exactly is cheaper. Let $\eta=\min\{\epsilon\lambda/(32n),1/10\}$. For $x_i\neq\pm x_j$, let $y^-_{ij}$ and $y^+_{ij}$ be the unit vectors along $x_i-x_j$ and $x_i+x_j$. By Lemma~\ref{thm:SRHT} applied to one vector at a time with failure probability $\delta/(3n^2)$, and a union bound over the at most $3n^2$ vectors $x_i$, $x_i\pm x_j$ and $y^-_{ij}\pm y^+_{ij}$, there is an {\srht} $T$ with $d'=\tilde O(\eta^{-2})=\tilde O(1+n^2/(\epsilon\lambda)^2)$ rows such that, with probability at least $1-\delta$, $(1-\eta)\|y\|_2^2\leq\|Ty\|_2^2\leq(1+\eta)\|y\|_2^2$ for each of these vectors $y$. Let $x_i'=Tx_i/\|Tx_i\|_2$ and $\k'_{i,j}=(\frac12-\frac{\arccos\langle x_i',x_j'\rangle}{2\pi})\langle x_i',x_j'\rangle$. We show below that on this event $(1-\epsilon)(\k+\lambda I_n)\preceq\k'+\lambda I_n\preceq(1+\epsilon)(\k+\lambda I_n)$. Computing $TX$ takes $O(n(d\log d+d'\log d'))$ time, and computing $\k'$ from it takes $O(n^2d')$ time. The total is $\tilde O(nd+n^2+n^4/(\epsilon\lambda)^2)$. When $\epsilon\lambda\leq n$, this is $\tilde O(nd+n^4/(\epsilon\lambda)^2)$, and its second term is smaller than the term $\epsilon^{-4}\lambda^{-4}n^6$ of Theorem~\ref{thm:ntk_banded} by a factor of $(n/(\epsilon\lambda))^2$. If a factor $W$ with $W^\top W=\k'$ is required, computing it adds $O(n^\omega)$ time, while the output of Theorem~\ref{thm:ntk_banded} is already such a factor. In that case Theorem~\ref{thm:ntk_banded} is faster when $n^\omega$ is larger than about $\epsilon^{-4}\lambda^{-4}n^6$.
\end{remark}
\begin{proof}[Proof of the guarantee in Remark~\ref{rem:ntk_jl}]
As a function of the angle $\theta\in[0,\pi]$ between two unit vectors, the {\ntk} is $k(\theta)=(\frac12-\frac{\theta}{2\pi})\cos\theta$, and $|k'(\theta)|\leq\frac{1}{2\pi}+\frac12<1$. Fix $i,j$. If $x_i=\pm x_j$, then $x_i'=\pm x_j'$ and $\k'_{i,j}=\k_{i,j}$. Otherwise, let $\theta\in(0,\pi)$ be the angle between $x_i$ and $x_j$, let $\theta'$ be the angle between $Tx_i$ and $Tx_j$, and let $\nu_i=\|Tx_i\|_2$ and $c_\pm=\|x_i\pm x_j\|_2$. Assume first that $\theta\leq\pi/2$. Since $\|x_i\|_2=\|x_j\|_2$, the vectors $y^-_{ij}$ and $y^+_{ij}$ are orthogonal, so $\|y^-_{ij}\pm y^+_{ij}\|_2^2=2$ and
\begin{align*}
|\langle Ty^-_{ij},Ty^+_{ij}\rangle|=\frac14\big|\|T(y^-_{ij}+y^+_{ij})\|_2^2-\|T(y^-_{ij}-y^+_{ij})\|_2^2\big|\leq\eta.
\end{align*}
Hence $|\nu_i^2-\nu_j^2|=|\langle T(x_i-x_j),T(x_i+x_j)\rangle|=c_-c_+|\langle Ty^-_{ij},Ty^+_{ij}\rangle|\leq 2\eta c_-$, and since $\nu_i+\nu_j\geq 2\sqrt{1-\eta}\geq 1$, also $|\nu_i-\nu_j|\leq 2\eta c_-$. By the law of cosines,
\begin{align*}
2\nu_i\nu_j(1-\cos\theta')=\|T(x_i-x_j)\|_2^2-(\nu_i-\nu_j)^2\in\big[(1-\eta-4\eta^2)c_-^2,\,(1+\eta)c_-^2\big].
\end{align*}
With $\nu_i\nu_j\in[1-\eta,1+\eta]$, $c_-^2=2(1-\cos\theta)$ and $\eta\leq 1/10$, this gives $(1-3\eta)(1-\cos\theta)\leq 1-\cos\theta'\leq(1+3\eta)(1-\cos\theta)$, so $\sin(\theta'/2)$ is within a factor $1\pm 3\eta$ of $\sin(\theta/2)$. Since $\sin(\theta/2)\leq\sin(\pi/4)$, both values lie in $[0,0.92]$, where the arcsine has derivative less than $3$. So $|\theta'-\theta|\leq 2\cdot 3\cdot 3\eta\sin(\pi/4)<14\eta$. If $\theta>\pi/2$, the same argument for $x_i$ and $-x_j$, which uses the same vectors with the roles of $x_i-x_j$ and $x_i+x_j$ exchanged, gives $|(\pi-\theta')-(\pi-\theta)|<14\eta$. In both cases $|\k'_{i,j}-\k_{i,j}|=|k(\theta')-k(\theta)|<14\eta$. So $\|\k'-\k\|_{\op}\leq\|\k'-\k\|_F\leq 14\eta n<\epsilon\lambda/2$, where the last step holds for both values of $\eta$. The guarantee follows from \eqref{eq:ridge_from_additive} with $A=\k'$.
\end{proof}
\section{Preconditioning to Solve a Kernel Linear System}\label{sec:precondition}
In this section, we illustrate how to construct a preconditioner for a kernel linear system. Specifically, we provide an algorithm to solve a Gaussian kernel linear system. Let $G=Z^\top Z$ be the Gaussian kernel. By Theorem~\ref{thm:gaussian_kernel:app}, we can compute an approximation to $G$, denoted $W_g(X)^\top W_g(X)$. In \cite{bpsw21} (see Algorithm 2 and Section 4.1 there), Brand, Peng, Song and Weinstein show that if we compute the QR decomposition of $W_g(X)=QR^{-1}$, where $Q$ has orthonormal columns and $R\in \R^{n\times n}$, then $R$ is a good preconditioner for $Z$, i.e., $ZR$ has constant condition number. However, in our setup where $d$ is large, it is not feasible to compute $Z$ directly, which takes $O(n^2d)$ time. Instead, we notice that $W_g(X)$ is fast to compute and has only an $\tilde O(n/\epsilon^2)$ number of rows. Our algorithm will sketch $W_g(X)$, and then use gradient descent to solve the optimization problem 
\begin{align*}
\underset{x\in \R^n}{\min}\|W_g(X)^\top W_g(X)x-y\|_2.
\end{align*} 
In our result, we follow a similar approach as in~\cite{bpsw21} and the proof is similar to the proof of Lemma 4.2 in their paper. The main novelty of our framework is that we use a spectral approximation to the kernel matrix and analyze the error and runtime under our approximation. For completeness, we include a proof in this setting.

\begin{algorithm}[ht]
\caption{Fast Regression for the Gaussian Kernel}
\label{alg2}
\begin{algorithmic}[1]
\Procedure{PreconditionedGradientDescent}{$X,y,\ov\kappa,\epsilon,\delta$} \Comment{Theorem~\ref{thm:precondition_main}}
\State $\wh\epsilon\gets\epsilon/8$, $\lambda\gets\wh\epsilon/\ov\kappa^2$, $\epsilon_0\gets 0.1$
\State $q\gets\Theta(1+\log(n\ov\kappa/\epsilon))$, $m \leftarrow \Theta(\wh\epsilon^{-2}nq^3\log^3(ndq/\wh\epsilon\delta))$
\State $l \leftarrow \Omega\left(n\log(mn/\epsilon_0\delta)\log(n/\delta)\right)$
\State Let $W_g(X)\in \R^{m\times n}$ be the approximate Gaussian kernel in Theorem~\ref{thm:gaussian_kernel:app} with radius $r=1$, accuracy $\wh\epsilon$, ridge parameter $\lambda$ and failure probability $\delta/2$
\State Let $S\in \R^{l/\epsilon_0^2\times m}$ be an {\srht} matrix. Compute $SW_g(X)$ 
\State Compute $R$ such that $SW_g(X)R$ has orthonormal columns via a QR decomposition \Comment{$R\in \R^{n\times n}$}
\State $z_0\gets {\bf 0}_n\in \R^n$, $t\gets\ceil{\log(6/\epsilon)/\log(10/9)}$
\For{$i=0$ to $t-1$}
\State $z_{i+1}\gets z_i-(R^\top W_g(X)^\top W_g(X)R)^\top (R^\top W_g(X)^\top W_g(X)Rz_i-R^\top y)$
\EndFor
\State return $Rz_t$
\EndProcedure
\end{algorithmic}
\end{algorithm}

\begin{theorem}[Sketching as a Preconditioner, formal version of Theorem~\ref{thm:precondition}]
\label{thm:precondition_main}
Let $G\in \R^{n\times n}$ be the Gaussian kernel matrix for $X\in \R^{d\times n}$, and assume that $\|x_i\|_2\leq 1$ for all $i\in [n]$. Write $G=Z^\top Z$, let $\kappa$ denote the condition number of $Z$, so that the condition number of $G$ is $\kappa^2$, and let $\ov\kappa\geq\kappa$ be an upper bound given as input. For $v\in\R^n$, write $\|v\|_{G^{-1}}=\sqrt{v^\top G^{-1}v}$. Let $\epsilon,\delta\in(0,1)$ and $y\in\R^n$. Then Algorithm~\ref{alg2}, with probability at least $1-\delta$, computes an $\wh x$ satisfying
\begin{align*}
\|G\wh{x}-y\|_{G^{-1}} \leq \epsilon\|y\|_{G^{-1}},
\end{align*}
and consequently $\|G\wh x-y\|_2\leq\epsilon\kappa\|y\|_2$. Moreover, $\wh{x}$ can be computed in time
\begin{align*}
\epsilon^{-2}n^2\cdot\poly(\log(nd\ov\kappa/\epsilon\delta))+n^\omega\log(mn/\delta)\log(n/\delta)+nd\log(nd/\epsilon\delta),
\end{align*}
where $\omega$ is the matrix multiplication exponent and $m$ is as in Algorithm~\ref{alg2}.
\end{theorem}

Before the proof, we define some notation for positive definite matrices.
For a positive definite matrix $B\in\R^{n\times n}$ and $v\in\R^n$, we write $\|v\|_B=\sqrt{v^\top Bv}=\|B^{1/2}v\|_2$. If $B\preceq cB'$ for positive definite matrices $B,B'$ and a scalar $c>0$, then $\|v\|_B\leq\sqrt{c}\|v\|_{B'}$ for all $v\in\R^n$, and $B'^{-1}\preceq cB^{-1}$.

We will make use of Lemma B.2 in \cite{bpsw21}.
\begin{lemma}[Lemma B.2 in \cite{bpsw21}]\label{lem:grad_desc}
Consider the regression problem: 
\begin{align*}
\underset{x\in \R^n}{\min}\|Bx-y\|_2^2. 
\end{align*}
Suppose $B$ is a PSD matrix for which $\frac{3}{4}\leq \|Bx\|_2\leq \frac{5}{4}$ holds for all $\|x\|_2=1$.
Using gradient descent for $t$ iterations, we obtain
\begin{align*}
\|B(x_t-x^*)\|_2\leq c^t \|B(x_0-x^*)\|_2, 
\end{align*}
where $x_0$ is our initial guess, $x^*$ is the optimal solution, and $c\in (0,0.9]$.
\end{lemma}

\begin{proof}[Proof of Theorem~\ref{thm:precondition_main}]
Throughout the proof, we set $\wh{\epsilon}=\epsilon/8$, $\lambda=\wh\epsilon/\ov\kappa^2$ and $\epsilon_0=0.1$, and we write $W=W_g(X)$ and $A=W^\top W$. Since $\kappa\leq\ov\kappa<\infty$, the matrix $G$ is invertible.

\paragraph{The sketched kernel.} Since $G_{i,i}=1$ for all $i\in[n]$, we have $\lambda_{\max}(G)\geq 1$. The condition number of $G$ is $\kappa^2$, so $\lambda_{\min}(G)=\lambda_{\max}(G)/\kappa^2\geq 1/\ov\kappa^2$. Therefore $\lambda\leq\wh\epsilon\lambda_{\min}(G)$, that is, $\lambda I_n\preceq\wh\epsilon G$. We apply Theorem~\ref{thm:gaussian_kernel:app} with radius $r=1$, accuracy $\wh\epsilon$, ridge parameter $\lambda$ and failure probability $\delta/2$. With probability at least $1-\delta/2$, we have $(1-\wh\epsilon)(G+\lambda I_n)\preceq A+\lambda I_n\preceq(1+\wh\epsilon)(G+\lambda I_n)$, and hence
\begin{align*}
A & \preceq (1+\wh\epsilon)G+\wh\epsilon\lambda I_n\preceq(1+\wh\epsilon+\wh\epsilon^2)G,\\
A & \succeq (1-\wh\epsilon)G-\wh\epsilon\lambda I_n\succeq(1-\wh\epsilon-\wh\epsilon^2)G.
\end{align*}
Since $\wh\epsilon\leq 1$, we have $\wh\epsilon+\wh\epsilon^2\leq 2\wh\epsilon$, so $A\approx_{2\wh\epsilon}G$. In particular, $A$ is positive definite. Since $\lambda\leq 1$, we have $\min(\lambda,1)=\lambda$ and $n/(\wh\epsilon\lambda)=n\ov\kappa^2/\wh\epsilon^2=64n\ov\kappa^2/\epsilon^2$, so the number of Taylor terms in Theorem~\ref{thm:gaussian_kernel:app} is
\begin{align*}
q=\Theta(1+\log(64n\ov\kappa^2/\epsilon^2))=\Theta(1+\log(n\ov\kappa/\epsilon)),
\end{align*}
and $q^3=O((1+\log(n\ov\kappa/\epsilon))^3)$ is polylogarithmic in $n\ov\kappa/\epsilon$. Since $(1-2\wh\epsilon)I_n\preceq G^{-1/2}AG^{-1/2}\preceq(1+2\wh\epsilon)I_n$, we can write
\begin{align*}
A=G^{1/2}(I_n+F)G^{1/2},
\end{align*}
where $F=G^{-1/2}AG^{-1/2}-I_n$ is symmetric and $\|F\|_{\op}\leq 2\wh\epsilon=\epsilon/4$.

\paragraph{The preconditioner.} By Lemma~\ref{thm:SRHT}, $S$ is an $(\epsilon_0,\delta/2,m,n)$-{\ose} (Definition~\ref{def:OSE}). So with probability at least $1-\delta/2$, for any $x\in \R^n$, we have
\begin{align*}
(1-\epsilon_0)\|Wx\|_2^2\leq\|SWx\|_2^2\leq(1+\epsilon_0)\|Wx\|_2^2.
\end{align*}
In particular, $SW$ has full column rank, so the matrix $R\in\R^{n\times n}$ computed via a QR decomposition, for which $SWR$ has orthonormal columns, is invertible. Let $M=R^\top AR$. For any $\|x\|_2=1$, we have $\|SWRx\|_2=1$, and therefore
\begin{align*}
\frac{1}{1+\epsilon_0}\leq x^\top Mx=\|WRx\|_2^2\leq\frac{1}{1-\epsilon_0}.
\end{align*}
For $\epsilon_0=0.1$, the eigenvalues of $M$ lie in $[10/11,10/9]\subseteq[3/4,5/4]$, so $\frac{3}{4}\leq\|Mx\|_2\leq\frac{5}{4}$ for all $\|x\|_2=1$, and the eigenvalues of $M^{-1}$ lie in $[1-\epsilon_0,1+\epsilon_0]=[9/10,11/10]$. Hence, for every $u\in\R^n$,
\begin{align*}
\frac{9}{10}\|u\|_2^2\leq\|u\|_{M^{-1}}^2\leq\frac{11}{10}\|u\|_2^2.
\end{align*}
Moreover, $M^{-1}=R^{-1}A^{-1}R^{-\top}$, so $A^{-1}=RM^{-1}R^\top$ and, for every $v\in\R^n$,
\begin{align*}
\|v\|_{A^{-1}}^2=v^\top RM^{-1}R^\top v=\|R^\top v\|_{M^{-1}}^2.
\end{align*}

\paragraph{Gradient descent.} The update of Algorithm~\ref{alg2} is $z_{i+1}=z_i-M^\top(Mz_i-R^\top y)$, which is gradient descent on the regression problem
\begin{align}
\label{eq:optimal_R_sol}
\underset{z\in \R^n}{\min}\|R^\top W^\top WRz-R^\top y\|_2=\underset{z\in \R^n}{\min}\|Mz-R^\top y\|_2,
\end{align}
whose optimal solution is $z^*=M^{-1}R^\top y$. Recalling that $z_0=0$, Lemma~\ref{lem:grad_desc} gives
\begin{align*}
\|Mz_t-R^\top y\|_2=\|M(z_t-z^*)\|_2\leq c^t\|M(z_0-z^*)\|_2=c^t\|R^\top y\|_2,
\end{align*}
where $c\leq 0.9$. Since $t=\ceil{\log(6/\epsilon)/\log(10/9)}$, we have $c^t\leq(9/10)^t\leq\epsilon/6$. Let $x_t=Rz_t$, so that $\wh x=x_t$. Then
\begin{align*}
\|Ax_t-y\|_{A^{-1}} & =\|R^\top(ARz_t-y)\|_{M^{-1}}=\|Mz_t-R^\top y\|_{M^{-1}}\\
& \leq\sqrt{11/10}\,\|Mz_t-R^\top y\|_2\\
& \leq\sqrt{11/10}\,c^t\|R^\top y\|_2\\
& \leq\sqrt{11/9}\,c^t\|R^\top y\|_{M^{-1}}\\
& =\sqrt{11/9}\,c^t\|y\|_{A^{-1}}.
\end{align*}
Since $(1-2\wh\epsilon)G\preceq A\preceq(1+2\wh\epsilon)G$, we have $A^{-1}\preceq(1-2\wh\epsilon)^{-1}G^{-1}$ and $G^{-1}\preceq(1+2\wh\epsilon)A^{-1}$. Using $\sqrt{11/9}\leq 6/5$ and $2\wh\epsilon=\epsilon/4\leq 1/4$, we get
\begin{align*}
\|Ax_t-y\|_{G^{-1}} & \leq\sqrt{1+2\wh\epsilon}\,\|Ax_t-y\|_{A^{-1}}\\
& \leq\frac{6}{5}\sqrt{1+2\wh\epsilon}\,c^t\|y\|_{A^{-1}}\\
& \leq\frac{6}{5}\sqrt{\frac{1+2\wh\epsilon}{1-2\wh\epsilon}}\,c^t\|y\|_{G^{-1}}\\
& \leq\frac{6}{5}\sqrt{\frac{5}{3}}\,c^t\|y\|_{G^{-1}}.
\end{align*}

\paragraph{From $A$ to $G$.} We write $Gx_t-y=GA^{-1}(Ax_t-y)+(GA^{-1}-I_n)y$. Since $\|v\|_{G^{-1}}=\|G^{-1/2}v\|_2$ and
\begin{align*}
G^{1/2}A^{-1}=G^{1/2}G^{-1/2}(I_n+F)^{-1}G^{-1/2}=(I_n+F)^{-1}G^{-1/2},
\end{align*}
we have
\begin{align*}
\|Gx_t-y\|_{G^{-1}} & \leq\|G^{1/2}A^{-1}(Ax_t-y)\|_2+\|(G^{1/2}A^{-1}-G^{-1/2})y\|_2\\
& =\|(I_n+F)^{-1}G^{-1/2}(Ax_t-y)\|_2+\|((I_n+F)^{-1}-I_n)G^{-1/2}y\|_2\\
& \leq\|(I_n+F)^{-1}\|_{\op}\cdot\|Ax_t-y\|_{G^{-1}}+\|(I_n+F)^{-1}-I_n\|_{\op}\cdot\|y\|_{G^{-1}}.
\end{align*}
Since $\|F\|_{\op}\leq 2\wh\epsilon\leq 1/4$, we have $\|(I_n+F)^{-1}\|_{\op}\leq\frac{1}{1-2\wh\epsilon}\leq\frac{4}{3}$. Since $(I_n+F)^{-1}-I_n=-(I_n+F)^{-1}F$, we also have $\|(I_n+F)^{-1}-I_n\|_{\op}\leq\frac{2\wh\epsilon}{1-2\wh\epsilon}\leq\frac{4}{3}\cdot\frac{\epsilon}{4}=\frac{\epsilon}{3}$. Putting everything together, and using $\frac{4}{3}\cdot\frac{6}{5}\sqrt{5/3}=\frac{8}{5}\sqrt{5/3}\leq 3$ and $c^t\leq\epsilon/6$, we get
\begin{align*}
\|G\wh x-y\|_{G^{-1}} & \leq 3c^t\|y\|_{G^{-1}}+\frac{\epsilon}{3}\|y\|_{G^{-1}}\\
& \leq\Big(\frac{\epsilon}{2}+\frac{\epsilon}{3}\Big)\|y\|_{G^{-1}}\\
& \leq\epsilon\|y\|_{G^{-1}}.
\end{align*}
By a union bound over the guarantees for $W$ and $S$, this holds with probability at least $1-\delta$.

For the Euclidean consequence, note that for every $v\in\R^n$, we have $\|v\|_2=\|G^{1/2}G^{-1/2}v\|_2\leq\lambda_{\max}(G)^{1/2}\|v\|_{G^{-1}}$ and $\|v\|_{G^{-1}}\leq\lambda_{\min}(G)^{-1/2}\|v\|_2$. Since $\lambda_{\max}(G)/\lambda_{\min}(G)=\kappa^2$, we get
\begin{align*}
\|G\wh x-y\|_2\leq\lambda_{\max}(G)^{1/2}\epsilon\|y\|_{G^{-1}}\leq\Big(\frac{\lambda_{\max}(G)}{\lambda_{\min}(G)}\Big)^{1/2}\epsilon\|y\|_2=\epsilon\kappa\|y\|_2.
\end{align*}

Now we analyze the runtime. Recall that $q=\Theta(1+\log(n\ov\kappa/\epsilon))$ and $\wh\epsilon=\epsilon/8$, so $m=\epsilon^{-2}n\cdot\poly(\log(nd\ov\kappa/\epsilon\delta))$.
\begin{itemize}
 \item Computing $W_g(X)$, by Theorem~\ref{thm:gaussian_kernel:app}, takes  time 
\begin{align*}
O(\wh\epsilon^{-2} n^2 q^3 \cdot \log^3(ndq/\wh\epsilon\delta)+nd\log(nd/\wh\epsilon\delta))=\epsilon^{-2} n^2 \cdot \poly (\log(nd\ov\kappa/\epsilon\delta) )+nd\log(nd/\epsilon\delta).
\end{align*}
 \item Applying $S$ to $W_g(X)$, using the FFT algorithm, takes time
\begin{align*}
O(nm\log m)=\epsilon^{-2} n^2 \cdot \poly(\log(nd\ov\kappa/\epsilon\delta) ).
\end{align*} 
 \item A QR decomposition of the $(l/\epsilon_0^2)\times n$ matrix $SW_g(X)$, due to \cite{ddh07}, takes $O((l/\epsilon_0^2)n^{\omega-1})=n^\omega\cdot O(\log(mn/\delta)\log(n/\delta))$ time.
\end{itemize}
  The cost of each iteration is bounded by the cost of computing $Mz_i=R^\top(W^\top(W(Rz_i)))$, which is $O(n^2+mn)=\epsilon^{-2}n^2\cdot\poly(\log(nd\ov\kappa/\epsilon\delta))$, and there are $t=O(1+\log(1/\epsilon))$ iterations in total. Thus, we obtain a final runtime of
\begin{align*}
\epsilon^{-2}n^2\cdot\poly(\log(nd\ov\kappa/\epsilon\delta))+n^\omega\log(mn/\delta)\log(n/\delta)+nd\log(nd/\epsilon\delta).
\end{align*}
\end{proof}

\section{Kernel Ridge Regression}
\label{sec:KRR}
In this section, we show how to compose our sketch with other sketches whose dimensions depend on the statistical dimension of $K$ instead of $n$. Before proceeding, we introduce the notion of the \emph{statistical dimension}.
\begin{definition}[Statistical Dimension]
Given $\lambda\geq 0$, for every positive semi-definite matrix $K\in \R^{n\times n}$, we define the $\lambda$-statistical dimension of $K$ to be 
\begin{align*}
s_{\lambda}(K) := & ~ \Tr[K(K+\lambda I_n)^{-1}].
\end{align*}
\end{definition}
Solving ridge regression with runtime depending on thhe statistical dimension is done in a number of works, for example~\cite{rr07,am15,akmmvz17, acw17b,mm17}. 

We state and prove our main result in this section below.
\begin{theorem}[Kernel Ridge Regression, formal version of Theorem~\ref{thm:KRR}]\label{thm:KRR:app}
Let $\epsilon\in (0,1/2]$, $\delta\in(0,1)$, $p>1$ be an integer, and $X\in \R^{d\times n}$. Let $K$ be the degree-$p$ polynomial kernel of $X$, assumed to have full rank, with statistical dimension $s_{\lambda}(K)$, where $0<\lambda<\epsilon^{-1}(1-\epsilon)\lambda_{\max}(K)$, and let
\begin{align*}
t=\tilde O(\max(\epsilon^{-2},\epsilon^{-1}\lambda_{\max}(K)/\lambda)\cdot p^2n),
\end{align*}
where $\tilde O$ also hides $\poly(\log(\lambda_{\max}(K)/\lambda))$ factors. Then we can compute $Z$ such that, with probability at least $1-\delta/2$, $Z$ has at most $t$ rows, is computed in $\tilde O(tn+nd+n^\omega)$ time, and $Z^\top Z\approx_{\epsilon'}K$ for some $\epsilon'=\Theta(\min(\epsilon,\sqrt{\epsilon\lambda/\lambda_{\max}(K)}))$.

Moreover, let $m=\tilde O(\epsilon^{-1}(s_\lambda(K)+1))$ and let $S_1,\ldots,S_r$ be $r=O(\log(1/\delta))$ independent {\srht} matrices with $m$ rows each. Let $\wh x_j$ be the optimal solution to $\|S_j(Z^\top Zx-y)\|_2^2+\lambda \|Zx\|_2^2$, and let $\wh x$ be the $\wh x_j$ with the smallest value of $\|Z^\top Zx-y\|_2^2+\lambda \|Zx\|_2^2$. Then with probability at least $1-\delta$,
\begin{align*}
\|K\wh x-y\|_2^2+\lambda\|X^{\otimes p}\wh x\|_2^2 \leq & ~ (1+\epsilon)\min_{x\in \R^n}~\|Kx-y\|_2^2+\lambda\|X^{\otimes p}x\|_2^2. 
\end{align*}

Finally, the time to compute $\wh x$ from $Z$ is $\tilde O(t(n+m^2)+n^\omega)$.
\end{theorem}

Larger $\lambda$ needs no sketch. The minimum of $\|Kx-y\|_2^2+\lambda\|X^{\otimes p}x\|_2^2$ is attained at $x=(K+\lambda I_n)^{-1}y$, where it equals $\mathrm{OPT}=\lambda y^\top(K+\lambda I_n)^{-1}y\geq\frac{\lambda}{\lambda_{\max}(K)+\lambda}\|y\|_2^2$, while the value at $x=0$ is $\|y\|_2^2$. So if $\lambda\geq\epsilon^{-1}(1-\epsilon)\lambda_{\max}(K)$ and $y\neq 0$, then $x=0$ is already $(1+2\epsilon)$-optimal, because $\|y\|_2^2/\mathrm{OPT}\leq 1+\lambda_{\max}(K)/\lambda\leq 1+\epsilon/(1-\epsilon)\leq 1+2\epsilon$.

Before starting the proof, we introduce a key lemma regarding using the {\srht} to approximate the solution of {\krr}.
\begin{lemma}[Corollary 15 of~\cite{acw17b}]
\label{lem:acw17}
Let $A\in \R^{n\times t}$ and $\epsilon\in (0,1/2]$, and write $s_\lambda(A)$ for $s_\lambda(A^\top A)$. Suppose $\lambda<\epsilon^{-1}\sigma^2_{\max}(A)$. Suppose 
\begin{align*}
m=\Omega(\epsilon^{-1}(s_{\lambda}(A)+\log(1/\epsilon))\log(s_{\lambda}(A)/\epsilon))
\end{align*}
 and $S\in \R^{m\times n}$ is a {\srht} matrix (Definition~\ref{def:SRHT}) and let $\wh x=\arg\min_{x\in \R^t}\|S(Ax-b)\|_2^2+\lambda \|x\|_2^2$. Then with probability at least $c_0$, where $c_0\in(0,1)$ is an absolute constant, we have
\begin{align*}
\|A\hat x-b\|_2^2+\lambda \|\hat x\|_2^2 \leq & ~ (1+\epsilon) \min_{x\in \R^t}~\|Ax-b\|_2^2+\lambda \|x\|_2^2.
\end{align*}
\end{lemma}

\begin{proof}[Proof of Theorem~\ref{thm:KRR:app}]
Recall that $K$ has full rank, and we write $\lambda_{\max}=\lambda_{\max}(K)$. We write
\begin{align*}
f(x)=\|Kx-y\|_2^2+\lambda x^\top Kx=\|Kx-y\|_2^2+\lambda\|X^{\otimes p}x\|_2^2
\end{align*}
for the objective and $\mathrm{OPT}=\min_{x\in\R^n}f(x)$. We run every step with accuracy $\epsilon_1=\epsilon/18$. Since $\epsilon\leq 1/2$, we have $\epsilon_1\leq 1/36$ and $\sqrt{\epsilon_1}\leq 1/6$.

The proof consists of 3 parts:
\begin{itemize}
\item Part 1: Provide a construction of matrix $Z$.
\item Part 2: Provide sketching matrices $S_j$ with the solution guarantee.
\item Part 3: Provide a runtime analysis for solving {\krr}.
\end{itemize}

\paragraph{Part 1: the construction of $Z$} We first estimate $\lambda_{\max}$ up to a factor of $3$. Let $Z_0\in\R^{m_0\times n}$ be the matrix of Theorem~\ref{thm:SRHT+TensorSRHT} with accuracy $1/4$ and failure probability $\delta/8$, so that $m_0=\tilde O(p^2n)$, and let $S_0$ be an {\srht} matrix with $O(n\log(nm_0/\delta)\log(n/\delta))$ rows, which is a $(1/4,\delta/8,m_0,n)$-{\ose} by Lemma~\ref{thm:SRHT}. With probability at least $1-\delta/4$, the matrix $B=(S_0Z_0)^\top S_0Z_0$ satisfies $\frac{9}{16}K\preceq B\preceq\frac{25}{16}K$, so $\Lambda=\frac{16}{9}\lambda_{\max}(B)$ satisfies $\lambda_{\max}\leq\Lambda\leq\frac{25}{9}\lambda_{\max}\leq 3\lambda_{\max}$. Computing $Z_0$ takes $\tilde O(p^2n^2+nd)$ time by Theorem~\ref{thm:SRHT+TensorSRHT}, applying $S_0$ to the $n$ columns of $Z_0$ takes $\tilde O(m_0n)=\tilde O(p^2n^2)$ time, and computing $B$ and its top eigenvalue \cite{ddh07} takes $\tilde O(n^\omega)$ time.

Let $\epsilon'=\min(\epsilon_1,\sqrt{\epsilon_1\lambda/\Lambda})$, and let $Z$ be the matrix of Theorem~\ref{thm:SRHT+TensorSRHT} with accuracy $\epsilon'$ and failure probability $\delta/4$. Then
\begin{align*}
\epsilon'^{-2}=\max(\epsilon_1^{-2},\Lambda/(\epsilon_1\lambda))\leq\max(324\epsilon^{-2},54\epsilon^{-1}\lambda_{\max}/\lambda),
\end{align*}
so $Z$ has $t=\tilde O(\epsilon'^{-2}p^2n)=\tilde O(\max(\epsilon^{-2},\epsilon^{-1}\lambda_{\max}/\lambda)p^2n)$ rows and is computed in $\tilde O(\epsilon'^{-2}p^2n^2+nd)=\tilde O(tn+nd)$ time. Since $\lambda_{\max}\leq\Lambda\leq 3\lambda_{\max}$ and $1/\sqrt{54}\geq 1/18$, we have $\frac{1}{18}\min(\epsilon,\sqrt{\epsilon\lambda/\lambda_{\max}})\leq\epsilon'\leq\min(\epsilon,\sqrt{\epsilon\lambda/\lambda_{\max}})$, so $\epsilon'=\Theta(\min(\epsilon,\sqrt{\epsilon\lambda/\lambda_{\max}}))$. With probability at least $1-\delta/2$, both $\Lambda$ and $Z$ have these guarantees (the bounds on the number of rows and on the running time hold on the event $\Lambda\leq 3\lambda_{\max}$), and in particular $A:=Z^\top Z\approx_{\epsilon'}K$. Since $K$ has full rank, so do $A$ and $Z$. Since $\Lambda\geq\lambda_{\max}$, we have
\begin{align}\label{eq:krr_key}
\epsilon'^2\lambda_{\max}\leq\epsilon_1\lambda\lambda_{\max}/\Lambda\leq\epsilon_1\lambda.
\end{align}
We write $A=K^{1/2}(I_n+F)K^{1/2}$, where $F=K^{-1/2}AK^{-1/2}-I_n$ is symmetric and $\|F\|_{\op}\leq\epsilon'\leq\epsilon_1$. By \eqref{eq:krr_key}, for every $u\in\R^n$,
\begin{align}\label{eq:krr_F}
\|K^{1/2}Fu\|_2^2\leq\lambda_{\max}\epsilon'^2\|u\|_2^2\leq\epsilon_1\lambda\|u\|_2^2.
\end{align}

\paragraph{Part 2: the solution guarantee} We define the surrogate objective
\begin{align*}
f_A(x)=\|Ax-y\|_2^2+\lambda x^\top Ax=\|Z^\top Zx-y\|_2^2+\lambda\|Zx\|_2^2,
\end{align*}
and $\mathrm{OPT}_A=\min_{x\in\R^n}f_A(x)$. Let $H=K^2+\lambda K$ and $H_A=A^2+\lambda A$, which are positive definite, and write $\|v\|_H=\sqrt{v^\top Hv}$ and $\|v\|_{H_A}=\sqrt{v^\top H_Av}$. Let $x_K=(K+\lambda I_n)^{-1}y$ and $x_A=(A+\lambda I_n)^{-1}y$. The function $f$ is quadratic with Hessian $2H$, and its gradient $2Hx-2Ky$ vanishes at $x_K$, since $Hx_K=K(K+\lambda I_n)x_K=Ky$. So $x_K$ minimizes $f$, and in the same way $x_A$ minimizes $f_A$. Hence, for every $x\in\R^n$,
\begin{align*}
f(x)-\mathrm{OPT}=\|x-x_K\|_H^2,\qquad f_A(x)-\mathrm{OPT}_A=\|x-x_A\|_{H_A}^2.
\end{align*}
Since $\mathrm{OPT}=\|Kx_K-y\|_2^2+\lambda x_K^\top Kx_K$ and $\mathrm{OPT}_A=\|Ax_A-y\|_2^2+\lambda x_A^\top Ax_A$, we have $\lambda x_K^\top Kx_K\leq\mathrm{OPT}$ and $\lambda x_A^\top Ax_A\leq\mathrm{OPT}_A$. We compare $x_K$, $x_A$ and $\wh x$ in five steps.

\paragraph{(i) The surrogate optimum} Let $a=\|Kx_K-y\|_2$ and $b=(\lambda x_K^\top Kx_K)^{1/2}$, so that $a^2+b^2=\mathrm{OPT}$, and let $u=K^{1/2}x_K$. By \eqref{eq:krr_F}, $\|(A-K)x_K\|_2=\|K^{1/2}Fu\|_2\leq\sqrt{\epsilon_1}\,b$. Hence $\|Ax_K-y\|_2\leq a+\sqrt{\epsilon_1}\,b$, and
\begin{align*}
\|Ax_K-y\|_2^2\leq a^2+2\sqrt{\epsilon_1}\,ab+\epsilon_1b^2\leq a^2+\sqrt{\epsilon_1}(a^2+b^2)+\epsilon_1b^2.
\end{align*}
Also, $\lambda x_K^\top Ax_K=\lambda u^\top(I_n+F)u\leq(1+\epsilon_1)b^2$. Adding the two bounds and using $\epsilon_1\leq\sqrt{\epsilon_1}\leq 1/6$, we get
\begin{align}\label{eq:krr_i}
\mathrm{OPT}_A\leq f_A(x_K)\leq(1+\sqrt{\epsilon_1}+2\epsilon_1)\mathrm{OPT}\leq(1+3\sqrt{\epsilon_1})\mathrm{OPT}\leq\frac{3}{2}\mathrm{OPT}.
\end{align}

\paragraph{(ii) The surrogate minimizer} Since $(K+\lambda I_n)x_K=y=(A+\lambda I_n)x_A$, we have $(K+\lambda I_n)(x_A-x_K)=(K-A)x_A$. Let $w=(K-A)x_A$. Then $x_A-x_K=(K+\lambda I_n)^{-1}w$, and since $K$ and $(K+\lambda I_n)^{-1}$ commute,
\begin{align*}
f(x_A)-\mathrm{OPT}=w^\top(K+\lambda I_n)^{-1}K(K+\lambda I_n)(K+\lambda I_n)^{-1}w=w^\top K(K+\lambda I_n)^{-1}w\leq\|w\|_2^2,
\end{align*}
because the eigenvalues of $K(K+\lambda I_n)^{-1}$ lie in $[0,1)$. By \eqref{eq:krr_F} with $u=K^{1/2}x_A$, we have $\|w\|_2^2\leq\epsilon_1\lambda x_A^\top Kx_A$. Since $A\succeq(1-\epsilon')K$ and $\epsilon'\leq 1/2$, we have $\lambda x_A^\top Kx_A\leq\lambda x_A^\top Ax_A/(1-\epsilon')\leq 2\mathrm{OPT}_A$. Therefore,
\begin{align}\label{eq:krr_ii}
\|x_A-x_K\|_H^2=f(x_A)-\mathrm{OPT}\leq 2\epsilon_1\mathrm{OPT}_A.
\end{align}

\paragraph{(iii) Comparing $H_A$ and $H$} Let $v\in\R^n$, $u=K^{1/2}v$, $\alpha=\|Kv\|_2$, $\beta=(\lambda v^\top Kv)^{1/2}=\sqrt{\lambda}\|u\|_2$ and $h=v^\top Hv=\alpha^2+\beta^2$. We have $Av=K^{1/2}(I_n+F)u=Kv+K^{1/2}Fu$, and by \eqref{eq:krr_F}, $\|K^{1/2}Fu\|_2^2\leq\epsilon_1\beta^2\leq\epsilon_1h$. So $\big|\|Av\|_2-\alpha\big|\leq\sqrt{\epsilon_1h}$, and since $\alpha\leq\sqrt h$,
\begin{align*}
\alpha^2-2\sqrt{\epsilon_1}\,h\leq\|Av\|_2^2\leq\alpha^2+2\sqrt{\epsilon_1}\,h+\epsilon_1h.
\end{align*}
For the lower bound, if $\alpha<\sqrt{\epsilon_1h}$, then the left side is negative. Otherwise $\|Av\|_2^2\geq(\alpha-\sqrt{\epsilon_1h})^2\geq\alpha^2-2\alpha\sqrt{\epsilon_1h}$. Also, $(1-\epsilon_1)\beta^2\leq\lambda v^\top Av=\lambda u^\top(I_n+F)u\leq(1+\epsilon_1)\beta^2$. Adding, and using $\epsilon_1\leq\sqrt{\epsilon_1}\leq 1/6$, we get
\begin{align}\label{eq:krr_iii}
\frac{1}{2}v^\top Hv\leq(1-3\sqrt{\epsilon_1})v^\top Hv\leq v^\top H_Av\leq(1+4\sqrt{\epsilon_1})v^\top Hv.
\end{align}

\paragraph{(iv) From $\|S(Z^\top Zx-y)\|_2^2+\lambda \|Zx\|_2^2$ to $\|Z^\top Zx-y\|_2^2+\lambda \|Zx\|_2^2$} For $z\in\R^t$, let $g_j(z)=\|S_j(Z^\top z-y)\|_2^2+\lambda\|z\|_2^2$ and $g(z)=\|Z^\top z-y\|_2^2+\lambda\|z\|_2^2$. Recall that $Z$ has full rank. Every $z\in\R^t$ can be written as $z=Zx+z'$ with $Z^\top z'=0$, and then $g_j(z)=g_j(Zx)+\lambda\|z'\|_2^2$ and $g(z)=g(Zx)+\lambda\|z'\|_2^2$. So the minimizers of $g_j$ and $g$ lie in the column span of $Z$, the minimizer of $g_j$ is $Z\wh x_j$, and $\min_z g(z)=\min_x g(Zx)=\mathrm{OPT}_A$, since $g(Zx)=f_A(x)$. Thus each sketched problem is the problem of Lemma~\ref{lem:acw17} with $Z^\top$ in place of $A$, $b=y$ and accuracy $\epsilon_1$, and we check the conditions of the lemma. First, $s_\lambda(Z^\top)=s_\lambda(ZZ^\top)=s_\lambda(Z^\top Z)$, since $ZZ^\top$ and $Z^\top Z$ have the same nonzero eigenvalues, and the statistical dimension of $A=Z^\top Z$ gives a good approximation to the statistical dimension of $K$:
\begin{align*}
s_\lambda(A) = & ~ \sum_{i=1}^n \frac{\lambda_i(A)}{\lambda_i(A)+\lambda} \\
\leq & ~ \sum_{i=1}^n \frac{(1+\epsilon')\lambda_i(K)}{(1-\epsilon')\lambda_i(K)+\lambda} \\
\leq & ~ \sum_{i=1}^n \frac{(1+\epsilon')\lambda_i(K)}{(1-\epsilon')(\lambda_i(K)+\lambda)}  \\
= & ~ \frac{1+\epsilon'}{1-\epsilon'}\cdot s_\lambda(K) \\
\leq & ~ (1+3\epsilon')\cdot s_\lambda(K),
\end{align*}
where the last step uses $\epsilon'\leq 1/3$. Since $s:=s_\lambda(A)\leq 2s_\lambda(K)\leq 2n$, we have
\begin{align*}
\epsilon_1^{-1}(s+\log(1/\epsilon_1))\log(s/\epsilon_1)\leq\epsilon_1^{-1}(s+1)\log(e/\epsilon_1)\log(2n/\epsilon_1),
\end{align*}
so $m=\tilde O(\epsilon^{-1}(s_\lambda(K)+1))$ rows suffice. Second, $\sigma_{\max}^2(Z^\top)=\lambda_{\max}(A)\geq(1-\epsilon')\lambda_{\max}\geq(1-\epsilon)\lambda_{\max}$, so
\begin{align*}
\lambda<\epsilon^{-1}(1-\epsilon)\lambda_{\max}\leq\epsilon^{-1}\sigma_{\max}^2(Z^\top)\leq\epsilon_1^{-1}\sigma_{\max}^2(Z^\top),
\end{align*}
and $\epsilon_1\leq 1/2$. By Lemma~\ref{lem:acw17}, for each $j$, conditioned on $Z$, with probability at least $c_0$ we have $f_A(\wh x_j)=g(Z\wh x_j)\leq(1+\epsilon_1)\mathrm{OPT}_A$. The $S_j$ are independent, so with $r=\ceil{\log(4/\delta)/\log(1/(1-c_0))}=O(\log(1/\delta))$, some $j$ has this property with probability at least $1-(1-c_0)^r\geq 1-\delta/4$. Since $\wh x$ has the smallest value of $f_A$ among the $\wh x_j$, we get $f_A(\wh x)\leq(1+\epsilon_1)\mathrm{OPT}_A$, that is,
\begin{align}\label{eq:krr_iv}
\|\wh x-x_A\|_{H_A}^2=f_A(\wh x)-\mathrm{OPT}_A\leq\epsilon_1\mathrm{OPT}_A.
\end{align}

\paragraph{(v) Putting everything together} By the triangle inequality for $\|\cdot\|_H$, then \eqref{eq:krr_iii}, \eqref{eq:krr_iv} and \eqref{eq:krr_ii}, and finally \eqref{eq:krr_i},
\begin{align*}
f(\wh x)-\mathrm{OPT}=\|\wh x-x_K\|_H^2 \leq & ~ 2\|\wh x-x_A\|_H^2+2\|x_A-x_K\|_H^2 \\
\leq & ~ 4\|\wh x-x_A\|_{H_A}^2+2\|x_A-x_K\|_H^2 \\
\leq & ~ 4\epsilon_1\mathrm{OPT}_A+4\epsilon_1\mathrm{OPT}_A \\
\leq & ~ 12\epsilon_1\mathrm{OPT}=\frac{2}{3}\epsilon\,\mathrm{OPT}.
\end{align*}
So $f(\wh x)\leq(1+\epsilon)\mathrm{OPT}$. By a union bound over parts 1 and 2, this holds with probability at least $1-\delta$. This completes our proof for part 2.

For the final part, part 1 already shows that $\Lambda$ and $Z$ take $\tilde O(p^2n^2+nd+n^\omega)$ and $\tilde O(tn+nd)$ time, which is $\tilde O(tn+nd+n^\omega)$ in total since $t\geq p^2n$. For each $j$, applying $S_j$ to $Z^\top$ takes $\tilde O(tn)$ time. To solve the regression problem, we instead solve:
\begin{align*}
\min_{z\in \R^t}~\|S_jZ^\top z-S_jy\|_2^2+\lambda \|z\|_2^2.
\end{align*}
By part (iv), the argument $z$ realizing the minimum is $Z\wh x_j$. To output an $x$, we can simply solve the linear system $Zx=z$, which takes $\tilde O(nt+n^\omega)$ time. Solving the above regression problem takes $\tilde O(m^2t)$ time~(see \cite{sgv98}). Evaluating $f_A(\wh x_j)=\|Z^\top(Z\wh x_j)-y\|_2^2+\lambda\|Z\wh x_j\|_2^2$ takes $O(tn)$ time. There are $r=O(\log(1/\delta))$ repetitions, and this factor goes inside $\tilde O$, so computing $\wh x$ from $Z$ takes $\tilde O(t(n+m^2)+n^\omega)$ time. This concludes our runtime analysis.
\end{proof}

\end{document}